\documentclass[11pt,a4paper]{article}

\usepackage{jcappub}
\usepackage{slashed}

\usepackage{multirow}
\usepackage{verbatim}
\usepackage{subcaption}

\newcommand{\fett}[1]{\boldsymbol{#1}}

\newcommand{\dd}{{\rm{d}}}
\newcommand{\ii}{{\rm{i}}}

\newcommand{\be}{\begin{equation}}
\newcommand{\ee}{\end{equation}}
\newcommand{\e}{{\rm e}}
\newcommand{\LP}{{\rm P}}
\newcommand{\D}{{\rm D}}

\definecolor{darkred}{rgb}{0.5,0,0}
\definecolor{darkgreen}{rgb}{0,0.5,0}
\definecolor{darkblue}{rgb}{0,0,0.5}

\usepackage{hyperref}
\hypersetup{ colorlinks,
linkcolor=darkblue,
filecolor=darkgreen,
urlcolor=darkred,
citecolor=darkblue }

\newcommand{\myArxiv}[1]{{\tt \href{http://arxiv.org/abs/#1}{{\color{black}[{\color{blue}#1}]}}} \href{http://inspirehep.net/search?p=find+EPRINT+#1}
 {{\color{black}[{\color{blue} {\small in}SPIRE}]}}}
\newcommand{\inspire}[1]{\href{http://inspirehep.net/search?p=find+J+#1}
 {{\color{black}[{\color{blue} {\small in}SPIRE}]}}}
\newcommand{\book}[1]{\href{http://inspirehep.net/search?p=#1}
 {{\color{black}[{\color{blue} {\small in}SPIRE}]}}}
\newcommand{\DOI}[2]{ 
 \href{http://dx.doi.org/#1}{\color{blue} #2}
}
\newcommand{\inspired}[1]{\href{http://inspirehep.net/search?p=#1}
 {{\color{black}[{\color{blue} {\small in}SPIRE}]}}}

\begin{document}

\title{Boltzmann hierarchy for interacting neutrinos I: formalism}

\date{\today}

\author[a,b]{Isabel M.~Oldengott,}
\emailAdd{ioldengott@physik.uni-bielefeld.de}

\author[c]{Cornelius Rampf}
\emailAdd{cornelius.rampf@port.ac.uk}

\author[d]{and Yvonne Y.~Y.~Wong}
\emailAdd{yvonne.y.wong@unsw.edu.au}

\affiliation[a]{Fakult\"at f\"ur Physik, Bielefeld University, D--33501 Bielefeld, Germany}
\affiliation[b]{Institut f\"ur Theoretische Teilchenphysik und Kosmologie, RWTH Aachen, D--52056 Aachen, Germany}
\affiliation[c]{Institute of Cosmology and Gravitation, University of Portsmouth, Portsmouth PO1 3FX, United Kingdom}
\affiliation[d]{School of Physics, The University of New South Wales, Sydney NSW 2015, Australia}

\abstract{Starting from the collisional Boltzmann equation, we derive for the first time and from first principles the Boltzmann hierarchy for neutrinos including interactions with a scalar particle. Such interactions appear, for example, in majoron-like models of neutrino mass generation. We study two limits of the scalar mass: (i)~An extremely massive scalar whose only role is to mediate an effective 4-fermion neutrino--neutrino interaction, and  (ii)~a massless scalar that can be produced in abundance and thus
demands its own Boltzmann hierarchy.
 In contrast to, e.g., the first-order Boltzmann hierarchy for Thomson-scattering photons, our interacting neutrino/scalar Boltzmann hierarchies contain additional momentum-dependent collision terms arising from a non-negligible energy transfer in the neutrino--neutrino and neutrino--scalar interactions.  This necessitates that we track each momentum mode of the phase space distributions individually, even if the particles were massless.
Comparing our hierarchy with the commonly used  $(c_{\rm eff}^2,c_{\rm vis}^2)$-parameterisation, we find no formal correspondence between the two approaches, which 
raises the question of whether the latter parameterisation even has an interpretation in terms of particle scattering.
Lastly, although we have invoked majoron-like models as a motivation for our study, our treatment is in fact generally applicable to all scenarios in which the neutrino and/or
other ultrarelativistic fermions interact with scalar particles.}

\maketitle   

\flushbottom
\section{Introduction}
\label{Introduction}

The nature of neutrinos and especially the mechanism of neutrino mass generation are some of the last unsolved puzzles in particle physics.  Neutrinos are exactly massless in the standard model (SM) of particle physics only as a consequence of the assumption that right-handed neutrinos do not exist.
But the observation of flavour oscillations in neutrinos from astrophysical as well as man-made sources has now unequivocally established that neutrinos are massive particles (see, e.g.,~\cite{Tortola:2012te} for a review).  This automatically implies that the SM is incomplete, and that any new physics introduced to account for neutrino masses will necessarily involve particles as yet unobserved and interactions as yet unmeasured.

Many different extensions of the SM have been proposed over the years to explain the origins of neutrino masses (see, e.g.,~\cite{Altarelli:2004za} for a review). 
One promising direction are majoron-like models, which give rise to neutrino masses through the spontaneous breaking of a global $U(1)_{B-L}$ symmetry~\cite{Chikashige:1980ui,Georgi:1981pg,Gelmini:1980re,Choi:1991aa,Acker:1992eh}. 
These models have the unique feature that the symmetry breaking is necessarily accompanied by the appearance of one or more new Goldstone bosons, the majorons, which couple primarily to neutrinos.  New interactions induced by this coupling have  interesting consequences for neutrinos of astrophysical origin (e.g.,~\cite{Ng:2014pca,Ioka:2014kca}), big bang nucleosynthesis (e.g.,~\cite{Aarssen:2012fx}),  neutrinoless double $\beta$-decay (e.g.,~\cite{Gando:2012pj}), and the decay widths of the $Z$ boson and certain mesons~(e.g., \cite{Laha:2013xua,Lessa:2007up,Abe:2013hdq})  alike.

On a different front, observations of the cosmic microwave background (CMB) an-isotropies have been providing useful insights into neutrino physics in the past decade
(see, e.g.,~\cite{Hannestad:2010kz,Wong:2011ip,Lesgourgues:2014zoa} for a review).  Notably, the most recent measurements of the temperature and polarisation fluctuations by the Planck mission have now constrained the neutrino mass sum to  $\sum m_{\nu}\lesssim 0.59$~eV (95\% C.L.),
 which can be further tightened  to $\sum m_{\nu}\lesssim 0.23$~eV using in addition non-CMB cosmological observations~\cite{Planck:2015xua}.  The same data combinations have likewise constrained the effective number of neutrino species to $N_{\rm eff} = 2.99\pm 0.20$ and $3.04 \pm 0.18$  (68\% C.I.) respectively~\cite{Planck:2015xua}.  Given that the CMB anisotropies are sensitive to these neutrino properties primarily because at the time of photon decoupling ($T \sim 0.3$~eV) neutrinos make up at least 10\% of the universe's energy content,
the question naturally arises as to whether we can also test new neutrino interactions using cosmological observations.

A number of recent works have attempted to address this question---some of which in direct relation to the majoron-like models discussed above---using a variety of heuristic methods to approximate the behaviour of interacting neutrinos around the epoch of photon decoupling, e.g.,~\cite{Hannestad:2004qu,Bell:2005dr,Friedland:2007vv,Basboll:2008fx,Cyr-Racine:2013jua,Archidiacono:2011gq,Smith:2011es,Diamanti:2012tg}.  We shall discuss these various approaches in detail in section~\ref{State}, along with their merits and shortcomings.  Suffice it to say here, however, that some aspects of these approximate treatments may have been too simplistic as to be entirely satisfactory.  Our take on the problem, therefore, is to go back to square one, and derive from first principles the Boltzmann hierarchy for interacting neutrinos by 
systematically folding into the Boltzmann equation the new neutrino scattering processes through the collision integral.  Such a first-principles derivation has, to our knowledge, never been undertaken in the literature.

For concreteness we focus on new neutrino scalar and pseudo-scalar interactions described by the Lagrangian density
\begin{align}
\mathcal{L}_{\rm int} &= \mathfrak{g}_{ij} \bar{\nu}_i \nu_j \phi +\mathfrak{h}_{ij} \bar{\nu}_i \gamma_5 \nu_j \phi  \,, \label{Lagrangian}
\end{align}
where $\mathfrak{g}_{ij}$ and $\mathfrak{h}_{ij}$ denote  the scalar and pseudo-scalar coupling constants respectively.
Simplicity aside, interactions of this form are also easily motivated from majoron-like models of neutrino mass generation, where~$\phi$ is identified with the majoron.
We implicitly assume that neutrinos are Majorana particles, since any interaction of Dirac neutrinos of the form~(\ref{Lagrangian}) will couple a left-handed to a (non-SM) right-handed state, where the abundance of the latter is strongly dependent on what other interactions it has besides~(\ref{Lagrangian}), which is beyond the scope of this pilot study.

Lastly, we emphasise that although we have motivated the study of new neutrino interactions by way of majoron-like models of neutrino mass generation, our investigation here is by no means restricted to majoron-like models; any scenario in which the neutrino and/or other ultrarelativistic fermions possess interactions of the form~(\ref{Lagrangian}), e.g.,~\cite{Chacko:2004cz}, is amenable to our treatment.  

The rest of the paper is organised as follows. In section~\ref{State} we briefly review and critique the state of the art in the treatment of neutrino interactions in the context of CMB anisotropies.
We introduce the collisional Boltzmann equation and our approach in section~\ref{Assumptions}, and discuss in section~\ref{sec:twolimits} the two limiting cases to be investigated in this work.
All relevant collision integrals for the neutrino--scalar system are computed and presented in section~\ref{sec:collisionterms}, while in section~\ref{sec:hierarchy} we derive the corresponding Boltzmann hierarchies.  We conclude in section~\ref{Summary}.  For those readers interested in the details of our derivations, three appendices following the main text have been included to serve that purpose.

\section{Interacting neutrinos in cosmology: state of the art}
\label{State}

Previous studies of new neutrino interactions and their effects on the CMB anisotropies model the phenomenology of the interactions following this argument: 
(i)~Ultrarelativistic neutrinos that interact at a per-particle rate~$\Gamma$ much smaller than the universe's expansion rate~$H$ are effectively non-interacting. These neutrinos free-stream, and generate anisotropies in the neutrino fluid elements, i.e., nonzero $\ell \geq 2$ moments in the neutrino Boltzmann hierarchy. This is the default behaviour of weakly-interacting SM neutrinos after $T \sim 1$~MeV (i.e., neutrino deocupling).
(ii)~If conversely the new interaction rate far exceeds the expansion rate, then the interaction will isotropise the fluid elements.  The $\ell \geq 2$ moments vanish in this so-called tightly-coupled limit, leaving the neutrino monopole and dipole moments to undergo acoustic oscillations.  The total absence of free-streaming damping in this limit also enhances the  monopole and dipole's contribution as a gravitational source, which in turn amplifies the acoustic oscillations in the photon--baryon fluid and ultimately the acoustic peaks ($\ell_{\rm CMB} \geq 200$) in the CMB temperature anisotropy spectrum.

Early investigations on the subject, e.g.,~\cite{Hannestad:2004qu,Bell:2005dr,Friedland:2007vv},  typically assume that the tightly-coupled limit
holds from the time the first observable length scale enters the horizon until well into the matter-domination era.  Such a treatment is certainly 
valid if the neutrinos interact at a sufficiently large rate, but clearly makes no provision for those intermediate scenarios in which $\Gamma$ should be comparable or below $H$ within the specified timeframe.  Three lines of strategies have emerged to deal with this possibility, which we outline below.

The first approach, pursued in~\cite{Basboll:2008fx}, finds its motivation in the same majoron-like models of neutrino mass generation discussed in the introduction. In this scenario, the neutrino self-interactions are mediated by a massless scalar so that $\Gamma \sim \mathfrak{g} T$.
Thus, the interaction rate is initially negligible, but can eventually overtake the expansion rate $H \sim T^2/m_{\rm Pl}$ (assuming radiation domination) as the temperature of the universe drops, causing the neutrinos to recouple. This prompts the authors of~\cite{Basboll:2008fx} to model the neutrino fluid as initially free-streaming, but switching instantaneously to the tightly-coupled limit at the moment $\Gamma=H$ is satisfied. The search for this new interaction in the CMB data then consists in looking for such a behavioural switch, the time at which it takes place, and the coupling strength to which the switching time maps. Obviously, this approach assumes that, after the switch, the interaction will instantaneously isotropise the neutrino fluid elements, which in reality is most likely not the case.

The second approach, adopted in~\cite{Cyr-Racine:2013jua}, moderates the amount of free-streaming damping through the introduction,
 for all $\ell \geq 2$ moments in the neutrino Boltzmann hierarchy, of a damping term 
proportional to the opacity~$\dot{\tau}_\nu \equiv -a G_{\rm eff}^2 T_\nu^5 \sim -a \Gamma$, where $a$ is the scale factor, $G_{\rm eff}$ the effective self-coupling constant, and $T_\nu$ the neutrino temperature. The choice of $\Gamma \sim G_{\rm eff}^2 T_\nu^5$ carries the implicit assumption that the mass of the mediating particle is much larger than the average energy of the neutrinos, and 
the self-interaction rate is initially large compared with the expansion rate~$H$.  Then, the role of the self-interaction is simply to postpone the neutrino decoupling epoch from the canonical  $T_\nu \sim 1$~MeV to a later time.  Once the neutrinos decouple, they free-stream to infinity, and the self-interaction does not provide for the possibility of recoupling (in contrast with the case of a massless mediator discussed immediately above).

Written out explicitly in standard notation~\cite{Ma:1995ey} in the synchronous gauge, the modified $\ell \geq 2$ evolution equations  proposed by~\cite{Cyr-Racine:2013jua} are
\begin{equation}
\begin{aligned}
\label{eq:cyr}
\dot{\cal F}_{\nu 2} & =\frac{8}{15} \theta_\nu  - \frac{3}{5} k {\cal F}_{\nu 3} +\frac{4}{15} \dot{h} + \frac{8}{5} \dot{\tilde \eta}+ \frac{9}{10} \alpha_2 \dot{\tau}_\nu {\cal F}_{\nu 2}\ , \\
\dot{\cal F}_{\nu \ell} &= \frac{k}{2 \ell + 1} \left[ \ell {\cal F}_{\nu (\ell -1)} - (\ell + 1) {\cal F}_{\nu (\ell +1)} \right] + \alpha_\ell \dot{\tau}_\nu {\cal F}_{\nu \ell}\ , \quad \ell \geq 3 \ ,
\end{aligned}
\end{equation}
where an overdot denotes a derivative with respect to conformal time~$\eta$,  $h$ and $\tilde{\eta}$ encapsulate the metric perturbations, 
and 
the $\alpha_\ell$s are model-dependent factors of order unity.  The search for new interactions then consists in testing the hypothesis of a nonzero $G_{\rm eff}$ against the $G_{\rm eff}=0$ null in the face of CMB data.  (The conclusion of~\cite{Cyr-Racine:2013jua} is that a $G_{\rm eff}$ orders of magnitude larger than the Fermi constant is allowed by Planck measurements.)

While the form of equation~(\ref{eq:cyr}) was not explicitly proved in~\cite{Cyr-Racine:2013jua}, it is clearly motivated by the structure of the photon Boltzmann hierarchy 
 in which a similar damping term appears for all $\ell \geq 2$ moments due to Thomson scattering.  As we shall show in this work, such a damping term does arise in the full treatment.  However, because ultrarelativistic neutrino--neutrino scattering is accompanied by significant energy transfer between particles
 while photon--electron scattering in the Thomson limit is not, 
 the final neutrino Boltzmann hierarchy formally has additional scattering contributions with no analogue in the photon hierarchy.
 
The third approach, adapted from the generalised dark matter model of~\cite{Hu:1998kj} and reinterpreted in, e.g.,~\cite{Archidiacono:2011gq,Smith:2011es,Diamanti:2012tg}, as a phenomenological model of interacting neutrinos, 
has, of the three avenues discussed here, arguably the weakest link to particle scattering.
 Here, an effective sound speed~$c_{\rm eff}^2$ and a viscosity parameter~$c_{\rm vis}^2$ 
 controlling respectively the fluid's acoustic oscillations and anisotropy
 are inserted by hand into the neutrino Boltzmann hierarchy, i.e., 
  \begin{equation}
 \begin{aligned}
 \dot{\delta}_{\nu} &= -\frac{4}{3} \theta_\nu- \frac{2}{3}\dot{h} + \frac{\dot{a}}{a}(1-3 c_{\text{eff}}^2) \left( \delta_{\nu}+ 4 \frac{\dot{a}}{a} \frac{\theta_{\nu}}{k^2} \right) \ ,  \\ 
 \dot{\theta}_{\nu}&=k^2 \left(\frac{1}{4} \delta_\nu- \sigma_\nu \right) - \frac{k^2}{4} (1-3 c_{\rm eff}^2)  \left( \delta_{\nu}+ 4 \frac{\dot{a}}{a} \frac{\theta_{\nu}}{k^2} \right) \ , \\
  \dot{\cal F}_{\nu2} &= 2 \dot{\sigma}_\nu= \frac{8}{15} \theta_\nu  - \frac{3}{5} k {\cal F}_{\nu 3} +\frac{4}{15} \dot{h} + \frac{8}{5} \dot{\tilde \eta}- (1-3 c_{\rm vis}^2) \left(\frac{8}{15} \theta_\nu +\frac{4}{15} \dot{h} + \frac{8}{5} \dot{\tilde \eta}  \right) \ ,\\
    \dot{\cal F}_{\nu \ell} &= \frac{k}{2 \ell + 1} \left[ \ell {\cal F}_{\nu (\ell -1)} - (\ell + 1) {\cal F}_{\nu (\ell +1)} \right] \ , \quad \ell \geq 3 \ , 
 \end{aligned}
\label{parametrisation} 
\end{equation}
recast here in standard notation~\cite{Ma:1995ey} in the synchronous gauge.

The choice of $c_{\text{eff}}^2 = c_{\text{vis}}^2=1/3$ recovers the standard case of non-interacting free-streaming neutrinos.  
Conversely, setting $c_{\text{vis}}^2=0$ effectively stops power from being transferred from the monopole and dipole to the $\ell \geq 2$ multipoles, and can, in a sense, reproduce the tightly-coupled limit {\it provided that} the $\ell \geq 2$ multipoles are unpopulated to begin with.  If the initial $\ell \geq 2$ moments should be nonzero, because, e.g., the neutrinos are initially free-streaming (as in the massless majoron models), then the choice of $c_{\text{vis}}^2=0$---which, unlike the collisional damping term in equation~(\ref{eq:cyr}),  does {\it not} drive ${\cal F}_{\nu2}$ to zero---at some later stage of the evolution will not completely eliminate free-streaming effects from the neutrino sector.

Indeed, the fact that $c_{\rm vis}^2$ does not act like a collisional damping term raises the question of whether the parameterisation~(\ref{parametrisation}) even has anything to do with particle scattering.  Furthermore, any choice of $c_{\rm eff}^2$ besides $1/3$ is likewise confounding, unless it is accompanied by an explicit coupling of the neutrinos to 
a nonrelativistic particle species, e.g., the dark matter.  If that is the case, then the equations of motion of the nonrelativistic species concerned must be modified {\it simultaneously}
 to reflect this coupling.   Such a modification has at least been incorporated into the study of~\cite{Diamanti:2012tg} (see also \cite{Boehm:2014vja,Wilkinson:2014ksa}). 
  Otherwise, a decoupled ultrarelativistic neutrino Boltzmann hierarchy with $c_{\rm eff}^2 \neq 1/3$ can have no interpretation in terms of particle interaction, as it would   violate energy--momentum conservation.  

In summary, the parameterisation~\eqref{parametrisation} may have been physically well motivated for other purposes.  But to reinterpret it as a phenomenological model of interacting neutrinos appears to us to be imprudent.
In fact, the neutrino Boltzmann hierarchy we shall derive from first principles in this work shows irreconcilable differences.


\section{Boltzmann equation}
\label{Assumptions}

The fundamental starting point of our calculation is the relativistic Boltzmann equation,
\be
\label{eq:liouville}
 P^{\alpha} \frac{\partial f}{\partial x ^{\alpha}}- \Gamma^\gamma_{\alpha \beta}   P^{\alpha} P^{\beta} \frac{\partial f}{\partial P^\gamma}=  \left(\frac{\partial f}{\partial \tau} \right)_{\rm coll} \,,
\ee
which acts on the phase space distribution $f(\fett{x},\fett{P},\eta)$ of a particle species, defined such that 
\be
  f(\fett{x},\fett{P},\eta) \,\dd x^1 \dd x^2 \dd x^3  \dd P_1 \dd P_2 \dd P_3 = \dd N 
\ee
gives the number of particles~$\dd N$ in a differential phase space volume  $\dd x^1 \dd x^2 \dd x^3  \dd P_1 \dd P_2 \dd P_3$, where
 $P_i$~is the canonical momentum conjugate to the comoving coordinates~$x^i$ (see, e.g.,~\cite{Ma:1995ey}). 
 The LHS of equation~(\ref{eq:liouville}) incorporates all gravitational physics through the Christoffel symbols~$\Gamma^\gamma_{\alpha \beta}$, which can be easily evaluated once the space-time line element~$\dd s^2$ has been specified.
The collision term~$(\partial f/\partial \tau)_{\rm coll}$ on the RHS, where $\dd \tau \equiv \sqrt{ -\dd s^2}$ denotes the incremental proper time,
encapsulates all non-gravitational scattering processes relevant for the particle species under consideration, and is the focus of this work.  As usual, greek indices $\alpha, \beta, \ldots$ run from 0 to 3, while latin indices $i,j,\ldots$ run from 1 to 3.  Summation over pairs of repeated upper and lower indices is implied.

Following the notation of~\cite{Ma:1995ey} and working in the synchronous gauge with line element
\be \label{eq:synch}
   \dd s^2 =  a^2(\eta) \left[ - \dd \eta^2 + \left( \delta_{ij}+ h_{ij} \right) \dd x^i \dd x^j \right] \,,
\ee
the conjugate momentum  can be expressed as $P_{\alpha} \dot= (E/a,a [\delta_{ij}+h_{ij}/2]p^j)$ in terms of the 4-momentum $p^{\alpha} \dot=(E,p^i)$
in the tetrad basis, i.e., the orthonormal basis of an observer comoving with the coordinates~(\ref{eq:synch}).  Since $E$ and $p^i$ generally redshift with the Hubble expansion, it is convenient to separate out this effect by  introducing the comoving momentum $\fett{q}= |\fett{q}| \hat{q} = a\fett{p}$ and the comoving energy $\epsilon = a E$.
It is likewise useful to split up the phase space distribution into a background and a perturbed component, i.e.,
\be
\label{eq:split}
  f_i(\fett{x},\fett{P},\eta)= \bar{f}_i(|\fett{q}|, \eta)+F_i(\fett{x}, \fett{q},\eta)\,.
\ee
For standard photons and non-interacting neutrinos (massless or massive), $\bar{f}_i(|\fett{q}|,\eta)=\bar{f}_i(|\fett{q}|)$ can be taken to be the ultrarelativistic Bose--Einstein or Fermi--Dirac distribution, because these particles are either always ultrarelativistic or decouple while ultrarelativistic, and suffer no particle production or decay/annihilation.
In the general case, however, $\bar{f}_i(|\fett{q}|,\eta)$ needs to be computed from the homogeneous limit of the Boltzmann equation~(\ref{eq:liouville}),
\begin{equation}
\label{eq:homogeneous}
\dot{\bar{f}}_i(|\fett{q}|, \eta) =  \left( \frac{\partial f_i}{\partial \eta}\right)^{(0)} _{\rm coll} \ ,
\end{equation}
where the superscript~``$(0)$'' indicates that the collision term contains only background quantities. 

Expanding  equation~(\ref{eq:liouville}) to first order in the perturbed quantities and performing a Fourier transform $\psi(\fett{k}) = \mathcal{F}[\psi(\fett{x})]$ on all functions of~$\fett{x}$, 
we also obtain an equation of motion for the perturbed component of the phase space perturbation~$F_i(\fett{k}, \fett{q},\eta)$~\cite{Ma:1995ey},
\begin{equation}
\dot F_i(\fett{k}, \fett{q},\eta)+ \ii \frac{|\fett{q}| |\fett{k}|}{\epsilon} ( \hat{k} \cdot \hat{q}) \, F_i(\fett{k}, \fett{q},\eta)+ \frac{\partial \bar{f}_i(|\fett{q}|,\eta)}{\partial \ln  |\fett{q}|} \left[ \dot{\tilde{\eta}} - (\hat{k} \cdot \hat{q})^2 \, \frac{\dot{h}+6 \dot{\tilde{\eta}}}{2} \right]  =\left( \frac{\partial f_i}{\partial \eta}\right)^{(1)} _{\rm coll} \ ,
\label{Boltzmann equation}
\end{equation}
where $h \equiv h^i{}_i(\fett{k},\eta)$ and $6\tilde{\eta} \equiv 6 \tilde{\eta}(\fett{k},\eta) = - 3/(2 k^4) k^i k^j h_{ij} +1/(2 k^2) h$ denote respectively the Fourier transforms of the 
trace and the traceless longitudinal perturbations in the space-space part of the synchronous metric~(\ref{eq:synch}). The superscript~``$(1)$'' accompanying the collision term indicates that it must likewise be first order in the perturbed quantities. Together the zeroth- and first-order Boltzmann equations~(\ref{eq:homogeneous}) and~(\ref{Boltzmann equation}) form a set to be solved for each individual particle species under consideration.

Turning our attention now to the RHS of equations~(\ref{eq:homogeneous}) and~(\ref{Boltzmann equation}), the collision term $(\partial f_i/\partial \eta)_{\rm coll}$
 counts the number of collisions a particle species~$i$ undergoes in a time interval~$\dd \eta$ per phase space volume.  For a  $CP$-invariant two-body scattering process $i(p) + j(n) \to k(p')+l(n')$ and its reverse process, this can be written as
\begin{equation}
\begin{aligned}
  \left( \frac{\partial f_i}{\partial t}\right) _{\rm coll} \left(\fett{k}, \fett{p}, \eta \right)
 = &\  \frac{1}{2 E(\fett{p})} \int \mathrm{d^{3}}\fett{\Pi}_j (\fett{n})
 \int \mathrm{d^{3}}\fett{\Pi}_k (\fett{p}')
 \int \mathrm{d^{3}}\fett{\Pi}_l(\fett{n}')
\;  (2 \pi)^4 \delta_\D^{(4)}(p+n-p'-n') \,\\ 
&\times  |{\cal M}_{i j \leftrightarrow k l}|^2 \Big( f_k(\fett{k},{p}', \eta)f_l(\fett{k},n', \eta)\left[ 1\pm f_i(\fett{k},{p},\eta) \right] \left[ 1 \pm f_j(\fett{k},{n},\eta) \right]  
\\ & \hspace{20mm}
-f_i(\fett{k},{p},\eta)f_j(\fett{k},{n},\eta) \left[ 1 \pm f_k(\fett{k},{p}',\eta) \right] \left[ 1 \pm f_l(\fett{k},{n}',\eta) \right] \!\Big) 
\label{General Collision Integral}
\end{aligned}
\end{equation}
in the tetrad basis, where 
\begin{equation}
\mathrm{d^3} \fett{\Pi}_j(\fett{n}) \equiv  g_j \frac{\mathrm{d^{3}}\fett{n}}{(2 \pi)^3 2 E(\fett{n})} \, 
\label{eq:dpi}
\end{equation}
and so on, $g_j$ is the number of internal degrees of freedom of particle species~$j$,  $p$, $n$, etc. are physical 4-momenta, $\delta_{\rm D}^{(4)}$ is the four-dimensional Dirac delta distribution,
and $|{\cal M}_{ij \leftrightarrow k l}|^2$ denotes the Lorentz-invariant squared matrix element for the interactions~$ij\leftrightarrow kl$, averaged over the initial and final spins of all particles, and including 
a symmeterisation factor $1/(N_{\rm i}! N_{\rm f}!)$ for $N_{\rm i}$ and $N_{\rm f}$ identical particles in the initial and final states.
The terms in square brackets, i.e., $[1\pm f_j]$, etc., arise from quantum statistics, and represent 
Pauli blocking (``$-$'') in case $j$ is a fermion and Bose enhancement (``$+$'') if $j$ should be a boson.

Note that the collision term has been defined in equation~(\ref{General Collision Integral}) with respect to the time interval~$\dd t$ measured by the comoving observer.  To convert it to $(\partial f_i/\partial \eta) _{\rm coll}$ for use in equation~(\ref{Boltzmann equation}), we employ the relation
\begin{equation}
\label{eq:convert}
\left( \frac{\partial f_i}{\partial \eta}\right) _{\rm coll} = a \left( \frac{\partial f_i}{\partial t}\right) _{\rm coll} \  ,
\end{equation}
which is exact in the synchronous gauge.

Equations~(\ref{eq:homogeneous}) to (\ref{eq:convert}) then form the starting point of this paper.  We shall evaluate the collision integral~\eqref{General Collision Integral}
for a neutrino--scalar interaction up to first order in the perturbed quantities.  Embedding especially the first-order result into equation~(\ref{Boltzmann equation}) and then performing a Legendre decomposition will then give us a Boltzmann hierarchy that describes the anisotropies in each participating ensemble of interacting particles.


\section{Two limits of the scalar mass}
\label{sec:twolimits}

\begin{figure}[t]
\begin{minipage}[t]{0.2\textwidth}
 \vspace{0.5cm}
 \begin{equation*}
 \nu \nu \leftrightarrow \nu \nu :
 \end{equation*}
 \vspace{1.7cm}
 \begin{equation*}
 \nu \nu \leftrightarrow \phi \phi :
 \end{equation*} 
 \vspace{1.7cm}
 \begin{equation*}
 \nu \phi \leftrightarrow \nu \phi :
 \end{equation*}
\end{minipage}
\begin{minipage}[t]{0.7\textwidth}
 \vspace{0pt}
 \centering
 \includegraphics[width=\linewidth]{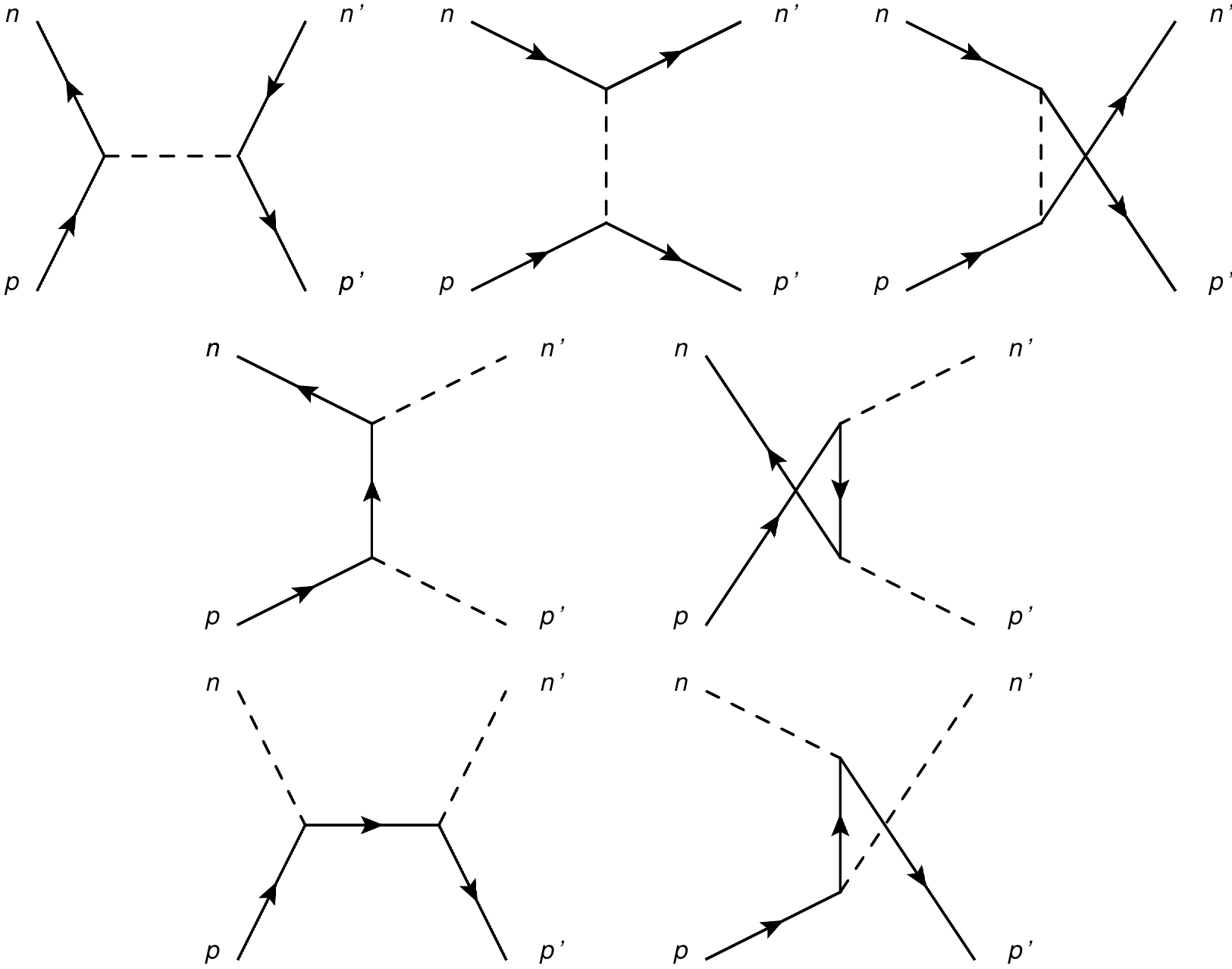} 
\end{minipage}
\caption{Tree-level diagrams for the $2 \to 2$ scattering processes  described by the Lagrangian~\eqref{Lagrangian} in the $s$, $t$, and $u$-channels.  The neutrinos have been explicitly assumed to be Majorana particles.}
\label{Feynman}
\end{figure}

Figure~\ref{Feynman} shows all relevant tree-level diagrams for the $2 \to 2$ scattering processes described by the Lagrangian~\eqref{Lagrangian}.
Given that the mass of the scalar particle is an unknown quantity, it is useful to focus our analysis on two limiting cases:
\begin{itemize}
\item[(i)] The scalar mass $m_\phi$ far exceeds the energies of the incoming neutrinos, and 
\item[(ii)] $m_\phi=0$.
\end{itemize}
These simplifications are motivated in part by the complex angular dependence 
of the general $m_\phi$-dependent matrix elements  for those scattering processes in which $\phi$  serves has a mediator. See table~\ref{Matrix elements}.
Phenomenologically, these two limits of $m_\phi$ also represent two vastly different thermal histories for the neutrinos as well as for the scalar particles, 
and determine which scattering processes in figure~\ref{Feynman} need to be taken into account in the analysis.  For simplicity we shall assume  in the following flavour-independent and 
diagonal coupling, i.e., $\mathfrak{g}_{ij} = 0$ for $i \neq j$, where $i,j$ label the neutrino flavour, and  $\mathfrak{g}_{ii}=\mathfrak{g}$.   Since scalar and pseudo-scalar interactions are phenomenologically identical in cosmological settings, we shall also neglect the latter by setting $\mathfrak{h}_{ij}=0$.

\subsection{Massive scalar limit}
\label{sec:massivelimit}

By assuming an extremely massive scalar in scenario~(i), the self-interaction $\nu \nu \to \nu \nu$ mediated by a virtual scalar described by the Lagrangian~(\ref{Lagrangian})
becomes effectively
a four-fermion interaction.  The same limit also guarantees that the process $\nu\nu \to \phi \phi$ is kinematically suppressed, and that 
any pre-existing $\phi$ population will have completely annihilated or decayed into neutrinos well before the timeframe of interest, thereby rendering the processes  $\nu \phi \to \nu \phi$ and $\phi \phi \to \nu \nu$ likewise irrelevant.  It therefore suffices in this case to consider only neutrino self-interaction and its impact on the neutrino Boltzmann hierarchy, while treating the
$\phi$ population as essentially non-existent.
The interaction rate per neutrino can be estimated to be $\Gamma_{\rm m} \sim  \mathfrak{g}^4 T^5_\nu/m_\phi^4$, i.e., the same temperature dependence as weak interactions, so that self-interaction with $\mathfrak{g}^4/m_\phi^4 \gg G_F^2$, where $G_F$ is the Fermi constant, only serves to {\it delay the epoch of neutrino decoupling}. We note also that opacity parameter~$\dot{\tau}= -a G_{\rm eff}^2 T_\nu^5$  adopted in the neutrino Boltzmann hierarchy of~\cite{Cyr-Racine:2013jua}
(see also equation~(\ref{eq:cyr})) implicitly assumes this scenario.

Because the neutrinos are initially in equilibrium in this scenario,  self-interaction does not alter the equilibrium form of the background phase space distribution even when $ \Gamma_{\rm m}$ drops below the expansion rate $H \sim T^2/m_{\rm Pl}$, {\it  provided that} the neutrinos have no other interaction in the timeframe of interest and remain ultrarelativistic (i.e., effectively massless) throughout the decoupling process.  The second condition imposes an upper limit on the neutrino mass, $m_\nu \lesssim  3\, T_{\nu,{\rm dec}}$, with $T_{\nu,{\rm dec}}$  as the new neutrino decoupling temperature, for which the statement holds.    We shall circumvent this issue by simply assuming that the neutrinos are exactly massless, regardless of $T_{\nu,{\rm dec}}$, which can be   justified as follows.

Phenomenologically, we expect the new neutrino interactions to have the strongest impact on the CMB anisotropies if they should remain operational down to the epoch of photon decoupling, $T_{\gamma,{\rm dec}} \sim 0.2$~eV.  Thus, for the assumption of ultrarelativistic neutrinos to hold we require $m_\nu \lesssim  0.6$~eV, which is easily satisfied by neutrinos with the minimum masses established by flavour oscillation experiments ($m_{\nu,1}  \simeq 0, m_{\nu, 2} \simeq 0.007$~eV, and $m_{\nu, 3} \simeq 0.05$~eV~\cite{Tortola:2012te}), and is 
in any case consistent with current laboratory limits on the effective Majorana mass, $|m_{\beta\beta}| \lesssim 0.3$~eV, from neutrinoless double $\beta$-decay (e.g.,~\cite{Bilenky:2012qi}).  The~$m_\nu=0$ approximation therefore appears to be reasonable, and $\bar{f}_\nu(|\fett{q}|, \eta)$ can be assumed to take a time-independent ultrarelativistic equilibrium form $\bar{f}^{\rm eq}(|\fett{q}|)$ at all times, thereby eliminating the need to solve the zeroth order Boltzmann equation~(\ref{eq:homogeneous}).

{\renewcommand{\arraystretch}{1.7}
\begin{table}
\centering
\begin{tabular}[t]{|c|c |c|c|}
\hline
$i j \to k l$ &$\ii \mathcal{M}_{ij \to kl}$ & $|\mathcal{M}_{ij \leftrightarrow k l}|_0^2$ & $|\mathcal{M}_{ij\leftrightarrow k l}|^2_{\rm m}$\\
 \hline
 \hline
  \multirow{3}{*}{$\nu \nu \rightarrow \nu \nu$}	  & 
 $s$: \;  $\bar{v}(n)(\ii 2\mathfrak{g}) u(p)\frac{\ii}{s-m_\phi^2} \bar{u}(p') (\ii2\mathfrak{g}) v(n') $ &
\multirow{3}{*}{$6 \mathfrak{g}^4$} 
& \multirow{3}{*}{$\frac{\mathfrak{g}^4}{2 m^4_{\phi}}  \left[s^2+t^2+u^2\right]$}	\\
&$t$: \;  $ \bar{u}(p') (\ii2\mathfrak{g}) u(p) \frac{\ii}{t-m_\phi^2}\bar{u}(n') (\ii2\mathfrak{g})u(n)$ & &\\
&$u$:  $- \bar{u}(n') (\ii2\mathfrak{g}) u(p)  \frac{\ii}{u-m_\phi^2}\bar{u}(p') (\ii2\mathfrak{g})u(n)$ && \\
\hline
  \multirow{2}{*}{$\nu \nu \rightarrow \phi \phi$}	&$t$: \; $\bar{v}(n)(\ii 2\mathfrak{g}) \frac{\ii(\slashed{p}-\slashed{p}'+m_\nu)}{t-m_\nu^2}  (\ii2\mathfrak{g}) u(p)$
    &  \multirow{2}{*}{$2 \mathfrak{g}^4 \left( \frac{s^2}{tu}-4 \right)$} &  \multirow{2}{*}{--} \\
&$u$: \; $\bar{v}(n)(\ii 2\mathfrak{g}) \frac{\ii(\slashed{p}-\slashed{n}'+m_\nu)}{u-m_\nu^2}  (\ii2\mathfrak{g}) u(p)$&&\\  
\hline
\multirow{2}{*}{$\nu \phi \rightarrow \nu \phi$} & $s$: \; $u(n')(\ii 2\mathfrak{g}) \frac{\ii(\slashed{p}+\slashed{n}+m_\nu)}{s-m_\nu^2}  (\ii2\mathfrak{g}) u(p)$
&   \multirow{2}{*}{$8 \mathfrak{g}^4 \left( 4- \frac{t^2}{su} \right)$} &   \multirow{2}{*}{--}\\
&$u$: \; $u(n')(\ii 2\mathfrak{g}) \frac{\ii(\slashed{p}-\slashed{n}'+m_\nu)}{u-m_\nu^2}  (\ii2\mathfrak{g}) u(p)$ && \\
\hline
\end{tabular}
\caption{Scattering amplitudes and spin-averaged squared matrix elements for all relevant processes shown in figure~\ref{Feynman}, where $s \equiv (p+n)^2= (p'+n')^2$, $t \equiv(p-p')^2=(n-n')^2$, and $u \equiv (p-n')^2=(n-p')^2$ are the Mandelstam variables.  For the squared matrix elements, we assume a vanishing neutrino mass, and report only results in the massless (subscript ``0'') and extremely massive (``m'') limits of the scalar particle.   In the latter case, it is sufficient to consider only the neutrino self-interaction $\nu \nu \leftrightarrow \nu \nu$, since the population of scalar particles is thermally suppressed.}
\label{Matrix elements}
\end{table}


\subsection{Massless scalar limit} 

In scenario~(ii), production of real $\phi$ particles via the annihilation process $\nu \nu \to \phi \phi$ in the timeframe of interest becomes kinematically possible. This immediately calls for a separate Boltzmann hierarchy for the scalar particles in addition to that for the neutrinos.  Furthermore,  the interaction rate per particle can be estimated to be 
$\Gamma_0  \sim   \mathfrak{g}^4 T_\nu$ at $T_\nu \gg m_\nu$.   Thus, for suitable choices of coupling $\mathfrak{g}$, the neutrinos can decouple at the canonical weak decoupling temperature of $T_{\nu, {\rm dec}} \sim 1$~MeV, free-stream for some time, and then {\it recouple} when $\Gamma_0$ overtakes  the expansion rate~$H$.   
From the model-building perspective, viable models containing massless majorons have been reported in, e.g.,~\cite{Choi:1991aa,Acker:1992eh}.

The recoupling epoch also signals the beginning of scalar particle production  via  $\nu \nu \to \phi \phi$ and its reverse process.  Assuming no pre-existing thermal $\phi$ population, 
these inelastic interactions will populate the scalar particle background phase space distribution~$\bar{f}_\phi(|\fett{q}|,\eta)$ from scratch, and alter the neutrino background $\bar{f}_\nu(|\fett{q}|, \eta)$ in the process.  It is thus necessary in this case to track the evolution of~$\bar{f}_\nu(|\fett{q}|,\eta)$ and  $\bar{f}_\phi(|\fett{q}|,\eta)$ using the homogeneous Boltzmann equation~(\ref{eq:homogeneous}).

For nondegenerate neutrino masses and nonzero flavour off-diagonal couplings, i.e., $\mathfrak{g}_{ij}, \mathfrak{h}_{ij} \neq 0$ for $i \neq j$,  the Lagrangian~(\ref{Lagrangian})  also provides for the possibility of neutrino decay, $\nu_i \to \nu_j + \phi$, in the $m_\phi=0$ limit.  We shall however ignore this possibility by assuming no flavour off-diagonal coupling and massless neutrinos, as discussed above.


\subsection{Matrix elements}

The scattering amplitudes~$\mathcal{M}_{ij\to kl}$ of the processes displayed in figure~\ref{Feynman} can be constructed 
following the Feynman rules of, e.g., \cite{Gluza:1991wj,Denner:1992me}, for Majorana fermions.  These are shown in table~\ref{Matrix elements}.
The corresponding spin-averaged squared matrix elements have been evaluated  in the two limits of $m_\phi$, under the assumptions  discussed above. 

In averaging the squared matrix elements over initial and final spins we have implicitly assumed that the initial phase space distributions of the two neutrino helicity states are identical, or equivalently, that the initial neutrino chemical potential is vanishing,~$\mu_{\nu}=0$. 
This is a logical  choice in the case of an extremely massive scalar particle, since the self-interaction process $\nu \nu \to \nu \nu$ which occurs at a rate $\Gamma_{\rm m} \sim \mathfrak{g}^4 T_\nu^5/m^2_\phi$ per particle always ensures that any preexisting asymmetry between the neutrino helicity states will have already evolved to zero well before the timeframe of interest.

In the case of a massless scalar, where the new interactions become operational only in the timeframe of interest, 
the assumption of  a vanishing initial $\mu_\nu$ needs to be justified by external reasons.
The most commonly invoked argument is  that  nonperturbative SM processes before electroweak symmetry breaking always render any nonzero lepton asymmetry the same order of magnitude as the baryon asymmetry, i.e., $\sim 10^{-9}$.  While we shall also work within this framework, we note on the side that neutrino asymmetries generated after electroweak symmetry breaking are not thus restricted.  In fact, present observational constraints stand at 
 ${\cal O}(-0.4) \lesssim \mu_{\nu}/T_\nu \lesssim {\cal O}(0.1)$  from an analysis of the CMB anisotropies~\cite{Schwarz:2012yw,Caramete:2013bua}, and
${\cal O}(-0.1) \lesssim \mu_\nu/T_\nu \lesssim {\cal O} (0.05)$ based on the Helium-4 abundance established in extragalactic HII regions~\cite{Schwarz:2012yw}.
Relaxing the assumption of $\mu_\nu=0$ would require that we track the two neutrino helicity states separately.  The scattering matrix element would likewise have to  be recomputed, this time without summation and averaging over the initial and final spins.


\section{Solutions of the collision integrals}
\label{sec:collisionterms}

Having specified the relevant processes in the two limits of the scalar mass and computed their corresponding squared matrix elements, we are now in position to solve the collision integral~(\ref{General Collision Integral}).   In order to reduce the number of terms we must deal with we shall neglect for now the important quantum statistical effects of Pauli blocking and Bose enhancement, i.e., for all occurrences of $1-f_\nu$ and $1+f_\phi$ we set  $1-f_\nu \simeq 1$ and $1+f_\phi \simeq 1$.  This assumption must be accompanied by a corresponding modification of the equilibrium distribution from Fermi--Dirac or Bose--Einstein to Maxwell--Boltzmann.  In particular, for the neutrino population, we have
 \begin{equation}
\bar{f}^{\rm eq}_\nu(|\fett{q}|) \simeq {\rm N} \exp \left(-\frac{|\fett{q}|}{a T_\nu} \right) =  {\rm N} \exp \left(-\frac{|\fett{q}|}{T_{\nu,0}} \right)
 \,, \label{eq:fbar}
\end{equation}
where $T_{\nu,0}=1.95$~K is the present-day neutrino temperature,%
\footnote{Because the new interactions are confined exclusively in the $\nu$-$\phi$ sector, entropy transfer from $e^+ e^-$-annihilation at $T \sim 0.2$~MeV affects the post-annihilation neutrino--photon temperature relation in the same way as in the standard case, i.e., $T_\nu = (4/11)^{1/3} T_\gamma$.}
and  ${\rm N}= 3 \zeta(3)/4 \simeq 0.9$ is a normalisation factor we have inserted by hand to ensure that 
$a^{-3} \int \dd^3 \fett{q}\ \bar{f}(|\fett{q}|)$ integrates to the same number density as the ultrarelativistic Fermi--Dirac distribution.

Then, splitting the phase space distribution into a background and a perturbed component 
as per equation \eqref{eq:split}, the zeroth- and first-order collision integrals to be solved for particle species~$i$ due to $ij \to kl$ and its reverse process are, respectively,
\begin{equation}
\begin{aligned} 
\left(\frac{\partial f_i}{\partial \eta} \right)^{(0)}_{ij \leftrightarrow kl}(\fett{q},\eta)  = & \ \frac{  g_j g_k g_l}{2 |\fett{q}|  (2 \pi)^{5}} \int \frac{\mathrm{d^{3}}\fett{q}'}{2 | \fett{q}'|}  \int \frac{\mathrm{d^{3}}\fett{l}}{2 | \fett{l}|}  \int \frac{\mathrm{d^{3}} \fett{l}'}{2 | \fett{l}'|} \, \delta_\D^{(4)}(q+l-q'-l')  \\
 &\hspace{15mm}\times   |{\cal M}_{ij\leftrightarrow kl}|^2 \Big( \bar{f}_k(| \fett{q}'|,\eta) \, \bar{f}_l(|\fett{l}'|,\eta)  - 
 \bar{f}_i(| \fett{q}|,\eta) \, \bar{f}_j(|\fett{l}|,\eta) \Big) \,,
\label{C[f]0zeroth}
\end{aligned}
\end{equation}
and
\begin{equation}
\begin{aligned} 
 \left(\frac{\partial f_i}{\partial \eta} \right)^{(1)}_{ij \leftrightarrow kl}(\fett{k},\fett{q},\eta)  = &  \ \frac{g_j g_k g_l}{2 |\fett{q}|  (2 \pi)^{5}} \int \frac{\mathrm{d^{3}}\fett{q}'}{2 | \fett{q}'|}  \int \frac{\mathrm{d^{3}}\fett{l}}{2 | \fett{l}|}  \int \frac{\mathrm{d^{3}} \fett{l}'}{2 | \fett{l}'|} \, \delta_\D^{(4)}(q+l-q'-l')  \\
 &\times  |{\cal M}_{ij\leftrightarrow kl}|^2 \Big( \bar{f}_k(| \fett{q}'|,\eta) \, F_l(\fett{k},\fett{l}',\eta)+  \bar{f}_l(| \fett{l}'|,\eta) \, F_k(\fett{k},\fett{q}',\eta) 
 \\&\hspace{25mm}
 -\bar{f}_i(| \fett{q}|,\eta) \, F_j(\fett{k},\fett{l},\eta)+\bar{f}_j(| \fett{l}|,\eta) \, F_i(\fett{k},\fett{q},\eta)  \Big),
\label{C[f]0}
\end{aligned}
\end{equation}
where $q,l,q',l'$ are comoving 4-momenta so that the four-dimensional Dirac delta distribution obeys $\delta^{(4)}_\D(p)=\delta^{(4)}_\D(q/a)=a^4 \delta^{(4)}_\D(q)$. 

We present the solutions of equations~(\ref{C[f]0zeroth}) and~(\ref{C[f]0}) in the two limits of the scalar mass below.  These constitute the first main results of this work.
The interested  reader can find the full derivation in appendices~\ref{app:massless} and~\ref{app:massive}.


\subsection{Massive scalar limit}  

As argued in section~\ref{sec:massivelimit}, the only relevant process in this scenario is the self-interaction $\nu \nu \rightarrow \nu \nu$.   Furthermore, because the background neutrino phase space distribution assumes the equilibrium form~(\ref{eq:fbar}), the zeroth-order collision term $(\partial f_\nu/\partial \eta)^{(0)}_{\rm coll, m}$
always evaluates to zero under the assumptions discussed above.   For the first-order term we find,
\begin{equation}
\begin{aligned}
& \left(\frac{\partial f_\nu}{\partial \eta} \right)^{(1)}_{\rm coll,m} (\fett{k},\fett{q},\eta)=-\frac{2 \mathrm{N} \mathfrak{g}^4 }{a^4 m_\phi^4(2\pi)^{3}}
\Bigg\{\frac{40}{3}\, |\fett{q}|\, T_{\nu,0}^4 \; F_{\nu}(\fett{k},\fett{q},\eta)   \\
&\hspace{5mm}   + \int  \mathrm{d \cos\theta} \, \mathrm{d|\fett{q}'|} \frac{|\fett{q}'|}{|\fett{q}|}
\left[
\frac{5 |\fett{q}|^2 |\fett{q}'|^2}{6} \e^{-| \fett{q}|/T_{\nu,0}}   (1-\cos \theta)^2
-K^{\rm m}(|\fett{q}|, |\fett{q}'|, \cos \theta)  \right]  F_{\nu}(\fett{k},\fett{q}',\eta) \Bigg\} , 
 \label{massive}
\end{aligned}
\end{equation}
where the time-independent integration kernel is given by
\begin{equation}
\begin{aligned}
K^{\rm m}(|\fett{q}|, |\fett{q}'|, \cos \theta) \equiv&
 \frac{1}{16 P^5} \, \e^{-(Q_-+P)/(2 T_{\nu,0})} \; T_{\nu,0}  \left(Q_-^2-P^2 \right)^2  \\
& \times \Big[ P^2 \left(3 P^2 - 2 PT_{\nu,0}-4 T^2_{\nu,0}\right)+ Q_+^2 \left( P^2 + 6 P T_{\nu,0} + 12 T_{\nu,0}^2 \right)
 \Big],
 \label{eq:km}
\end{aligned}
\end{equation}
with variables $P \equiv |\fett{q}-\fett{q}'|$ and $Q_\pm \equiv |\fett{q}|\pm |\fett{q}'|$.


\subsection{Massless scalar limit}\label{sec:masslesscollision}

All processes shown in figure~\ref{Feynman} contribute to modifying  both the neutrino and the scalar phase space distributions. 
For the neutrinos, we find the total collision term to be
\begin{equation}
\left(\frac{\partial f_{\nu}}{\partial \eta} \right)_{\rm coll,0}= \left(\frac{\partial f_{\nu}}{\partial \eta} \right)_{\nu \nu \leftrightarrow \nu \nu}+\left(\frac{\partial f_{\nu}}{\partial \eta} \right)_{\nu \nu \leftrightarrow \phi \phi}+ \left(\frac{\partial f_{\nu}}{\partial \eta} \right)_{\nu \phi \leftrightarrow \nu \phi}
\end{equation} 
at both zeroth and first order in the phase space perturbations.

At zeroth order in the phase space perturbations, the contributions of the individual processes to the total collision term are
\begin{equation}
\begin{aligned}
\left(\frac{\partial f_\nu}{\partial \eta} \right)^{(0)}_{{\nu \nu \leftrightarrow \nu \nu}} (\fett{q},\eta) 
=   \frac{3 \mathfrak{g}^4}{(2\pi)^{3}|\fett{q}|} & \left\lbrace -2 \, \bar{f}_{\nu}(|\fett{q}|,\eta)
\int \mathrm{d|\fett{l}|} \,  |\fett{l}|\,  \bar{f}_\nu(|\fett{l}|,\eta) \right. \\
& \left.  + \frac{1}{|\fett{q}|} \sum_{n=1}^4 \int_{ {\cal I}_n} \dd |\fett{q}'| \, \dd |\fett{l}'| \, \bar{f}_{\nu}(|\fett{q}'|,\eta)\, \bar{f}_{\nu}(|\fett{l}'|,\eta) \, k^0_n(|\fett{q}|,|\fett{q}'|,|\fett{l}'|) \right\rbrace,
\label{eq:0self}
\end{aligned}
\end{equation}

\begin{equation}
\begin{aligned}
\left(\frac{\partial f_\nu}{\partial \eta} \right)^{(0)}_{{\nu \nu \leftrightarrow \phi \phi}} 
= & \frac{\mathfrak{g}^4}{(2\pi)^{3} |\fett{q}|} \left\lbrace\, 2  \bar{f}_{\nu}(|\fett{q}|,\eta) 
\int \dd |\fett{l}| \, |\fett{l} |\left[
\left( \frac{3}{2} - \frac{1}{2} \log(4) \right)- \frac{1}{2} \log\left(\frac{|\fett{l}| |\fett{q}|}{\tilde{m}_\nu^2}\right)
\right] \bar{f}_\nu (|\fett{l}|,\eta)  \right. \\
& \left. - \frac{1}{|\fett{q}|} \sum_{n=1}^4 \int_{ {\cal I}_n} \dd |\fett{q}'| \, \dd |\fett{l}'| \, \bar{f}_{\phi}(|\fett{q}'|,\eta)\, \bar{f}_{\phi}(|\fett{l}'|,\eta) \, \left( \frac{5}{8} k^0_n -k^u_n -k^t_n \right)(|\fett{q}|,|\fett{q}'|,|\fett{l}'|) \right\rbrace,
\label{eq:0ann}
\end{aligned}
\end{equation}
 \begin{equation}
 \begin{aligned}
 \left(\frac{\partial f_\nu}{\partial \eta} \right)^{(0)}_{{\nu \phi \leftrightarrow \nu \phi}} 
=& - \! \frac{4\mathfrak{g}^4}{(2\pi)^{3}|\fett{q}|} \left\lbrace  \bar{f}_\nu (|\fett{q}|,\eta) \!
\int \dd |\fett{l}|\, |\fett{l}|
\left[ \left( \frac{13}{4} - \frac{1}{2} \log(4) \right)- \frac{1}{2}
 \log\left(\frac{|\fett{l}| |\fett{q}|}{\tilde{m}_\nu^2}\right)
\right] \bar{f}_\phi(|\fett{l}|,\eta) \right. \\
& \left. + \frac{1}{|\fett{q}|} \sum_{n=1}^4 \int_{ {\cal I}_n} \dd |\fett{q}'| \, \dd |\fett{l}'| \, \bar{f}_{\nu}(|\fett{q}'|,\eta)\, \bar{f}_{\phi}(|\fett{l}'|,\eta) \, \left( k^0_i -k^u_i +k^s_i \right)(|\fett{q}|,|\fett{q}'|,|\fett{l}'|) \right\rbrace ,
\label{eq:0elastic}
\end{aligned}
\end{equation}
where the integral domains ${\cal I}_n$ are given by 
\begin{equation}
\begin{aligned}
&\int_{ {\cal I}_1}+\int_{ {\cal I}_2}+\int_{ {\cal I}_3}+\int_{ {\cal I}_4} \\ 
& \hspace{1cm} \equiv \int_0^{|\fett{q}|} \dd |\fett{q}'|  \left(\int_{|\fett{q}|-|\fett{q}'|}^{|\fett{q}|} \dd |\fett{l}'| + \int_{|\fett{q}|}^\infty \dd |\fett{l}'| \right)+ \int_{|\fett{q}|}^\infty  \dd |\fett{q}'| \left( \int_0^{|\fett{q}|} \dd |\fett{l}'| +\int_{|\fett{q}|}^\infty \dd |\fett{l}'| \right)  \; , 
\label{eq:domains}
\end{aligned}
 \end{equation}
and the functions $k^0_n, k^u_n, k^s_n, k^t_n$ can be found in  equation~\eqref{eq:IntKernelsZero}.  

At first order we find the contributions to be
 \begin{equation}
\begin{aligned}
&\left(\frac{\partial f_\nu}{\partial \eta} \right)^{(1)}_{{\nu \nu \leftrightarrow \nu \nu}} (\fett{k},\fett{q},\eta)
 =-  \frac{6 \mathfrak{g}^4}{(2\pi)^{3}}
\Bigg\{ X_{\nu}(|\fett{q}|,\eta)  F_{\nu}(\fett{k},\fett{q},\eta) \\
& \hspace{15mm}+ \int  \dd\cos\theta \, \dd|\fett{q}'| \, \frac{|\fett{q}'|}{|\fett{q}|}  \left[
\frac{1}{2} \bar{f}_{\nu}(|\fett{q}|,\eta) -K_{\nu}^0
(|\fett{q}|,|\fett{q}'|,\cos \theta,\eta) \right] F_{\nu}(\fett{k},\fett{q}',\eta) \Bigg\} ,
\label{eq:coll}
\end{aligned}
\end{equation}
\begin{equation}
\begin{aligned}
& \left( \frac{\partial f_{\nu}}{\partial \eta}\right)^{(1)}_{\nu \nu \leftrightarrow \phi \phi}(\fett{k},\fett{q},\eta)
  = 
 \frac{2 \mathfrak{g}^4}{(2\pi)^{3}} \left\{
\left[\left(\frac{3}{2} - \frac{1}{2} \log(4) \right)
X_\nu(|\fett{q}|,\eta) - Z_\nu(|\fett{q}|,\eta) \right] F_{\nu}(\fett{k},\fett{q},\eta) \right.\\
& \hspace{2mm}+ \bar{f}_{\nu}(|\fett{q}|,\eta)\int  \dd\cos\theta \, \dd|\fett{q}'| \,
\frac{ |\fett{q}'|}{|\fett{q}|} \left[ \frac{1}{2} - \kappa(|\fett{q}|,|\fett{q}'|,\cos \theta) - \kappa(|\fett{q}'|,|\fett{q}|,\cos \theta)   \right] 
F_{\nu}(\fett{k},\fett{q}',\eta) \\
& \hspace{2mm} \left. -
\int  \dd\cos\theta \, \dd|\fett{q}'| \, \frac{|\fett{q}'|}{|\fett{q}|} 
\left(\frac{5}{8} K_\phi^0-K_\phi^u-K_\phi^t \right)(|\fett{q}|,|\fett{q}'|,\cos \theta,\eta) \;
F_{\phi}(\fett{k},\fett{q}',\eta) \right\},
\label{eq:nunuphiphifinal}
\end{aligned}
\end{equation}
\begin{equation}
\begin{aligned}
&\left( \frac{\partial f_{\nu}}{\partial \eta}\right)^{(1)}_{\nu \phi \leftrightarrow \nu \phi}(\fett{k},\fett{q},\eta) 
=  - \frac{8\mathfrak{g}^4}{(2\pi)^{3}} \left\{
\left[ \left(\frac{13}{8} - \frac{1}{4} \log(4) \right)
X_\phi(|\fett{q}|,\eta) -\frac{1}{2} Z_\phi(|\fett{q}|,\eta) \right] F_{\nu}(\fett{k},\fett{q},\eta) \right. \\
&\hspace{15mm}  -\frac{1}{2} \int  \dd\cos\theta \, \dd|\fett{q}'| \, \frac{|\fett{q}'|}{|\fett{q}|} \left(K_{\phi}^0-K_\phi^u + K_\phi^s \right)(|\fett{q}|,|\fett{q}'|,\cos \theta,\eta)\; F_{\nu}(\fett{k},\fett{q}',\eta) \\
& \hspace{15mm}+ \int  \dd\cos\theta \, \dd|\fett{q}'| \,  \frac{|\fett{q}'|}{ |\fett{q}|} \left[\bar{f}_{\nu}(|\fett{q}|,\eta)\, 
\left( \frac{11}{16} - \kappa(|\fett{q}|,|\fett{q}'|,\cos \theta)  \right) \right. \\
&\hspace{45mm} \left. \left. -\frac{1}{2}   \left( \frac{11}{8} K_{\nu}^0-K_\nu^s - K_\nu^t \right)(|\fett{q}|,|\fett{q}'|,\cos \theta,\eta) \right]\;
F_{\phi}(\fett{k},\fett{q}',\eta) \right\}.
\label{eq:elasticfinal}
\end{aligned}
\end{equation}
The $|\fett{q}|$-dependent functions in the first-order collision terms have the general form
\begin{equation}
\begin{aligned}
X_{i=\nu, \phi} (|\fett{q}|,\eta) & \equiv \frac{1}{|\fett{q}|}\int \mathrm{d|\fett{l}|} \bar{f}_i(|\fett{l}|,\eta) |\fett{l}|  , \\
Z_{i=\nu,\phi} (|\fett{q}|,\eta)&\equiv \frac{1}{2 |\fett{q}|} \int  \mathrm{d |\fett{l}|} \, \bar{f}_i (|\fett{l}|,\eta) \, |\fett{l}| \;
\log\left(\frac{|\fett{l}| |\fett{q}|}{\tilde{m}_\nu^2}\right),
\end{aligned}
\end{equation}
and can be easily evaluated once the time-dependent background phase space distributions~$\bar{f}_i(|\fett{l}|,\eta)$ have been obtained from solving the zeroth-order Boltzmann equation~(\ref{eq:homogeneous}) using the collision integrals~(\ref{eq:0self}) to~(\ref{eq:0elastic}).

The first-order collision terms~(\ref{eq:coll}) to~(\ref{eq:elasticfinal}) also contain five forms of integration kernels.  One of these is time-independent:
\begin{equation}
\begin{aligned}
\kappa(|\fett{q}|,|\fett{q}'|,\cos \theta) & \equiv  \frac{1}{8} \log 
\left[\frac{\sqrt{(|\fett{q}'|-|\fett{q}|+|\fett{q}'+\fett{q}|)^2+4 \tilde{m}_\nu^2} +|\fett{q}'|-|\fett{q}|+|\fett{q}'+\fett{q}|} 
{\sqrt{(|\fett{q}'|-|\fett{q}|-|\fett{q}'+\fett{q}|)^2+4 \tilde{m}_\nu^2} +|\fett{q}'|-|\fett{q}|-|\fett{q}'+\fett{q}|}
\right],
\label{eq:kappatext}
\end{aligned}
\end{equation}
while the rest are given by
\begin{equation}
\begin{aligned}
K_{i=\nu,\phi}^0(|\fett{q}|,|\fett{q}'|,\cos \theta,\eta)
& \equiv \frac{1}{|\fett{q}-\fett{q}'|} 
\int_{R_+}^\infty \dd |\fett{l}'| \, \bar{f}_{i}(|\fett{l}'|,\eta), \\
K_{i=\nu,\phi}^u (|\fett{q}|,|\fett{q}'|,\cos \theta,\eta) & \equiv 
\frac{1}{4}   \int_{R_+}^{\infty} \mathrm{d |\fett{l}'|} \, \bar{f}_{i}(|\fett{l}'|,\eta)  \: \frac{1}{\sqrt{(|\fett{l}'|-|\fett{q}|)^2+\tilde{m}_\nu^2}}, \\
K_{i=\nu,\phi}^t (|\fett{q}|,|\fett{q}'|,\cos \theta,\eta) 
& \equiv 
\frac{1}{8}   \frac{|\fett{q}|+|\fett{q}'|}
  { |\fett{q}-\fett{q}'|^3}
 \int_{R_+}^{\infty} \mathrm{d |\fett{l}'|} \, \bar{f}_{i}(|\fett{l}'|,\eta) \:
 (2 |\fett{l}'| -|\fett{q}|+|\fett{q}'|),\\
  K_{i=\nu,\phi}^s (|\fett{q}|,|\fett{q}'|,\cos \theta,\eta) & \equiv 
  \frac{1}{4}   \int_{R_+}^{\infty} \mathrm{d |\fett{l}'|} \, \bar{f}_{i}(|\fett{l}'|,\eta) \: \frac{1}{|\fett{l}'|+|\fett{q}'|},
  \label{eq:ktext}
  \end{aligned}
\end{equation}
which are again time-dependent through the background phase space distribution~$\bar{f}_i(|\fett{l}|,\eta)$, and $R_+ \equiv (|\fett{q}-|\fett{q}'|+|\fett{q}|-\fett{q}'
|)/2$.

The total scalar collision integral receives contributions from two processes, 
\begin{equation}
\left(\frac{\partial f_{\phi}}{\partial \eta} \right)_{\rm coll,0}= \left(\frac{\partial f_{\phi}}{\partial \eta} \right)_{\phi \phi \leftrightarrow \nu \nu}+\left(\frac{\partial f_{\phi}}{\partial \eta} \right)_{\phi \nu \leftrightarrow \phi \nu} .
\end{equation}
An inspection of the general forms of the zeorth- and first-order collision integrals~(\ref{C[f]0zeroth}) and~(\ref{C[f]0}) for $ij \leftrightarrow kl$ tells us that 
the collision terms for the scalar particle from $\phi \phi \leftrightarrow \nu \nu$ and $\phi \nu \leftrightarrow \phi \nu$ are formally identical to their neutrino counterparts given respectively  in equations~(\ref{eq:nunuphiphifinal}) and ~(\ref{eq:elasticfinal}), save for an exchange of the labels $\nu \leftrightarrow \phi$, and an additional factor of $2$ originating from the two internal degrees of freedom of the neutrino.  Thus, without explicit calculations, we can write
\begin{equation}
\begin{aligned}
\left( \frac{\partial f_{\phi}}{\partial \eta}\right)_{\phi\phi \leftrightarrow \nu \nu}(\fett{k},\fett{q},\eta)
  &= 2  \left( \frac{\partial f_{\nu}}{\partial \eta}\right)_{\nu\nu \leftrightarrow \phi \phi}(\fett{k},\fett{q},\eta) \, (\nu \leftrightarrow \phi),\\
\left( \frac{\partial f_{\phi}}{\partial \eta}\right)_{\phi \nu \leftrightarrow \phi\nu}(\fett{k},\fett{q},\eta) 
&=  2 \left( \frac{\partial f_{\nu}}{\partial \eta}\right)_{\nu \phi \leftrightarrow \nu \phi}(\fett{k},\fett{q},\eta)  \, (\nu \leftrightarrow \phi),
\end{aligned}
\end{equation}
and these relations apply at both zeroth and first order in the phase space perturbations

\section{Boltzmann hierarchy for interacting neutrinos}
\label{sec:hierarchy}

In the absence of interactions, the coefficients on the LHS of the first-order Boltzmann equation~\eqref{Boltzmann equation} are functions only of $|\fett{k}|$, $|\fett{q}|$, and  $\cos \epsilon = \hat{k} \cdot \hat{q}$. It follows that the phase space perturbation~$F(\fett{k},\fett{q})$ of the particle species must also depend only on these parameters, i.e., 
$F(\fett{k},\fett{q})=F(|\fett{k}|,|\fett{q}|, \cos \epsilon)$, in which case it is convenient (and a standard practice) to expand~$F(|\fett{k}|,|\fett{q}|, \cos \epsilon)$  into a Legendre series,
\begin{equation}
\begin{aligned}
\label{eq:legendre}
  F(|\fett{k}|,|\fett{q}|,\cos \epsilon) &= \sum_{\ell =0}^\infty (-\ii)^\ell (2 \ell+1) \,F_{\ell} (|\fett{k}|,|\fett{q}|) \,\LP_\ell (\cos \epsilon) \,, \\
  F_{\ell}(|\fett{k}|,|\fett{q}|) &= \frac{\ii^\ell}{2} \int_{-1}^{1} \dd \cos \epsilon\, F(|\fett{k}|,|\fett{q}|,\cos \epsilon) \,\LP_\ell(\cos \epsilon) \,,
\end{aligned}
\end{equation}
where $F_\ell(|\fett{k}|,|\fett{q}|)$ denotes the $\ell$th multipole moment of the phase space perturbation, and $\LP_\ell(\cos \epsilon)$ is a Legendre polynomial of order~$\ell$.
The LHS of the first-order Boltzmann equation~\eqref{Boltzmann equation} can then be rewritten in terms of these multipole moments, resulting in an infinite hierarchy of equations commonly known as the Boltzmann hierarchy.  

Note that in general there is one hierarchy for each  combination of $\{|\fett{k}|,|\fett{q}|\}$.  However,  in the case of massless neutrinos, the $|\fett{q}|$-direction can be eliminated with an 
integration $\int \dd |\fett{q}| \, |\fett{q}|^3$, because the equality $\epsilon = |\fett{q}|$ removes any $|\fett{q}|$-dependence from the homogeneous part of the hierarchy 
(see also the homogeneous part of equation~\eqref{Boltzmann equation});%
\footnote{In the case $\epsilon = |\fett{q}|$, the time-independent $\partial \bar{f}/\partial \ln |\fett{q}|$ factor preceding the gravitational source term in the first-order Boltzmann
equation~\eqref{Boltzmann equation} alters only the normalisation of the individual~$F_\ell(|\fett{k}|,\fett{q}|)$, and could equally have been absorbed into the definition of~$F_\ell(\fett{k},\fett{q}')$ in the same way that some authors choose to work with $\Psi_\ell(|\fett{k}|,|\fett{q}|) \equiv F_\ell(|\fett{k}|,\fett{q}|)/\bar{f}(|\fett{q}|)$.}
the factor $|\fett{q}|^3$ in the integration simply ensures that the integrated zeroth moment can be identified with the energy density  perturbation~$\delta_\nu$  in the neutrino fluid, the first moment with the velocity divergence~$\theta_\nu$, and the second with the anisotropic stress~$\sigma_\nu$.
 Boltzmann hierarchies for non-interacting massive and massless neutrinos can be found in, e.g.,~\cite{Ma:1995ey}.

The situation complicates somewhat in two respects in the presence of a non-vanishing collision term. The first complication concerns a residual dependence on the 
 azimuthal angle between $\fett{k}$ and $\fett{q}$ in the collision terms presented in section~\ref{sec:collisionterms}. This originates from $F_{\nu,\phi}(\fett{k},\fett{q}')$ in the 
 $\cos \theta$- and $|\fett{q}'|$-integrals, and can be 
 easily seen using the parameterisation
 \begin{equation}
 \begin{aligned} 
\fett{q} &= |\fett{q}|(0,0,1) \,,  \\
\fett{q}' &= |\fett{q}'|(0, \sin\theta, \cos\theta) \,,  \\
 \fett{k} &= |\fett{k}|( \sin\psi \sin\epsilon , \cos\psi \sin\epsilon,\cos\epsilon) \, ,
 \end{aligned}
 \end{equation}
where $\psi$ is the said azimuthal angle.   We handle this $\psi$-dependence  by way of an additional integration $\int_0^{2 \pi} \dd \psi/(2 \pi)$ over the first-order Boltzmann equation~\eqref{Boltzmann equation}; such an averaging procedure will not affect the coefficients on the LHS of the equation, but will remove the unwanted $\psi$-variable from the collision term on the RHS.
 Averaging is also perfectly acceptable from a phenomenological viewpoint, since it is the integrated effect of $F(\fett{k},\fett{q}')$ we observe, not 
$F(\fett{k},\fett{q}')$ {\it per se}. The second complication concerns a residual momentum dependence, the discussion of which we defer to section~\ref{sec:discussion}.


\subsection{Massive scalar limit}

Applying the aforementioned $\psi$-averaging together with the decomposition \eqref{eq:legendre} to the collision term~\eqref{massive} in the limit of an extremely massive scalar, we find the $\ell$th moment of the collision integral to be 
\begin{equation} 
\begin{aligned} \label{eq:colldecomposed_mass}
&\frac{\ii^\ell}{2}  \int_0^{2 \pi} \frac{\dd \psi}{2 \pi}\int^1_{-1}  \dd \cos \epsilon \, \LP_\ell (\cos \epsilon)\,  \left( \frac{\partial f_{\nu}}{\partial \eta}\right)_{\rm coll,m}^{(1)} 
=-\frac{2 \mathrm{N} \mathfrak{g}^4 }{a^4 m_\phi^4(2\pi)^{3}}
\Bigg\{\frac{40}{3}\, |\fett{q}|\, T_{\nu,0}^4 \; F_{\nu,\ell}(|\fett{k}|,|\fett{q}|,\eta)   \\
&\hspace{5mm}   + \int   \mathrm{d|\fett{q}'|}  \, \frac{|\fett{q}'|}{|\fett{q}|}
\left[
\frac{2 |\fett{q}^2| |\fett{q}'|^2}{9} \e^{-| \fett{q}|/T_{\nu,0}}  \left(10 \delta_{\ell 0} - 5 \delta_{\ell 1} + \delta_{\ell 2} \right)
-K^{\rm m}_\ell (|\fett{q}|, |\fett{q}'|)  \right]  F_{\nu,\ell}(|\fett{k}|,|\fett{q}'|,\eta) \Bigg\} , 
\end{aligned}
\end{equation}
where 
\begin{equation}
K^{\rm m}_{\ell}(|\fett{q}|, |\fett{q}'|)= \int_{-1}^1 \dd \cos \theta \, K^{\rm m}(|\fett{q}|, |\fett{q}'|,\cos \theta) \, \LP_{\ell}(\cos \theta),
\label{Kernel_Legendre}
\end{equation}
with $K^{\rm m}(|\fett{q}|, |\fett{q}'|,\cos \theta)$ given in equation~(\ref{eq:km}).  The interested reader may consult appendix~\ref{app:legendre} for the derivation of equation~(\ref{Kernel_Legendre}).

Observe the $\cos \theta$ variable is hidden in $P \equiv \sqrt{|\fett{q}|^2 - 2 |\fett{q}| |\fett{q}'| \cos \theta + |\fett{q}'|^2}$ in the function~$K^{\rm m}(|\fett{q}|, |\fett{q}'|,\cos \theta)$.  It is therefore computationally more convenient to change the integration variable from $\cos \theta$ to $P$.  Thus,  equation~(\ref{Kernel_Legendre}) is equivalently
\begin{equation}
K^{\rm m}_{\ell}(|\fett{q}|, |\fett{q}'|)=\frac{1}{|\fett{q}| |\fett{q}'|} \int_{||\fett{q}|-|\fett{q}'||}^{|\fett{q}|+|\fett{q}'|} \dd P  \, P\,  K^{\rm m}(|\fett{q}|, |\fett{q}'|,P) \, 
\LP_{\ell}\left(\frac{|\fett{q}|^2 + |\fett{q}'|^2 - P^2}{2 |\fett{q}| |\fett{q}'|} \right),
\label{eq:massivekernelP}
\end{equation}
where, in this case, the integrand is simple enough (a polynomial in $P$, multiplied by an exponential whose exponent is linear in $P$) that (lengthy) analytical solutions exist in principle to all orders~$\ell$.  We shall not write out these expressions however, but simply show numerical evaluations of $K^{\rm m}_{\ell}(|\fett{q}|, |\fett{q}'|)$ for $\ell = 0,1,2,3$
in figure \ref{Fig:MassiveKernel}. 

\begin{figure}[t]
    \centering
    \begin{subfigure}[b]{0.49\textwidth}
        \centering
        \caption{$K^{\rm m}_0$}
        \includegraphics[width=\textwidth]{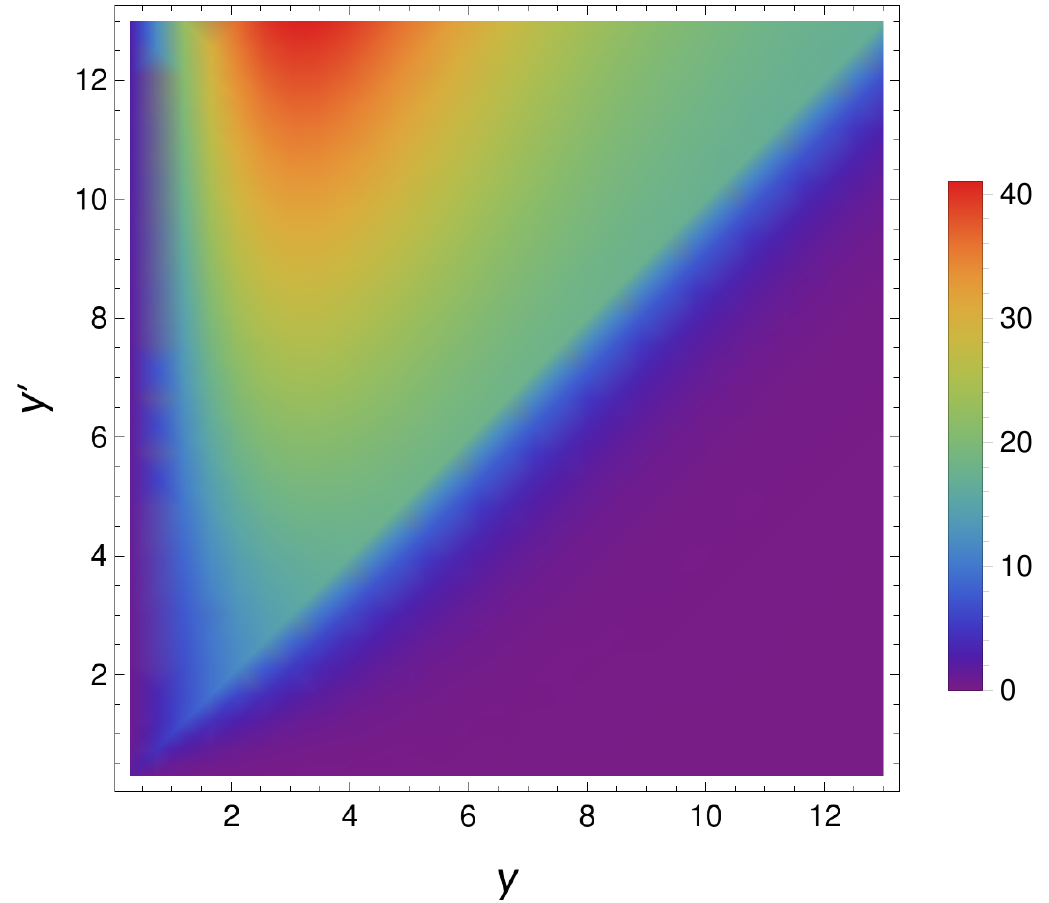}
    \end{subfigure}%
    ~ 
    \begin{subfigure}[b]{0.49\textwidth}
        \centering
        \caption{$K^{\rm m}_1$}
        \includegraphics[width=\textwidth]{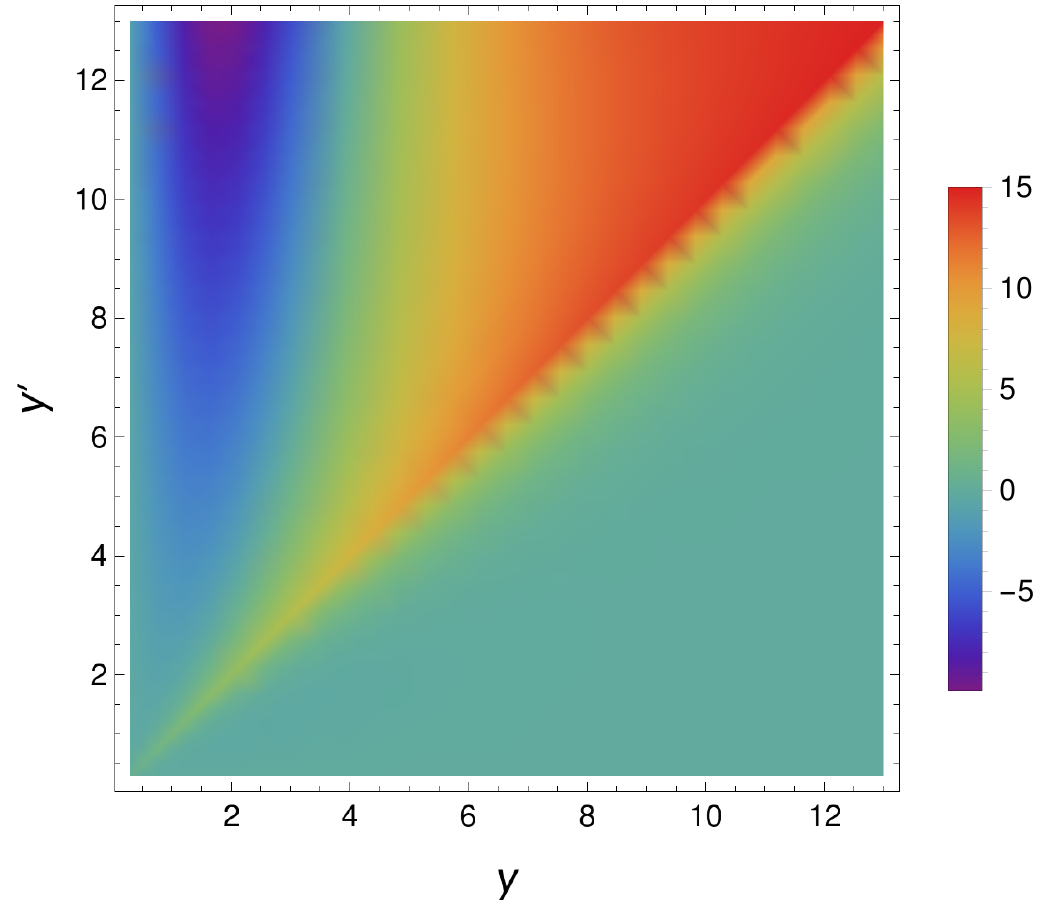}
    \end{subfigure}
    
    \begin{subfigure}[b]{0.49\textwidth}
        \centering
        \caption{$K^{\rm m}_2$}
        \includegraphics[width=\textwidth]{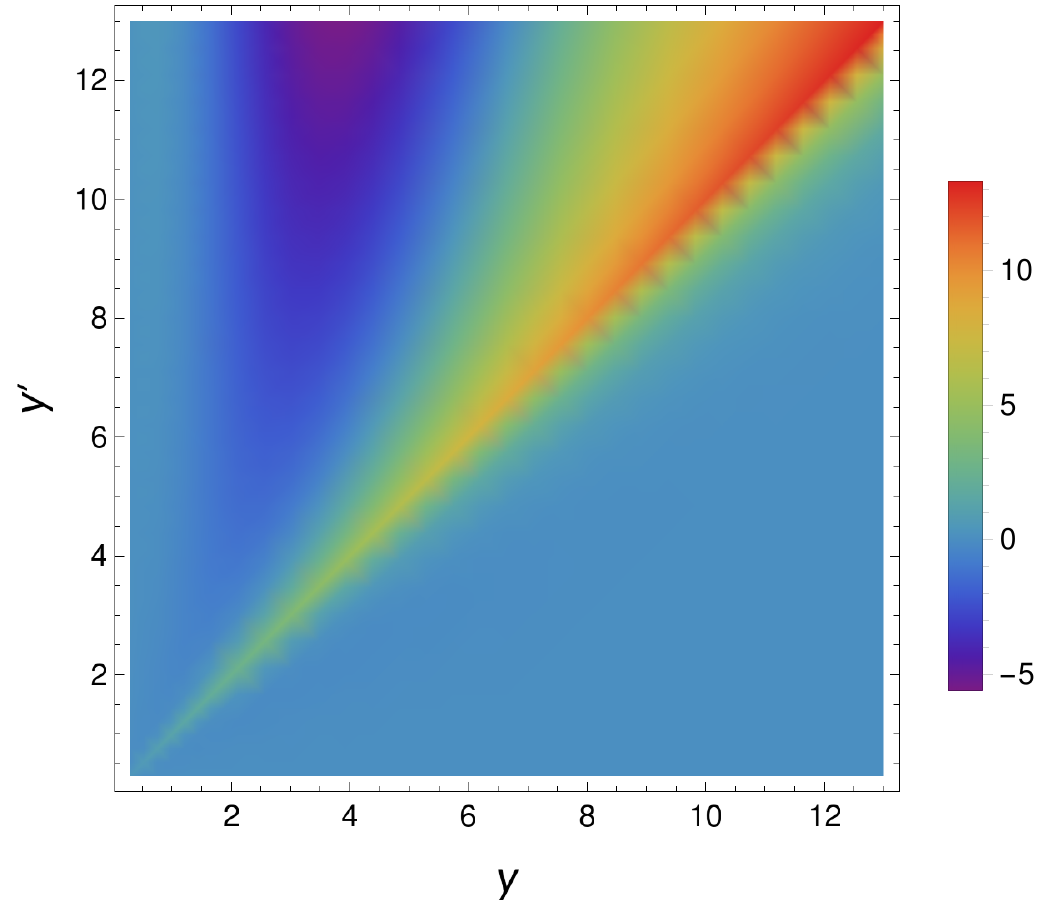}
    \end{subfigure}%
    ~ 
    \begin{subfigure}[b]{0.49\textwidth}
        \centering
        \caption{$K^{\rm m}_3$}
        \includegraphics[width=\textwidth]{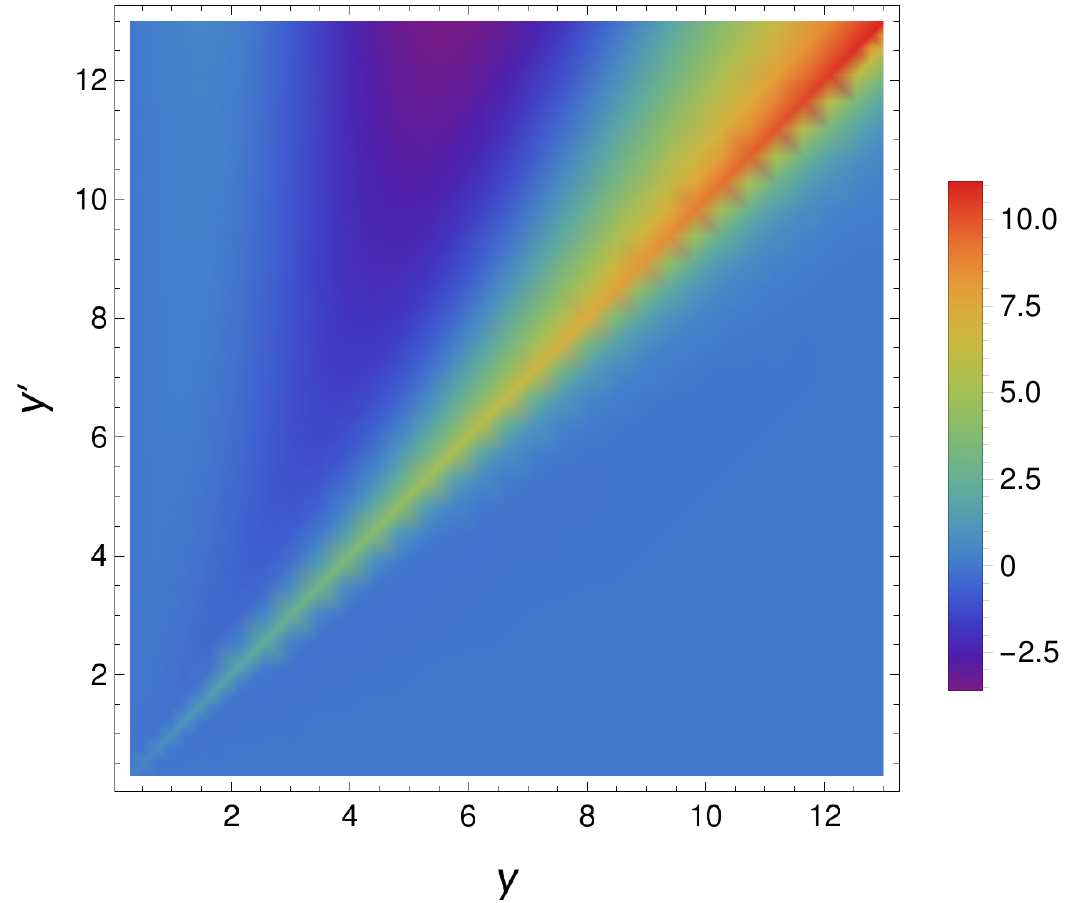}
    \end{subfigure}
    \caption{Integral kernels $K^{\rm m}_{0,1,2,3}$, defined in equation \eqref{eq:massivekernelP} and plotted here in units of $T_{\nu,0}^4$, 
    as functions of $y\equiv |\fett{q}|/ T_{\nu,0}$ and $y' \equiv |\fett{q}'|/ T_{\nu,0}$.}
    \label{Fig:MassiveKernel}
\end{figure}

It is now straightforward to write down the Boltzmann hierarchy for self-interacting neutrinos using the Legendre-decomposed collision term~(\ref{eq:colldecomposed_mass}). 
By matching up the $\ell$th moment of the collision term with the corresponding (well-known~\cite{Ma:1995ey}) decomposition of the LHS of equation~\eqref{Boltzmann equation}, we find
\begin{equation}
\begin{aligned} 
\label{finalBoltzmann_mass}
 \dot{F}_{\nu,0}(q) =&  - k F_{\nu,1}(q) + \frac{1}{6} \frac{\partial \bar{f_{\nu}}}{\partial \ln q} \dot{h} - \frac{40}{3} G^{\rm m} q \,T_{\nu,0}^4 \,
 F_{\nu,0}(q)  \\
 &\hspace{15mm}+ G^{\rm m} \int  \dd q' \, \frac{q'}{q} \left[ K^{\rm m}_0(q,q') - \frac{20}{9} q^2\, {q'}^2 e^{-q/T_{\nu,0}}
  \right] \, F_{\nu,0}(q') \, , \\ 
\dot{F}_{\nu,1}(q) =&  - \frac{2}{3}k F_{\nu,2}(q)+ \frac{1}{3}k F_{\nu,0}(q) - \frac{40}{3} G^{\rm m} q \,T_{\nu,0}^4 \, F_{\nu,1}(q) \\
&\hspace{15mm}+ G^{\rm m} \int  \dd q' \, \frac{q'}{q} \left[ K^{\rm m}_1(q,q')
+\frac{10}{9} q^2 \, {q'}^2 e^{-q/T_{\nu,0}} \right] \, F_{\nu,1}(q') \, , \\ 
\dot{F}_{\nu,2} (q) =&  - \frac{3}{5}k F_{\nu,3}(q) + \frac{2}{5}k F_{\nu,1}(q)- \frac{\partial \bar{f}_{\nu}}{\partial \ln q} \left(\frac{2}{5} \dot{\tilde{\eta}}+ \frac{1}{15} \dot{h} \right) -  \frac{40}{3} G^{\rm m} q \,T_{\nu,0}^4 \, F_{\nu,2}(q) \\
&\hspace{15mm} + G^{\rm m}  \int  \dd q' \, \frac{q'}{q} \left[K^{\rm m}_2(q,q') - \frac{2}{9} q^2 \, {q'}^2 e^{-q/T_{\nu,0}}\right] \, F_{\nu,2}(q') \, , \\
\dot{F}_{\nu,\ell>2}(q) =& \frac{k}{2\ell+1} \left[ \ell F_{\nu,\ell-1}(q)-(\ell+1)F_{\nu,\ell+1}(q) \right] 
  - \frac{40}{3} G^{\rm m} q \,T_{\nu,0}^4 \, F_{\nu,\ell}(q) \\
&\hspace{15mm} + G^{\rm m} \int  \dd q' \, \frac{q'}{q} K^{\rm m}_\ell(q,q') \, F_{\nu,\ell}(q') \, ,
\end{aligned}
\end{equation}
where we have defined $G^{\rm m}  \equiv 2 \mathrm{N} \mathfrak{g}^4 /(a^4 m_\phi^4(2\pi)^{3})$,
and used the shorthand notation $q \equiv |\fett{q}|$, $q'\equiv |\fett{q}'|$, and $k \equiv |\fett{k}|$.

\subsection{Massless scalar limit}\label{sec:masslesshierarchy}

In the opposite limit of a massless scalar, we find that the three interaction channels make the following contributions to the $\ell$th multipole moment of the neutrino collision integral:
\begin{equation}
\begin{aligned}
&\frac{\ii^\ell}{2}  \int_0^{2 \pi} \frac{\dd \psi}{2 \pi}\int^1_{-1}  \dd \cos \epsilon \, \LP_\ell (\cos \epsilon)\,  \left(\frac{\partial f_\nu}{\partial \eta} \right)^{(1)}_{{\nu \nu \leftrightarrow \nu \nu}} 
 =-  \frac{ 6 \mathfrak{g}^4}{(2\pi)^{3}}
\Bigg\{ X_{\nu}(|\fett{q}|,\eta)  F_{\nu,\ell}(|\fett{k}|,|\fett{q}|) \\
& \hspace{15mm}+ \int   \dd|\fett{q}'| \,\frac{|\fett{q}'|}{|\fett{q}|}  \left[
\bar{f}_{\nu}(|\fett{q}|,\eta) \delta_{\ell 0} -K_{\nu,\ell}^0
(|\fett{q}|,|\fett{q}'|,\eta) \right] F_{\nu,\ell}(|\fett{k}|,|\fett{q}'|,\eta) \Bigg\} ,
\end{aligned}
\end{equation}
\begin{equation}
\begin{aligned}
&\frac{\ii^\ell}{2}  \int_0^{2 \pi} \frac{\dd \psi}{2 \pi}\int^1_{-1}  \dd \cos \epsilon \, \LP_\ell (\cos \epsilon)\, \left( \frac{\partial f_{\nu}}{\partial \eta}\right)^{(1)}_{\nu \nu \leftrightarrow \phi \phi}
 =  \\
& \hspace{12mm} \frac{2 \mathfrak{g}^4}{(2\pi)^{3}} \left\{
\left[\left(\frac{3}{2} - \frac{1}{2} \log(4) \right)
X_\nu(|\fett{q}|,\eta) - Z_\nu(|\fett{q}|,\eta) \right] F_{\nu,\ell}(|\fett{k}|,|\fett{q}|,\eta) \right.\\
& \hspace{22mm}+ \bar{f}_{\nu}(|\fett{q}|,\eta)\int  \dd|\fett{q}'| \, 
\frac{ |\fett{q}'|}{|\fett{q}|} \left[\delta_{\ell 0} - \kappa_\ell(|\fett{q}|,|\fett{q}'|) - \kappa_\ell(|\fett{q}'|,|\fett{q}|)   \right] 
F_{\nu,\ell}(|\fett{k}|,|\fett{q}'|,\eta) \\
& \hspace{22mm} \left. -
\int   \dd|\fett{q}'| \, \frac{|\fett{q}'|}{|\fett{q}|} 
\left(\frac{5}{8} K_{\phi,\ell}^0-K_{\phi,\ell}^u-K_{\phi,\ell}^t \right)(|\fett{q}|,|\fett{q}'|,\eta) \;
F_{\phi,\ell}(|\fett{k}|,|\fett{q}'|,\eta) \right\},
\end{aligned}
\end{equation}
\begin{equation}
\begin{aligned}
&\frac{\ii^\ell}{2}  \int_0^{2 \pi} \frac{\dd \psi}{2 \pi}\int^1_{-1}  \dd \cos \epsilon \, \LP_\ell (\cos \epsilon)\,  
\left( \frac{\partial f_{\nu}}{\partial \eta}\right)^{(1)}_{\nu \phi \leftrightarrow \nu \phi}
= \\
& \hspace{10mm}- \frac{8\mathfrak{g}^4}{(2\pi)^{3}} \left\{
\left[ \left(\frac{13}{8} - \frac{1}{4} \log(4) \right)
X_\phi(|\fett{q}|,\eta) -\frac{1}{2} Z_\phi(|\fett{q}|,\eta) \right] F_{\nu,\ell}(|\fett{k}|,|\fett{q}|,\eta) \right. \\
&\hspace{25mm}  -\frac{1}{2} \int   \dd|\fett{q}'| \, \frac{|\fett{q}'|}{|\fett{q}|} \left(K_{\phi,\ell}^0-K_{\phi,\ell}^u + K_{\phi,\ell}^s \right)(|\fett{q}|,|\fett{q}'|,\eta)\; F_{\nu,\ell}(\fett{k},\fett{q}',\eta) \\
& \hspace{25mm}+ \int   \dd|\fett{q}'| \, \frac{|\fett{q}'|}{|\fett{q}|} \left[\bar{f}_{\nu}(|\fett{q}|,\eta)\,  
\left(\frac{11}{8} \delta_{\ell 0} - \kappa_\ell(|\fett{q}|,|\fett{q}'|)  \right) \right. \\
&\hspace{40mm} \left. \left. -\frac{1}{2}   \left( \frac{11}{8}  K_{\nu,\ell}^0-K_{\nu,\ell}^s - K_{\nu,\ell}^t \right)(|\fett{q}|,|\fett{q}'|,\eta) \right]\;
F_{\phi,\ell}(|\fett{k}|,|\fett{q}'|,\eta) \right\},
\end{aligned}
\end{equation}
where the $\ell$th-order kernels $\kappa_\ell$ and $K_{i,\ell}^{0,s,t,u}$ are related to $\kappa$ and $K_{i}^{0,s,t,u}$ of equations~(\ref{eq:kappatext}) and~(\ref{eq:ktext}) 
in a manner analogous to  equation~(\ref{Kernel_Legendre}).

To evaluate the $\ell$th order kernels, we find it convenient to change integration variables from $\cos \theta$ to $P \equiv \sqrt{|\fett{q}|^2 - 2 |\fett{q}| |\fett{q}'| \cos \theta + |\fett{q}'|^2}$.  Thus, for the time-independent $\kappa_\ell$, we can immediately generalise equation~(\ref{eq:massivekernelP}) and write
\begin{equation}
\kappa_{\ell}(|\fett{q}|, |\fett{q}'|)=\frac{(-1)^\ell}{|\fett{q}| |\fett{q}'|} \int_{||\fett{q}|-|\fett{q}'||}^{|\fett{q}|+|\fett{q}'|} \dd P  \, P\,  \kappa(|\fett{q}|, |\fett{q}'|,P) \, 
\LP_{\ell}\left(\frac{|\fett{q}|^2 + |\fett{q}'|^2 - P^2}{2 |\fett{q}| |\fett{q}'|} \right),
\label{eq:kappaP}
\end{equation}
where the prefactor $(-1)^\ell$ follows from $\LP_\ell\, (-x) = (-1)^\ell \, \LP_\ell\, (x)$, and 
comes about because it is in fact $|\fett{q}+\fett{q}'|$, rather than $|\fett{q}-\fett{q}'|$, that appears in $\kappa$ kernel. 
Again, analytical solutions exist in principle to all orders~$\ell$, but we shall not write them out here because of their length.

The same generalisation of  equation~(\ref{eq:massivekernelP}) applies in principle also to the time-dependent kernels~$K_{i,\ell}^{0,s,t,u}$.  However, because the variable~$P$ now appears also in the lower integration limit~$R_+$ of the $|\fett{l}'|$-integral, instead of a brute force application of~(\ref{eq:massivekernelP}) a better alternative is to split the double $(|\fett{l}'|,P)$-integral  into four parts, along with new integration limits, according to
\begin{equation}
\begin{aligned}
\int_{||\fett{q}|-|\fett{q}'||}^{|\fett{q}|+|\fett{q}'|} \dd P \! \int_{R_+}^\infty \dd |\fett{l}'| & \to 
 \Theta\left(|\fett{q}|-|\fett{q}'| \right) \left[\int_{|\fett{q}|-|\fett{q}'|}^{|\fett{q}|} \dd |\fett{l}'| 
 \int_{|\fett{q}|-|\fett{q}'|}^{2|\fett{l}'|-|\fett{q}|+|\fett{q}'|}  \dd P
+\int_{|\fett{q}|}^\infty \dd |\fett{l}'| 
 \int_{|\fett{q}|-|\fett{q}'|}^{|\fett{q}|+|\fett{q}'|}  \dd P \right] \\
& +\Theta\left(|\fett{q}'|-|\fett{q}| \right) \left[ \int_0^{|\fett{q}|} \dd |\fett{l}'| 
 \int_{|\fett{q}'|-|\fett{q}|}^{2|\fett{l}'|-|\fett{q}|+|\fett{q}'|}  \dd P +\int_{|\fett{q}|}^\infty \dd |\fett{l}'| 
 \int_{|\fett{q}'|-|\fett{q}|}^{|\fett{q}|+|\fett{q}'|}  \dd P \right] \\
 & \equiv\int_{ {\cal R}_1}+ \int_{{\cal R}_2} + \int_{{\cal R}_3} + \int_{{\cal R}_4}.
 \label{eq:splitintegral}
 \end{aligned}
 \end{equation}
The interested reader can find the justification for this splitting in appendix~\ref{app:zeroth}.  Applying the splitting to  the kernels~$K_{i,\ell}^{0,s,t,u}$, we find
\begin{equation}
\begin{aligned}
K_{i,\ell}^0(|\fett{q}|,|\fett{q}'|,\eta) & = \frac{1}{|\fett{q}| |\fett{q}'|} \sum_{n=1}^4 \int_{{\cal R}_n} \dd |\fett{l}'| \, \bar{f}_i(|\fett{l}'|,\eta) \int \dd P \, 
\LP_{\ell}\left(\frac{|\fett{q}|^2 + |\fett{q}'|^2 - P^2}{2 |\fett{q}| |\fett{q}'|} \right), \\
K_{i,\ell}^u (|\fett{q}|,|\fett{q}'|,\eta) & =
\frac{1}{4|\fett{q}| |\fett{q}'|}   \sum_{n=1}^4 \int_{{\cal R}_n}  \dd |\fett{l}'| \frac{\bar{f}_{i}(|\fett{l}'|,\eta)}{\sqrt{(|\fett{l}'|-|\fett{q}|)^2+\tilde{m}_\nu^2}} \, \int  \dd P \, P \, \LP_{\ell}\left(\frac{|\fett{q}|^2 + |\fett{q}'|^2 - P^2}{2 |\fett{q}| |\fett{q}'|} \right), \\
K_{i,\ell}^t (|\fett{q}|,|\fett{q}'|,\eta) 
& =
\frac{|\fett{q}|+|\fett{q}'|}{8 |\fett{q}| |\fett{q}'|}   \sum_{n=1}^4\int_{{\cal R}_n}  \mathrm{d |\fett{l}'|} \, \bar{f}_{i}(|\fett{l}'|,\eta) \:
 (2 |\fett{l}'| -|\fett{q}|+|\fett{q}'|) \\
 & \hspace{30mm}\times  \int   \dd P \, \frac{1}{P^2} \, \LP_{\ell}\left(\frac{|\fett{q}|^2 + |\fett{q}'|^2 - P^2}{2 |\fett{q}| |\fett{q}'|} \right),\\
  K_{i,\ell}^s (|\fett{q}|,|\fett{q}'|,\eta) &= \frac{1}{4|\fett{q}| |\fett{q}'|} 
   \sum_{n=1}^4 \int_{{\cal R}_n}  \mathrm{d |\fett{l}'|}\,  \frac{\bar{f}_{i}(|\fett{l}'|,\eta)}{|\fett{l}'|+|\fett{q}'|}
  \int \dd P \, P \, \LP_{\ell}\left(\frac{|\fett{q}|^2 + |\fett{q}'|^2 - P^2}{2 |\fett{q}| |\fett{q}'|} \right), 
  \label{eq:klegendreP}
  \end{aligned}
\end{equation}
where ${\cal R}_n$ denotes the $n$th integration domain given in equation~(\ref{eq:splitintegral}).

Because the $|\fett{l}'|$-integration limits are now independent of $P$, the $P$-integration can be performed first over time-independent integrands (whose polynomial nature makes the operation very easy) before we integrate over $|\fett{l}'|$.   For the latter, the  integrand still contains an unknown, time-dependent background  phase space distribution~$\bar{f}_i(|\fett{l}'|,\eta)$, which has to be obtained from solving the zeroth-order Boltzmann equation~(\ref{eq:homogeneous}) in conjunction  with the collision integrals~(\ref{eq:0self}) to~(\ref{eq:0elastic}).  Note also that the $K_{i,\ell}^t$ kernel is divergent at $|\fett{q}|=|\fett{q}'|$ due to an approximately massless mediating neutrino. As for the $K_{i,\ell}^u$ kernel (see appendix \ref{Computing C3_ann}), we regularise this divergence by introducing a comoving neutrino mass in the lower limit of the $P$-integration in equation~\eqref{eq:splitintegral}, i.e., $||\fett{q}|-|\fett{q}'||\rightarrow \sqrt{(|\fett{q}|-|\fett{q}'|)^2+\tilde{m}^2_{\nu}}$.

Then, collecting the collision kernels from all relevant processes, the  Boltzmann hierarchy for a neutrino component interacting with itself and with a scalar particle population can be written as
\begin{equation}
\begin{aligned} 
\label{finalBoltzmann_neutrino}
 \dot{F}_{\nu,0}(q) =&  - k F_{\nu,1}(q) + \frac{1}{6} \frac{\partial \bar{f_{\nu}}}{\partial \ln q} \dot{h} - G^{\rm 0} \mathcal{X}^{\nu}(q)\, F_{\nu,0}(q) \\
 &\hspace{15mm} + G^{\rm 0} \int  \dd q' \, \frac{q'}{q} \, \mathcal{K}^{\nu}_0(q,q') \, F_{\nu,0}(q') + G^{\rm 0} \int  \dd q' \, \frac{q'}{q} \, \mathfrak{K}^{\nu}_0(q,q') \, F_{\phi,0}(q') \, , \\ 
\dot{F}_{\nu,1}(q) =&  - \frac{2}{3}k F_{\nu,2}(q)+ \frac{1}{3}k F_{\nu,0}(q) - G^{\rm 0} \mathcal{X}^{\nu}(q)\, F_{\nu,1}(q) \\
&\hspace{15mm} + G^{\rm 0} \int  \dd q' \, \frac{q'}{q} \, \mathcal{K}^{\nu}_1(q,q') \, F_{\nu,1}(q')  + G^{\rm 0} \int  \dd q' \,\frac{q'}{q}\,  \mathfrak{K}^{\nu}_1(q,q') \, F_{\phi,1}(q')  , \\ 
\dot{F}_{\nu,2} (q) =&  - \frac{3}{5}k F_{\nu,3}(q) + \frac{2}{5}k F_{\nu,1}(q)- \frac{\partial \bar{f}_{\nu}}{\partial \ln q} \left(\frac{2}{5} \dot{\tilde{\eta}}+ \frac{1}{15} \dot{h} \right) - G^{\rm 0} \mathcal{X}^{\nu}(q)\, F_{\nu,2}(q) \\
& \hspace{15mm} + G^{\rm 0} \int  \dd q' \,  \frac{q'}{q}\, \mathcal{K}^{\nu}_2(q,q') \, F_{\nu,2}(q')  + G^{\rm 0} \int  \dd q' \,  \frac{q'}{q}\, \mathfrak{K}^{\nu}_2(q,q') \, F_{\phi,2}(q')\, , \\
\dot{F}_{\nu,\ell>2}(q) =& \frac{k}{2\ell+1} \left[ \ell F_{\nu,\ell-1}(q)-(\ell+1)F_{\nu,\ell+1}(q) \right] 
- G^{\rm 0} \mathcal{X}^{\nu}(q)\, F_{\nu,\ell}(q) \\
& \hspace{15mm}  + G^{\rm 0} \int  \dd q' \, \frac{q'}{q}\, \mathcal{K}^{\nu}_\ell(q,q') \, F_{\nu,\ell}(q')  + G^{\rm 0} \int  \dd q' \,  \frac{q'}{q}\, \mathfrak{K}^{\nu}_{\ell}(q,q') \, F_{\phi,\ell}(q')\, ,
\end{aligned}
\end{equation}
where we have defined $G^{\rm 0} \equiv 2 \mathfrak{g}^4/(2 \pi)^3$,
and
\begin{equation}
\begin{aligned}
{\cal X}^\nu (q,\eta)  \equiv & \;  Z_{\nu}(q,\eta)-2 Z_{\phi}(q,\eta) +\left[ \frac{13}{2}-\log(4) \right] X_{\phi}(q,\eta) +\left[ \frac{3}{2}+ \frac{1}{2}\log(4) \right] X_{\nu}(q,\eta), \\
{\cal K}^\nu_\ell (q,q',\eta)  \equiv & \vphantom{\frac11} \; 3 K^0_{\nu,\ell}(q,q',\eta)+2K^0_{\phi,\ell}(q,q',\eta) -2K^u_{\phi,\ell}(q,q',\eta)+2K^s_{\phi,\ell}(q,q',\eta)   \\
& \vphantom{\frac11} \; -\bar{f}_{\nu}(q,\eta) \left[ 2\delta_{\ell 0}+ \kappa_{\ell}(q,q')+\kappa_{\ell}(q,q') \right],  \\
\mathfrak{K}^\nu_\ell (q,q') \equiv & \; \bar{f}_{\nu}(q,\eta) \left[ 4 \,\kappa_{\ell}(q,q')-\frac{11}{2} \delta_{\ell 0} \right] -\frac{5}{8} K^0_{\phi,\ell}(q,q',\eta)+K^u_{\phi,\ell}(q,q',\eta)+K^t_{\phi,\ell}(q,q',\eta)  \\
&  \; + \frac{11}{4} K^0_{\nu,\ell}(q,q',\eta)- 2 K^s_{\nu,\ell}(q,q',\eta)-2K^t_{\nu,\ell}(q,q',\eta) 
\label{SumMassless}
\end{aligned}
\end{equation}
sum up contributions from all three scattering processes involving the neutrino. 
To give the reader a rough idea of how the functions ${\cal X}^{\nu}(q,\eta)$, ${\cal K}^{\nu}_{\ell}(q,q',\eta)$, and $\mathfrak{K}^{\nu}_{\ell}(q,q',\eta)$ might look, we show in figures~\ref{Fig:Damping}, \ref{Fig:MasslessKernel1}, and \ref{Fig:MasslessKernel2} the solutions of~(\ref{SumMassless}) for $\ell= 0, 1, 2, 3$, assuming a Maxwell--Boltzmann background distribution for both neutrinos and scalars, and a neutrino mass of $m_{\nu}=0.05$~eV, all evaluated at the time of recombination.

Note that because the $t$ and $u$-channel processes mediated by an approximately massless neutrino always incur an infrared divergence and the regularisation process requires that we introduce a small neutrino mass~$m_\nu$, the exact choice of the $m_\nu$ will have an impact on the kernel values.
In particular, for very small $m_\nu$ values, the presence of the $-2 Z_\phi (q, \eta)$ term in the expression for  ${\cal X}^{\nu}(q,\eta)$ in equation~(\ref{SumMassless}) may cause 
 ${\cal X}^{\nu}(q,\eta)$ to turn negative, which would ruin the interpretation of  $- G^{\rm 0} \mathcal{X}^{\nu}(q)\, F_{\nu,\ell}(q)$ in equation~\eqref{finalBoltzmann_neutrino} as a damping term.  We have investigated this possibility, and found that, fortunately, for $m_{\nu}$ values as small as 0.005~eV (roughly the minimum $m_{\nu, 2} \sim 0.007$~eV
 established by oscillation experiments~\cite{Tortola:2012te}), the function ${\cal X}^{\nu}(q,\eta)$ is always positive within the momentum range $0.3 < q/T_{\nu,0} < 13$. Thus,  the interpretation of $- G^{\rm 0} \mathcal{X}^{\nu}(q)\, F_{\nu,\ell}(q)$ as a damping term is valid for a wide range of neutrino masses.

\begin{figure}[t]
\centering
\includegraphics[width=0.45\textwidth]{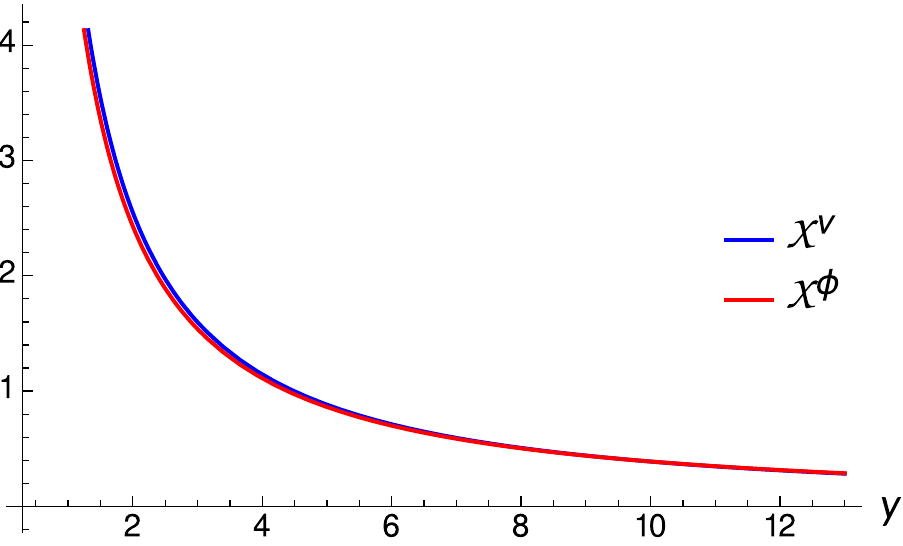}~~
\caption{The functions~${\cal X}^{\nu}(|\fett{q}|,\eta)$ (blue) and ${\cal X}^{\phi}(|\fett{q}|,\eta)$ (red), 
defined in equation~\eqref{SumMassless} and shown here in units of $T_{\nu,0}$, as functions of $y \equiv |\fett{q}|/T_{\nu,0}$, assuming a Maxwell-Boltzmann background distribution for both the neutrinos and scalar particles. Both functions have been evaluated at the time of recombination  for a neutrino mass of $m_{\nu}=0.05$ eV. }
\label{Fig:Damping}
\end{figure}

\begin{figure}[t]
    \centering
    \begin{subfigure}[b]{0.49\textwidth}
        \centering
        \caption{$\log_{10}(|{\cal K}^{\nu}_0|)$}
        \includegraphics[width=\textwidth]{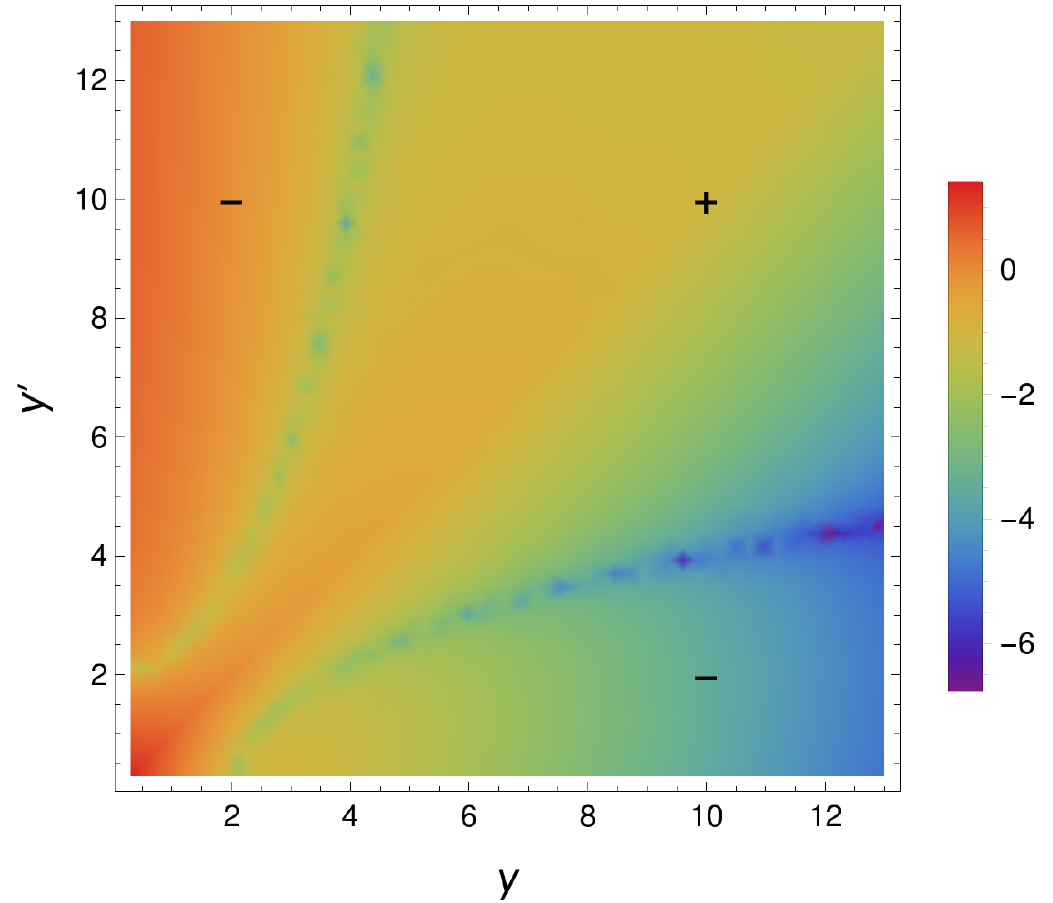}
    \end{subfigure}%
    ~ 
    \begin{subfigure}[b]{0.49\textwidth}
        \centering
        \caption{$\log_{10}(|{\cal K}^{\nu}_1|)$}
        \includegraphics[width=\textwidth]{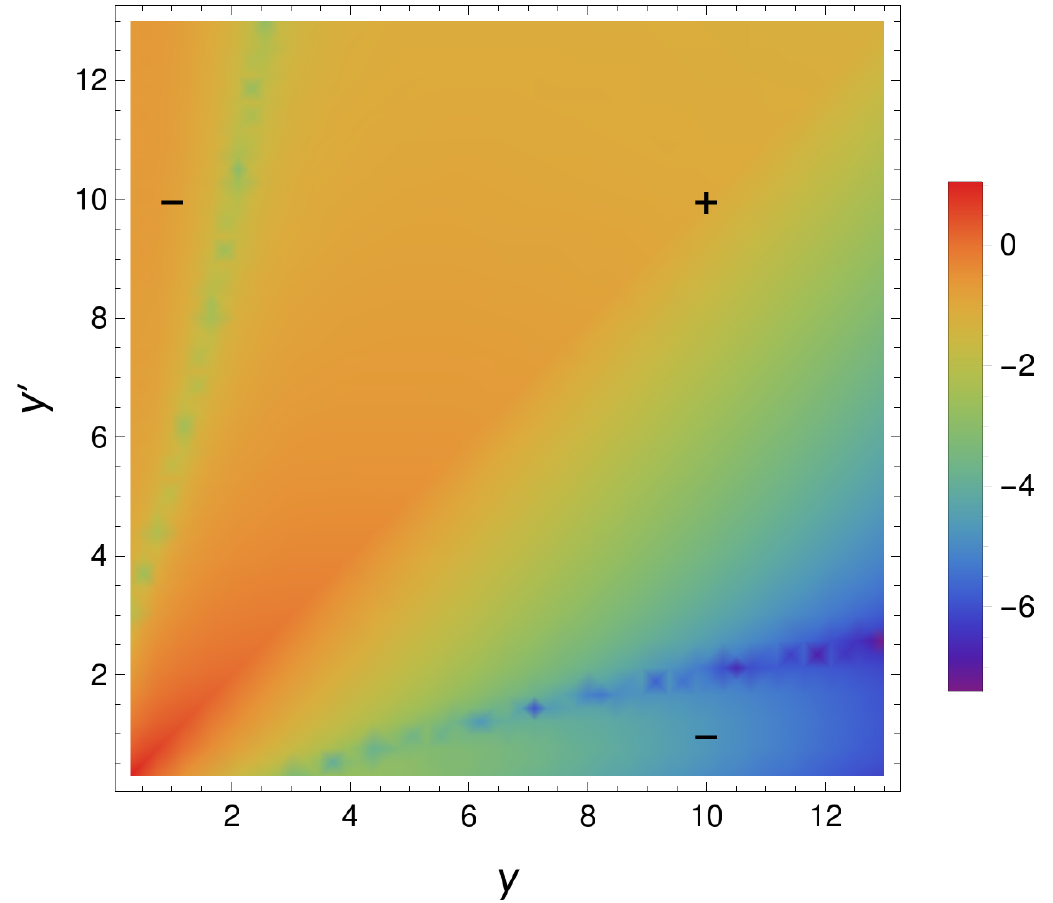}
    \end{subfigure}
    
    \begin{subfigure}[b]{0.49\textwidth}
        \centering
        \caption{$\log_{10}(|{\cal K}^{\nu}_2|)$}
        \includegraphics[width=\textwidth]{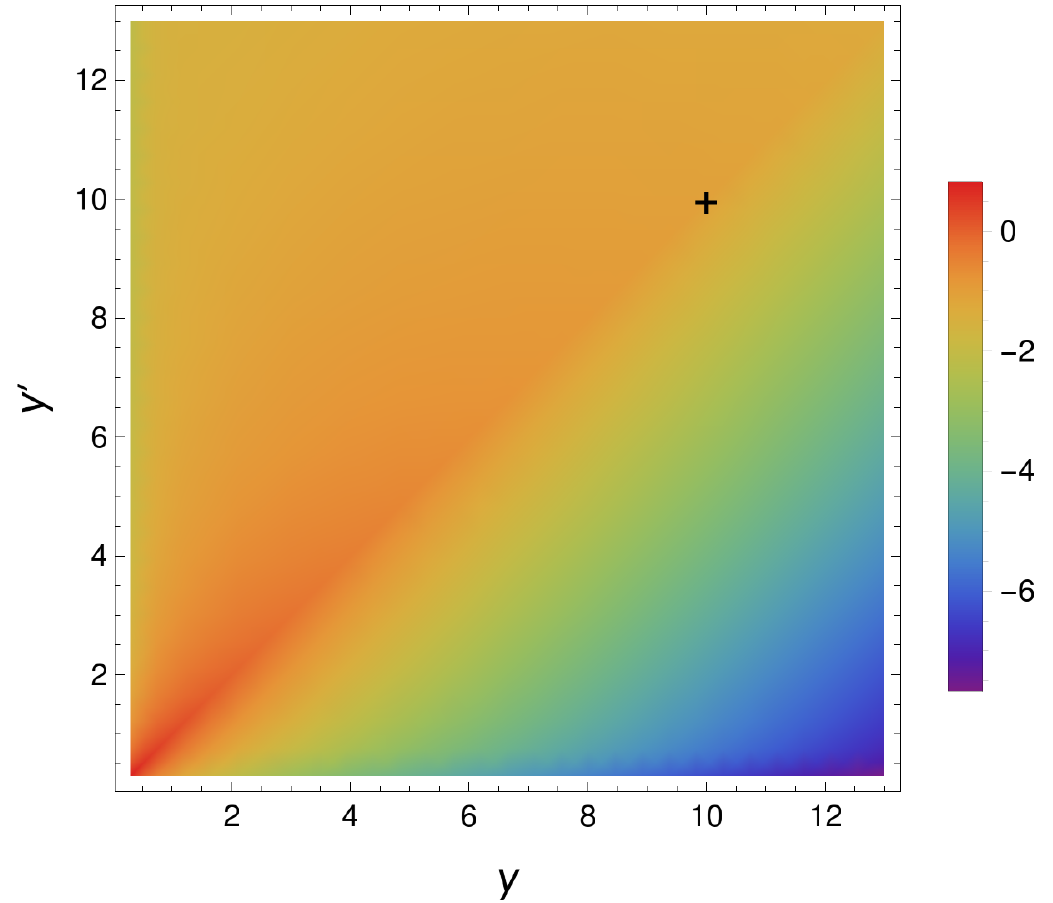}
    \end{subfigure}%
    ~ 
    \begin{subfigure}[b]{0.49\textwidth}
        \centering
        \caption{$\log_{10}(|{\cal K}^{\nu}_3|)$}
        \includegraphics[width=\textwidth]{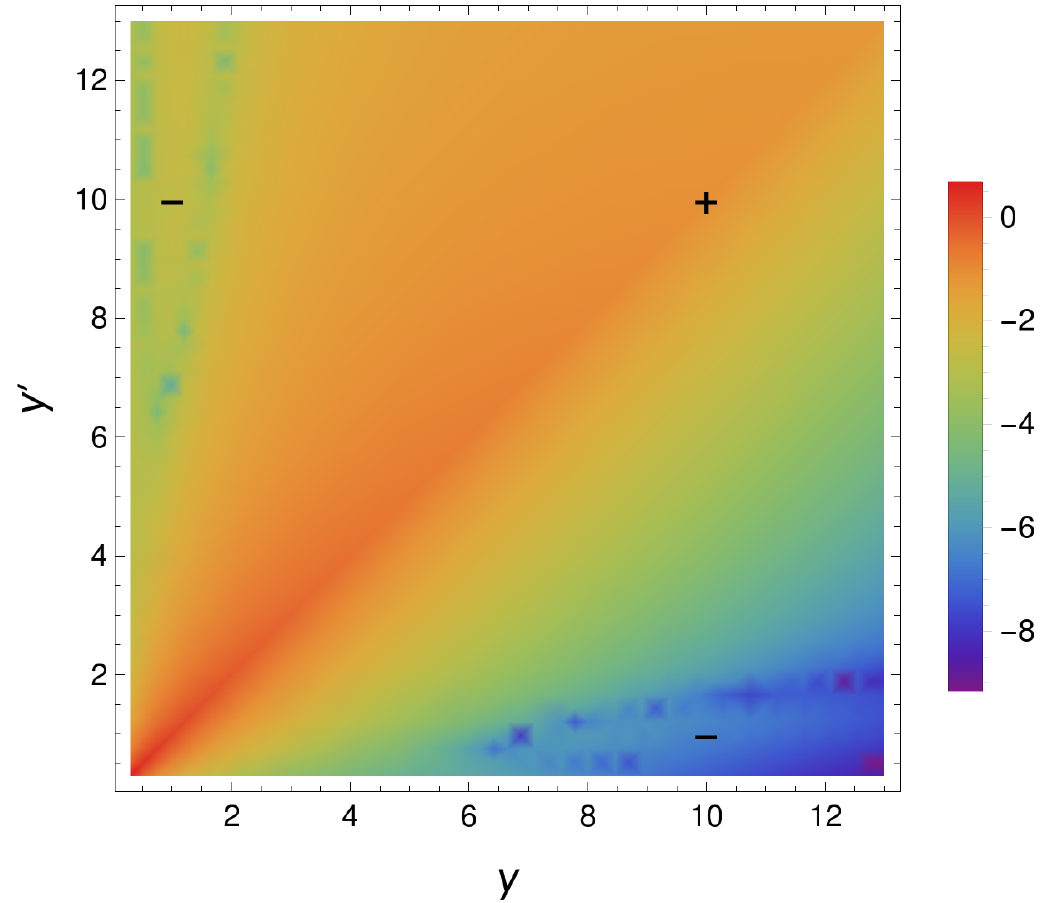}
    \end{subfigure}
    \caption{Integral kernels ${\cal K}^{\nu}_{0,1,2,3}$, defined in equation~\eqref{SumMassless} and shown here as $\log_{10}(|{\cal K}^{\nu}_{0,1,2,3}|)$, as functions of $y\equiv |\fett{q}|/ T_{\nu,0}$ and $y' \equiv |\fett{q}'|/ T_{\nu,0}$, assuming a Maxwell--Boltzmann background distribution for the neutrinos and scalar particles. The functions have been evaluated at the time of recombination for a neutrino mass of $m_{\nu}=0.05$ eV.  For ${\cal K}^{\nu}_{0}$, ${\cal K}^{\nu}_{1}$ and ${\cal K}^{\nu}_{3}$ the two green--blue lines mark the points where the functions flip sign.}
       \label{Fig:MasslessKernel1}
\end{figure}

\begin{figure}[t]
    \centering
    \begin{subfigure}[b]{0.49\textwidth}
        \centering
        \caption{$\log_{10}(|\mathfrak{K}^{\nu}_0|)$}
        \includegraphics[width=\textwidth]{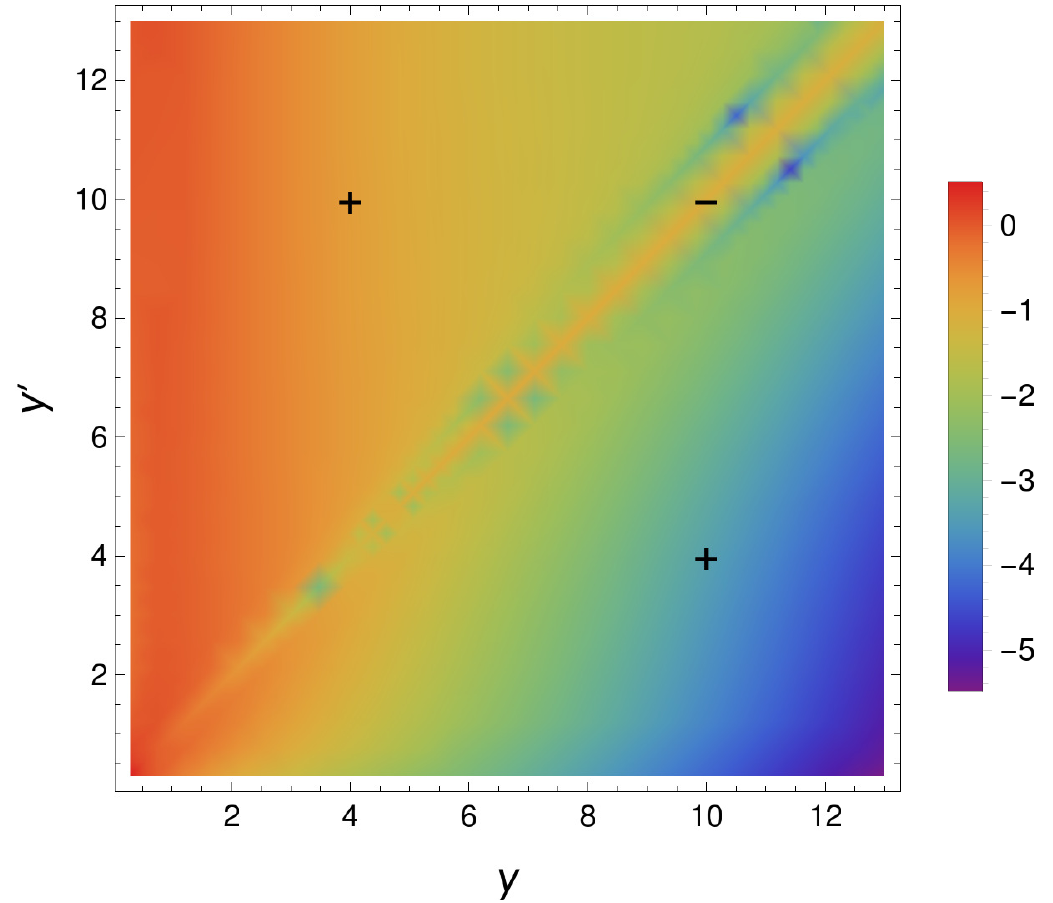}
    \end{subfigure}%
    ~ 
    \begin{subfigure}[b]{0.49\textwidth}
        \centering
        \caption{$\log_{10}(|\mathfrak{K}^{\nu}_1|)$}
        \includegraphics[width=\textwidth]{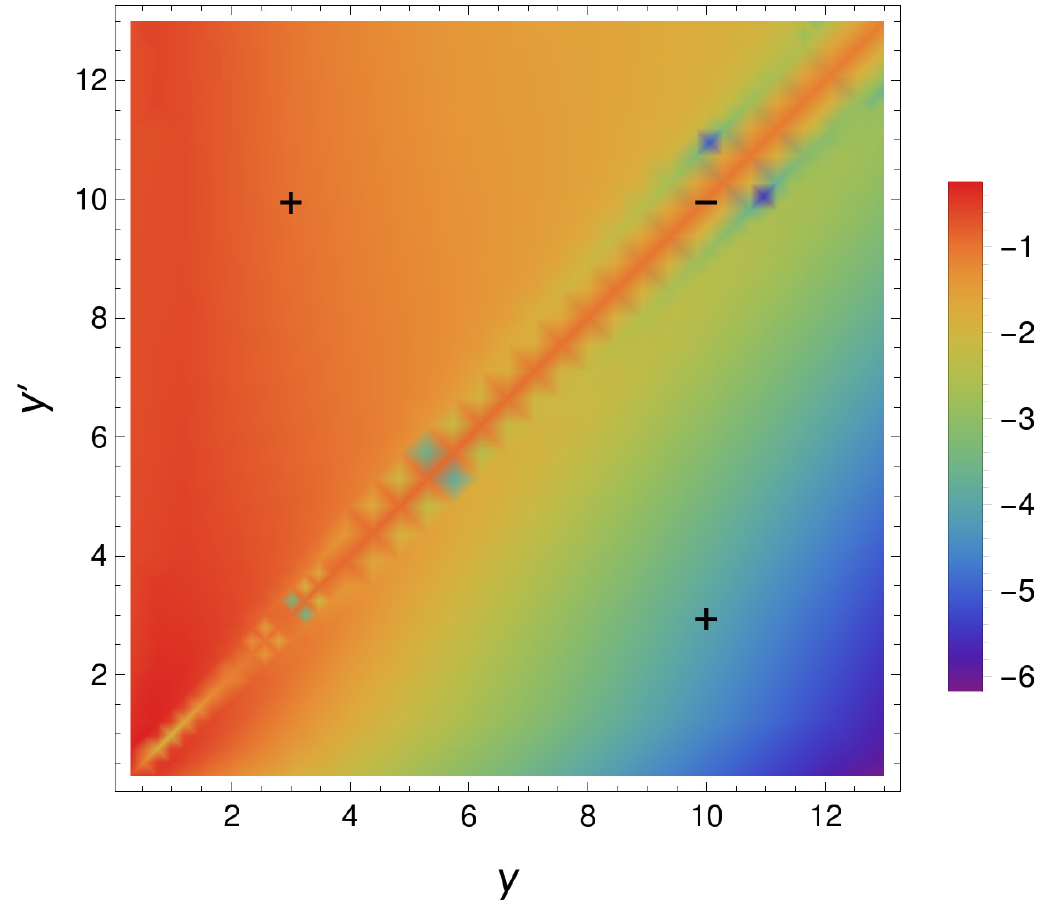}
    \end{subfigure}
    
    \begin{subfigure}[b]{0.49\textwidth}
        \centering
        \caption{$\log_{10}(|\mathfrak{K}^{\nu}_2|)$}
        \includegraphics[width=\textwidth]{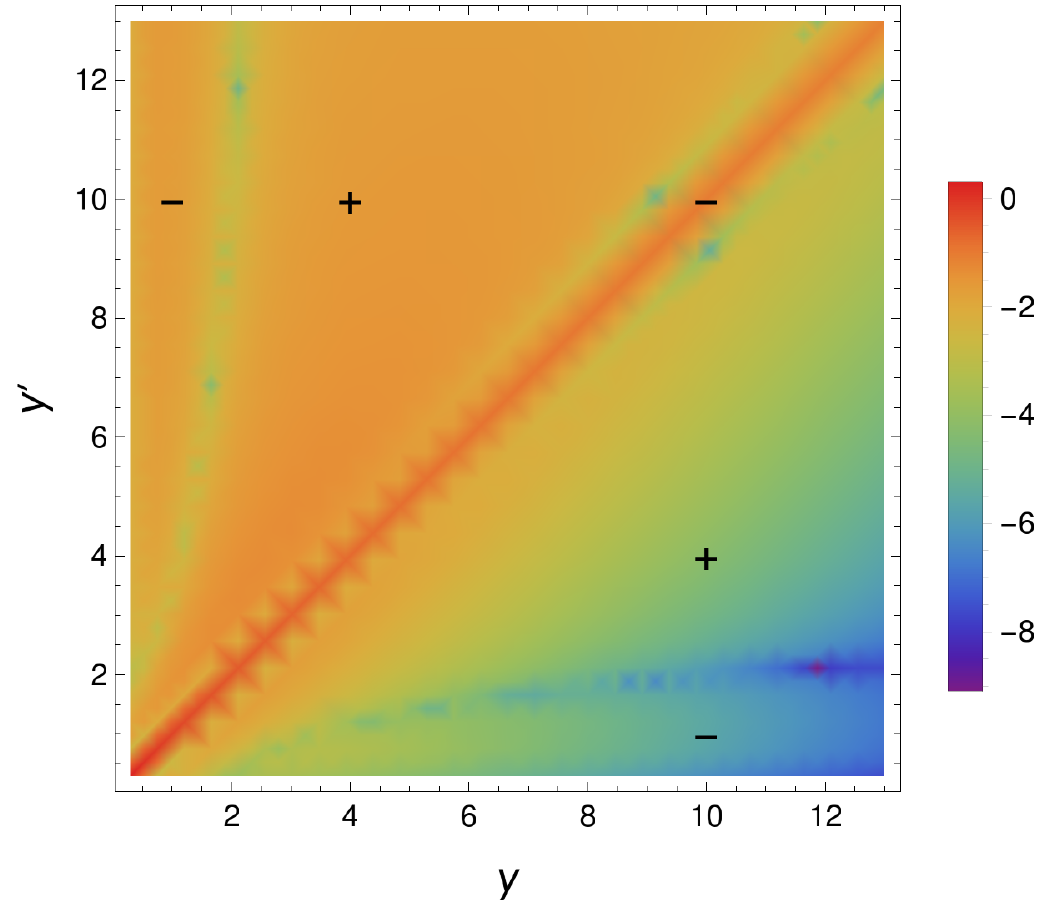}
    \end{subfigure}%
    ~ 
    \begin{subfigure}[b]{0.49\textwidth}
        \centering
        \caption{$\log_{10}(|\mathfrak{K}^{\nu}_3|)$}
        \includegraphics[width=\textwidth]{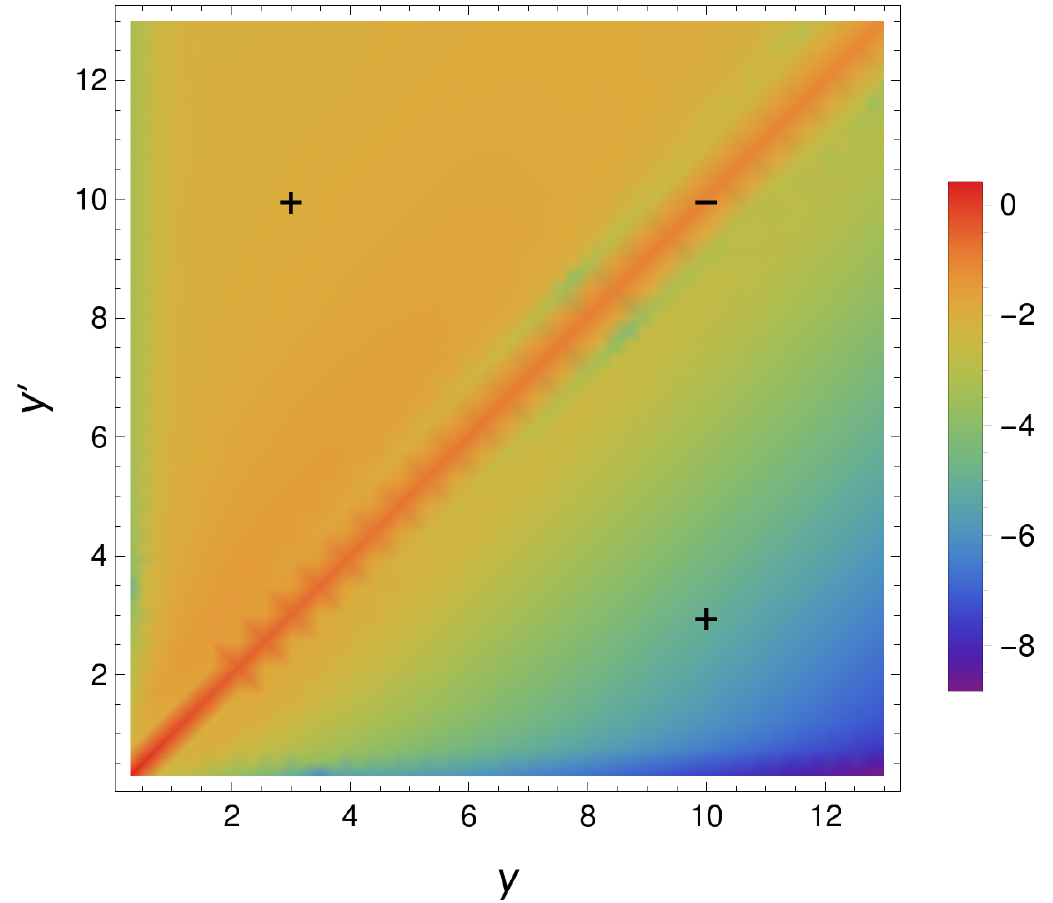}
    \end{subfigure}
\caption{Same as figure~\ref{Fig:MasslessKernel1}, but for   the kernels $\mathfrak{K}^{\nu}_{0,1,2,3}$.}
    \label{Fig:MasslessKernel2}
\end{figure}

The corresponding hierarchy for the interacting scalar particles is formally identical to equation~(\ref{finalBoltzmann_neutrino}), except for the replacements
\begin{equation}
\begin{aligned}
F_{\nu, \ell} & \to F_{\phi, \ell}, \\
\bar{f}_\nu   & \to \bar{f}_\phi, \\
{\cal X}^\nu & \to {\cal X}^\phi (q, \eta) \equiv 2 {\cal X}^\nu (q,\eta) \,(\nu \leftrightarrow \phi) - 6 X_{\phi}(q,\eta), \\
{\cal K}^\nu_\ell &\to {\cal K}_{\ell}^\phi (q,q',\eta) \equiv 2 {\cal K}^\nu_\ell(q,q',\eta) \, (\nu \leftrightarrow \phi) +6 \bar{f}_{\phi}(q,\eta) \delta_{\ell 0}- 6 K^0_{\phi, \ell}(q,q',\eta),\\
\mathfrak{K}^\nu_\ell &\to \mathfrak{K}_{\ell}^\phi (q,q',\eta) \equiv 2 \mathfrak{K}^\nu_\ell (q,q',\eta) (\nu \leftrightarrow \phi).
\end{aligned}
\end{equation}
Figure~\ref{Fig:Damping} shows ${\cal X}^\phi_\ell$ evaluated at the time of recombination, assuming a Maxwell--Boltzmann background distribution for both neutrinos and scalars, and a neutrino mass of $m_\nu = 0.05$~eV.  We do not plot the kernels  ${\cal K}_{\ell}^\phi$ and $\mathfrak{K}^\phi_\ell$, because they are structurally very similar to their neutrino counterparts ${\cal K}_{\ell}^\nu$ and $\mathfrak{K}^\nu_\ell$ already shown in figures~\ref{Fig:MasslessKernel1} and ~\ref{Fig:MasslessKernel2}.
Note that the neutrino hierarchy~\eqref{finalBoltzmann_neutrino} and its scalar counterpart are coupled to one another {\it at all orders $\ell$} via the kernels~$\mathfrak{K}^{\nu,\phi}_\ell$.  We discuss this coupling  in more detail in the next section.


\section{Discussion}\label{sec:discussion}

The neutrino Boltzmann hierarchies~\eqref{finalBoltzmann_mass} and~\eqref{finalBoltzmann_neutrino} and the latter's scalar counterpart are the second main result of this work.  The hierarchy~\eqref{finalBoltzmann_mass} for the case of a massive mediating scalar, in particular, is  in a form that  can be immediately embedded into a standard CMB Boltzmann solver such as {\sc CAMB}~\cite{Lewis:1999bs} or {\sc CLASS}~\cite{Blas:2011rf} without further evaluation.  In the massless scalar case, the hierarchy~\eqref{finalBoltzmann_neutrino} and its scalar counterpart in principle contain kernels that depend on the unknown background phase space distributions of the particles, $\bar{f}_\nu$ and $\bar{f}_\phi$.  However, as demonstrated in section~\ref{sec:masslesscollision},  $\bar{f}_\nu$ and $\bar{f}_\phi$ can be obtained from  the zeroth-order Boltzmann equation~(\ref{eq:homogeneous}) in conjunction with the collision integrals~(\ref{eq:0self}) to~(\ref{eq:0elastic}), such as in detailed calculations of neutrino decoupling, e.g.,~\cite{Hannestad:1995rs}, and leptogenesis~\cite{HahnWoernle:2009qn}.

In contrast with  non-interacting massless neutrinos, collision terms in the interacting scenario---in both limits of the scalar mass---render the corresponding Boltzmann hierarchies mo\-men\-tum-dependent.   We can attempt to integrate out this momentum dependence.  Focussing for the moment on the hierarchy~\eqref{finalBoltzmann_mass} in the massive scalar scenario, and defining
\begin{equation}
\begin{aligned}
\delta_\nu(k) &= \frac{4 \pi}{\bar{\rho}_\nu a^4} \int \dd q \, q^3 \, F_0 (k,q) = {\cal F}_{\nu 0}\, ,\\
\theta_\nu(k) &= \frac{3 \pi k}{\bar{\rho}_\nu a^4} \int \dd q \, q^3 \, F_1 (k,q)= \frac{3}{4} k {\cal F}_{\nu 1} \, ,\\
\sigma_\nu(k) &= \frac{2 \pi}{\bar{\rho}_\nu a^4} \int \dd q \, q^3 \, F_2 (k,q) = \frac{1}{2} {\cal F}_{\nu 2} \, ,\\
\end{aligned}
\end{equation}
where $\bar{\rho}_\nu$ is the background neutrino energy density, we find that integrating the $\ell=0$ and $\ell=1$ equations 
 over $\int \dd |\fett{q}| \, |\fett{q}|^3$ 
gives, respectively,
\begin{equation}
\begin{aligned}
\dot{\delta}_{\nu} &=  -\frac{4}{3} \theta_\nu - \frac{2}{3} \dot{h} \, , \\
\dot{\theta}_{\nu} &= k^2 \left(\frac{1}{4}  \delta_{\nu}- \sigma_{\nu} \right) \,,
\label{eq:integrated01}
\end{aligned}
\end{equation}
which are identical to those for non-interacting massless neutrinos~\cite{Ma:1995ey}.  This is not surprising, because a vanishing integrated collision term in a self-interacting but otherwise decoupled fluid simply follows from  the conservation of energy  ($\ell=0$) and of momentum ($\ell=1$).

Comparing equation~(\ref{eq:integrated01}) with the $(c_{\rm eff}^2, c_{\rm vis}^2)$-parameterisation~\eqref{parametrisation}, we find that the only viable value for the effective sound speed is~$c_{\text{eff}}^2= 1/3$, if equation~\eqref{parametrisation} is to have an  interpretation in terms of the scattering of identical massless particles amongst themselves; all other values are forbidden as they cannot be mapped to an underlying particle self-interaction scenario without violating momentum conservation.

For $\ell \geq 2$, there are no conserved quantities to ensure the total cancellation of the collision terms after integration over momentum. 
This immediately argues against the $(c_{\rm eff}^2, c_{\rm vis}^2)$-parameterisation~\eqref{parametrisation} as a model of particle interaction, which, for $\ell > 2$, has no collisional damping terms at all.
Likewise, we cannot reproduce the structure of the $\ell=2$ ``damping term'' in equation~\eqref{parametrisation}, with its curious proportionality  to $\theta_\nu$ and the gravitational source term.  As already discussed in section~\ref{State}, contrary to a genuine collisional damping term, the ``damping term'' of equation~\eqref{parametrisation} does not in fact drive ${\cal F}_{\nu 2}$ to zero; it merely stops power from being passed from the $\ell=0,1$ to the $\ell \geq 2$ multipoles, but not vice versa.  We therefore conclude that the viscosity parameter~$c_{\text{vis}}^2$ has no interpretation whatsoever in terms of particle scattering.

Contrasting equation~\eqref{finalBoltzmann_mass}  with the approach of~\cite{Cyr-Racine:2013jua} (see also equation~(\ref{eq:cyr})), we find that
the nontrivial $|\fett{q}|$-dependence of our $\ell \geq 2$ collision terms,
\begin{equation}
-  \frac{40}{3} G^{\rm m}  q \, T_{\nu,0}^4 \, F_{\nu,\ell}(q) +  \int  \dd q' \, \frac{q'}{q} \, K^m_\ell(q,q') \, F_{\nu,\ell}(q') \, ,
\end{equation}
renders it formally impossible to bring the $\ell \geq 2$ portion of the hierarchy~(\ref{finalBoltzmann_mass}) after momentum integration into the well-known form ${\cal F}_{\nu \ell} =  \ldots - |\dot{\tau}| {\cal F}_{\nu \ell}$, encountered in, e.g., Thomson scattering of CMB photons on electrons,
{\it unless} the phase space perturbation $F(|\fett{q}|, \cos \epsilon)$ separates into a purely $|\fett{q}|$-dependent and a purely angular part.  Such a separation is, however, strictly attainable only for scattering processes that do not alter the energy of the scattered particle because, e.g., the incident energy is much smaller than the mass of the scattering centre, which is obviously untrue for ultrarelativistic neutrinos scattering on other ultrarelativistic neutrinos.  Indeed, this is precisely the reason why, at leading order, the CMB photon hierarchy takes such a simple form (the mean energy of CMB photons around the decoupling epoch is much smaller than the electron mass).%
\footnote{Moving away from the Thomson limit, energy transfer between  photons and electrons does occur generally  in Compton scattering, and affects the form of the second-order photon Boltzmann hierarchy. See, e.g.,~\cite{Beneke:2010eg}.}
The appearance of unresolved  integral collision terms in the Boltzmann hierarchy at  $ \ell \geq 2$ merely reflects the presence of significant energy transfer in the scattering process concerned, irrespective of the interaction type, quantum statistics, or any other simplifying approximation we have adopted in the present calculation. 
This constitutes the second complication alluded to earlier in section~\ref{sec:hierarchy}

The same argument obviously holds also in the case of a massless scalar, i.e., hierarchy~\eqref{finalBoltzmann_neutrino} and its scalar counterpart, except with one small twist:
energy and momentum are now not only conserved within the neutrino sector itself, but across the combined neutrino--scalar system.
Consequently,  the coupling terms in the $\ell=0,1$ equations of the {\it  individual} neutrino or scalar hierarchy 
do not  in general vanish when integrated over $\int \dd |\fett{q}| \, |\fett{q}|^3$---they are however vanishing if the combined neutrino--scalar system is considered as a whole.
(Technically this means we need to consider the sum of the perturbations $2 F_{\nu, \ell} + F_{\phi, \ell}$, where the factor of 2 accounts for the neutrino's two internal degrees of freedom.)  Coupling terms between neutrinos and scalars are also present at $\ell \geq 2$.

It is interesting to contrast this neutrino--scalar coupling
with the case of CMB photon--electron scattering:  
the photon Boltzmann hierarchy is coupled to the baryon hierarchy only at $\ell=1~$\cite{Ma:1995ey}. The $\ell=0$ coupling vanishes because in the limit of Thomson scattering energy is
 effectively conserved within the individual photon and electron sectors, while at $\ell \geq 2$ 
 nonrelativistic particles such as baryons have no mutipole moments higher than the dipole, i.e., $F_{b,\ell \geq 2}=0$, so that all $\ell \geq 2$ couplings are automatically zero.

Lastly, the fact that the Boltzmann hierarchies derived in this work are formally significantly more complex
does not preclude the possibility that they could in the end have a very similar phenomenological impact on the CMB anisotropy spectrum as the approximate model~(\ref{eq:cyr}) and possibly even the arguably dubious $(c_{\rm eff}^2, c_{\rm vis}^2)$-parameterisation~\eqref{parametrisation}.  Indeed, in the case of a very massive mediating scalar, the large initial scattering rate could enable the separation of $F(|\fett{q}|, \cos \epsilon)$ into a purely $|\fett{q}|$-dependent and a purely angular part in an approximate way during the neutrino decoupling process, so that
the model~(\ref{eq:cyr}) can still  be a good approximation of the full hierarchy~(\ref{finalBoltzmann_mass}).%
\footnote{K.~Sigurdson, private communications.}
The phenomenology of the opposite case of a massless mediating scalar---in which the neutrinos decouple, free-stream, and then recouple---is however much less clear, and 
to investigate it would require that we implement equation~(\ref{finalBoltzmann_neutrino}) and its scalar counterpart into a CMB Boltzmann solver such as {\sc CAMB}~\cite{Lewis:1999bs} or {\sc CLASS}~\cite{Blas:2011rf}.  We defer this exercise to a later work.

\section{Conclusions}
\label{Summary}

We have computed in this work the  Boltzmann collision integral for neutrino--neutrino and neutrino--scalar interactions that arise in  majoron-type models, up to first order in spacetime perturbations and in two limits of the scalar mass---massless and extremely massive.

 Using the first-order results, we proceeded to derive the corresponding neutrino and scalar Boltzmann hierarchies, and these are presented in a form suitable for implementation in a CMB Boltzmann solver such as {\sc CAMB}~\cite{Lewis:1999bs} or {\sc CLASS}~\cite{Blas:2011rf}.
 In the case of a massless scalar, kinematics favours the production of real scalar particles; for this reason we have also derived  the zeroth-order collision Boltzmann equation, which can be used to track the neutrino and scalar background populations.
 This is the first time the Boltzmann hierarchy for interacting neutrinos has been derived from {\it from first principles}.  In comparison with various heuristic models of neutrino interactions in the literature, our results clearly reveal a richer---in some cases, vastly different---structure for the collision terms.

The first salient feature is the presence unresolved integral collision terms in the Boltzmann hierarchy that formally cannot be removed by integration over momentum, even in the limit of massless neutrinos.  This is a consequence of significant energy transfer in the ultrarelativistic neutrino--neutrino scattering process (and neutrino--massless scalar scattering), and a feature that is liable to be missed in a simple generalisation of the first-order photon Boltzmann hierarchy, where this feature is absent because photon--electron scattering in the Thomson limit preserves the photon's energy. The precise phenomenological impact of these additional terms, however, must be determined on a case-by-case basis, and it is not inconceivable that in some scenarios  they can be integrated out in an approximate way.

The second point pertains to the commonly used $(c_{\rm eff}^2,c_{\rm vis}^2)$-parameterisation as a model of neutrino interactions, where $c_{\rm eff}^2$ is the effective sound speed and $c_{\rm vis}^2$ the viscosity parameter.
Conservation of energy--momentum automatically restricts $c_{\rm eff}^2$ to $1/3$ in a self-interacting but otherwise decoupled ensemble of massless neutrinos; no other value of $c_{\rm eff}^2$ makes physical sense in this context.  The interpretation of $c_{\rm vis}^2$ in terms of particle scattering, on the other hand, is always tenuous, if not altogether meaningless; of the spectrum of behaviours described by various choices of $c_{\rm vis}^2$ values, none passes as collisional damping, a minimum phenomenon expected of particle scattering.  We therefore strongly argue against the $(c_{\rm eff}^2,c_{\rm vis}^2)$-parameterisation as a phenomenological description of particle scattering for CMB anisotropy calculations.

Lastly, while we have made a good number of simplifying assumptions and approximations in order to keep our problem tractable,
we emphasise that these approximations impact only on the precise forms of integration kernels, not on the gross structure of the collision terms in the Boltzmann hierarchy itself;
our conclusions above, drawn on the basis of kinematics arguments, remain valid even if the assumptions should be relaxed.  Nevertheless, some of our assumptions are certainly more in need of revision than others (e.g., zero neutrino masses, the absence of Pauli-blocking, etc.).  This, together with the implementation of the Boltzmann hierarchies  into a CMB Boltzmann solver, will be explored in a future publication.

\section*{Acknowledgements}

A part of this work contributed to the master thesis of IMO~at RWTH Aachen University. We thank D. Boriero, C.~Fidler, P.~Gondolo, and T.~Tram for useful discussions. CR~acknowledges the support of the individual fellowship RA 2523/1-1 from the Deutsche Forschungsgemeinschaft. IMO~furthermore acknowledges the support by Studienstiftung des Deutschen Volkes and the mobility grant provided by Bielefeld Graduate School in Theoretical Sciences.

\appendix



\section{Collision integral reduction: massless scalar limit}
\label{app:massless}

\subsection{First-order \texorpdfstring{$\nu \nu \leftrightarrow \nu \nu$}{nu nu <-> nu nu} collision integrals}
\label{Massless scalar}

Symmetry under the exchange $q' \leftrightarrow l'$ simplifies the first-order collision integral~\eqref{C[f]0} to
\begin{equation}
\begin{aligned}
\left( \frac{\partial f_\nu}{\partial \eta}\right)^{(1)}_{\nu \nu \leftrightarrow \nu \nu}(\fett{k},\fett{q},\eta) 
 &=    \frac{24 \mathfrak{g}^4}{(2 \pi)^{5}|\fett{q}|} \int \frac{\mathrm{d^{3}} \fett{q}'}{2 | \fett{q}'|}  \int \frac{\mathrm{d^{3}}\fett{l}}{2 | \fett{l}|}  \int \frac{\mathrm{d^{3}}\fett{l}'}{2 | \fett{l}'|} \, \delta_\D^{(4)}(q+l-q'-l')    \\
& \hspace{-2mm} \times \Big[  2\bar{f}_{\nu}(| \fett{l}'|,\eta) \, F_{\nu}(\fett{k},\fett{q}',\eta)-\bar{f}_{\nu}(| \fett{q}|,\eta) \, F_{\nu}(\fett{k},\fett{l},\eta)-\bar{f}_{\nu}(| \fett{l}|,\eta) \, F_{\nu}(\fett{k},\fett{q},\eta) \Big] \\
& \equiv \ {\cal C}_{1}^{{\nu \nu \leftrightarrow \nu \nu}}[f]+{\cal C}_{2}^{{\nu \nu \leftrightarrow \nu \nu}}[f]+{\cal C}_{3}^{{\nu \nu \leftrightarrow \nu \nu}}[f] \, ,
\label{Collision Integral}
\end{aligned}
\end{equation}
which splits into three parts. These  will be treated separately below, following in part the procedure outlined in the appendix of~\cite{Hannestad:1995rs}, itself adapted from~\cite{yueh}.
Note that {\it both} the background and perturbed components of the phase space distribution, $\bar{f}_\nu$ and $F_\nu$, are time-dependent in this scenario; for brevity, however, we shall sometimes omit the time label.


\subsubsection{Reduction of \texorpdfstring{${{\cal C}_{3}^{\nu \nu \leftrightarrow \nu \nu}[f]}$}{C3}}
\label{Computing C_3[f]}

Since the only dependence here
is in the Dirac delta, we first integrate in $\dd ^3 \fett{l}'$ to get
\begin{align}
 \!\!\int \frac{\mathrm{d^{3}}\fett{l}'}{2 |\fett{l}'|} \delta_\D^{(4)}(q+l-q'-l') = \delta_\D\!\left( \left[|\fett{q}|+|\fett{l}|-|\fett{q}'| \right]^{2}-|\fett{q}+\fett{l}-\fett{q}'|^{2}\right) \, \Theta(|\fett{q}|+|\fett{l}|-|\fett{q}'|) \,.
\label{delta function}
\end{align}
Choosing an angular parameterisation for the 3-vectors as follows,
\begin{equation}
\begin{aligned}
 \fett{q} &= |\fett{q}|(0,0,1)  \,, \\
 \fett{l} &= |\fett{l}|(0, \sin\alpha, \cos\alpha) \,,  \\
 \fett{q}' &= |\fett{q}'|(\sin\beta \sin\theta, \cos\beta \sin\theta, \cos\theta) \,,  
\label{k-parametrisation} 
\end{aligned}
\end{equation}
equation~(\ref{delta function}) can be equivalently expressed as
\be
 \int \frac{\mathrm{d^{3}}\fett{l}'}{2 |\fett{l}'|} \delta_\D^{(4)}(q+l-q'-l') \equiv  \delta_\D \left(g(|\fett{q}|,|\fett{l}|,|\fett{q}'|,\alpha,\beta,\theta) \right) \, \Theta(|\fett{q}|+|\fett{l}|-|\fett{q}'|) \,, 
\label{delta function2} 
\ee
where
\be
g(\cdots)/2 \equiv |\fett{q}||\fett{l}|(1-\cos\alpha)+|\fett{q}||\fett{q}'|(\cos\theta-1)+|\fett{l}||\fett{q}'|(\cos\alpha \cos\theta + \sin\alpha \sin\theta \cos\beta -1) \, .
\label{g(beta)}
\ee
Then, feeding equation~(\ref{delta function2}) back into equation~(\ref{Collision Integral}), we find
\begin{equation}
\begin{aligned}
{\cal C}_{3}^{\nu \nu \leftrightarrow \nu \nu}[f]=&- \frac{6 \mathfrak{g}^4}{(2\pi)^{4}|\fett{q}|}F_{\nu}(\fett{k},\fett{q}) \\
& \times
\int  \mathrm{d\beta} \, \mathrm{d \cos\theta} \, \mathrm{d \cos\alpha} \, \mathrm{d|\fett{l}|} \, \mathrm{d|\fett{q}'|} \, 
\delta_\D(g(\cdots)) 
\Theta(|\fett{q}|+|\fett{l}|-|\fett{q}'|) |\fett{q}'||\fett{l}| \bar{f}_{\nu}(| \fett{l}|).
\label{C4 after d'l}
\end{aligned}
\end{equation}
where we have used $\dd^3\fett{l} = 2 \pi  |\fett{l}|^2 \, \mathrm{d |\fett{l}|} \,\mathrm{d} \cos\alpha$ and  $\dd^3\fett{q}' =|\fett{q}'|^2 \, \mathrm{d |\fett{q}'|} \,\mathrm{d} \cos\theta \, \mathrm{d}\beta$.

For the $\beta$-integral, we use this well-known identity of the Dirac delta:
\be
\int_{0}^{2\pi} \mathrm{d\beta} \,\, \delta_\D(g(\ldots,\beta,\ldots))= \int_{0}^{2\pi} \mathrm{d\beta} \, \sum_{i} \frac{1}{\left|\frac{\partial g}{\partial \beta} \right|_{\beta_i}} \delta_\D(\beta-\beta_i) \,,
\label{beta-relation}
\ee
where $\beta_i$ are the simple roots of the real-valued function $g(\ldots,\beta,\ldots)$. In our case, we have
\begin{eqnarray}
\frac{\partial g}{\partial \beta} &= & -2|\fett{l}||\fett{q}'| \sin\alpha \, \sin\theta \, \sin\beta \,, \label{eq:dgdb}\\
\cos\beta_i &= & \frac{|\fett{q}||\fett{q}'|(1-\cos\theta)-|\fett{q}||\fett{l}|(1-\cos\alpha)+|\fett{
l}||\fett{q}'|(1-\cos\alpha \, \cos\theta)}{|\fett{l}||\fett{q}'|\sin\alpha \, \sin\theta} \,.
\label{cosbi}
\end{eqnarray}
Because $\cos \beta_i = \cos (-\beta_i)$, equation \eqref{cosbi} has one solution for $\beta_i$ in  $[0, \pi]$ and one in $[\pi,2 \pi]$.  
Thus, equation~(\ref{beta-relation}) can be trivially rewritten as
\begin{equation}
\int_{0}^{2\pi} \mathrm{d\beta} \, \delta_{\rm D}(g(\ldots,\beta,\ldots))=2 \int_{0}^{\pi} \mathrm{d\beta} \, \left|\frac{\partial g}{\partial \beta} \right|_{\cos\beta_i}^{-1} \delta_\D(\beta-\beta_i) = 2 \left|\frac{\partial g}{\partial \beta} \right|_{\cos\beta_i}^{-1} \, ,
\label{beta}
\end{equation}
provided any further dependence of the collision integrand on $\beta$ is a function of $\cos \beta$.

Observe that $\cos \beta_i$ must lie in $[-1,1]$, or, equivalently, $\cos^2 \beta_i \leq 1$.  When applied to equation~(\ref{cosbi}), this condition 
effectively limits the values the outgoing  parameters $|\fett{q}'|$ and $\cos \theta$ can take  in a physical process for a given combination of incoming parameters $|\fett{q}|, |\fett{l}|$ and $\cos \alpha$.
(Recall that equation~(\ref{cosbi}) originates from energy--momentum conservation.)
To ensure that this physical limitation is respected in subsequent integrations in $\dd |\fett{q}'|, \dd \cos \theta$, etc., we introduce in equation~(\ref{beta}) a step function
\begin{align}
\label{eq:theta}
\Theta(1-\cos^2\beta_i)= \Theta \left( \left| \frac{\partial g}{\partial  \beta} \right|^2_{\cos\beta_i} \right) = \Theta\left(  a_3 \, \cos^{2}\theta+b_3 \, \cos\theta + c_3 \right) \,,
\end{align}
where the first equality follows from the observation that the condition $\cos^2 \beta_i \leq 1$ translates  simply into $|\partial g/\partial \beta|^2_{\cos \beta_i} =  4 |\fett{l}|^2|\fett{q}'|^2 \sin^2\alpha \, \sin^2\theta \, (1-\cos^2\beta_i) \geq 0$ using equation~(\ref{eq:dgdb}), and for the second equality 
we have defined
\begin{equation}
\begin{aligned}
\label{eq:abc}
 a_3 &=  -4|\fett{q}'|^{2}(|\fett{l}|^{2}+2|\fett{l}||\fett{q}|\cos\alpha+|\fett{q}|^{2})=-4|\fett{q}'|^{2}|\fett{l}+\fett{q}|^{2}\leq 0 \,,\\
 b_3 &=  8 |\fett{q}'| (-|\fett{q}||\fett{l}| + |\fett{q}||\fett{l}| \cos\alpha +|\fett{q}||\fett{q}'|+ |\fett{l}||\fett{q}'|) (|\fett{q}|+|\fett{l}| \cos\alpha) \,, \\
 c_3 &=  4|\fett{l}|^2|\fett{q}'|^2(1-\cos^2\alpha)-4(-|\fett{q}||\fett{l}|(1-\cos\alpha)+|\fett{q}||\fett{q}'|+|\fett{l}||\fett{q}'|)^2 \, ,
\end{aligned}
\end{equation}
noting that $a_3$ is always negative.
Inserting~(\ref{beta}) and~(\ref{eq:theta}) into equation~(\ref{C4 after d'l}) then gives 
\begin{equation}
\begin{aligned}
{\cal C}_{3}^{\nu \nu \leftrightarrow \nu \nu}[f]&=-\frac{12 \mathfrak{g}^4}{(2\pi)^{4}|\fett{q}|}F_{\nu}(\fett{k},\fett{q})\int  \mathrm{d\cos\alpha} \, \mathrm{d|\fett{l}|} \, \mathrm{d|\fett{q}'|} \, \Theta(|\fett{q}|+|\fett{l}|-|\fett{q}'|) \, |\fett{q}'||\fett{l}| \, \bar{f}_{\nu}(| \fett{l}|) \\
&\hspace{0mm} \times \int \dd \cos \theta  \, \frac{1}{\sqrt{-|a_3| \, \cos^{2}\theta+b_3 \, \cos\theta +c_3}} \Theta(-|a_3| \, \cos^{2}\theta+b_3 \, \cos\theta +c_3) \, ,
\end{aligned}
\label{C4 after dbeta}
\end{equation}
and we remind the reader that $a_3,b_3,c_3$ are functions of $|\fett{q}|, |\fett{q}'|, |\fett{l}|$ and $\alpha$.  

Consider now  the $\cos \theta$-integral in the second line of equation~(\ref{C4 after dbeta}).
Since $a_3$ is always negative, the argument of the step function is a downward parabola.  If the condition $b_3^2 > 4a_3c_3$ is also satisfied, 
then the integrand is real and positive between the parabola's roots, 
\be
\label{eq:roots}
  x_{\pm} = \frac{b_3}{2|a_3|}\pm \sqrt{\left( \frac{b_3}{2|a_3|} \right) ^{2}+\frac{c_3}{|a_3|}} \, ,
\ee
with $x_+>x_-$.  Furthermore, because $\Theta(-|a_3| x^2 + b_3x+c_3)$ originates from energy--momentum conservation, once the real and nondegenerate roots~$x_{\pm}$ have been identified, they are automatically guaranteed to lie in the interval $(-1,1)$.
 Indeed, as we shall see later, the condition for real and nondegenerate roots, $b_3^2> 4a_3c_3$, corresponds to the physical limitation on the outgoing momentum~$|\fett{q}'|$ for a given combination of incoming parameters $|\fett{q}|, |\fett{l}|$ and $\cos \alpha$.  Thus, the  $\cos \theta$-integral in equation~(\ref{C4 after dbeta}).
 is equivalently,
\begin{equation}
\begin{aligned}
\frac{1}{\sqrt{|a_3|}} \int_{x_-}^{x_+}  \frac{\mathrm{d} x}{\sqrt{\, (x-x_-)(x_+-x)}}
\Theta(b_3^2-4a_3c_3)
= \frac{\pi}{\sqrt{|a_3|}} \Theta(b_3^2-4a_3c_3)\ ,
\label{IntTheta2}
\end{aligned}
\end{equation}
where $\Theta(b_3^2-4a_3c_3)$ has been inserted to ensure the realness of $x_\pm$.
Substituting into equation~(\ref{C4 after dbeta}) then gives 
\begin{equation}
\begin{aligned}
{\cal C}_{3}^{\nu \nu \leftrightarrow \nu \nu}[f]= & - \frac{3 \mathfrak{g}^4}{(2\pi)^{3}|\fett{q}|}F_{\nu}(\fett{k},\fett{q})\int \mathrm{d \cos\alpha} \, \mathrm{d|\fett{l}|} \bar{f}_{\nu}(|\fett{l}|)\frac{|\fett{l}|}{\sqrt{|\fett{l}|^{2}+2|\fett{q}||\fett{l}|\cos\alpha+|\fett{q}|^{2}}}   \\ 
&\hspace{40mm} \times \int \mathrm{d|\fett{q}'|} \, \Theta(|\fett{q}|+|\fett{l}|-|\fett{q}'|)  \, \Theta(b_3^2-4a_3c_3) \, ,
\label{C4 after dcostheta}
\end{aligned}
\end{equation}
with three remaining integrations.

Looking now at the step function~$\Theta(b_3^2-4a_3c_3)$, and evaluating its argument explicitly using the definitions~(\ref{eq:abc}), 
\begin{equation}
\label{eq:explicit}
b_3^2-4 a_3c_3 = 512 |\fett{q}'|^2 |\fett{l}|^3 |\fett{q}| \cos^2\left(\frac{\alpha}{2}\right) \sin^4 \left(\frac{\alpha}{2} \right) [-2 |\fett{q}'|^2
+2 |\fett{q}'| ( |\fett{l}| + |\fett{q}|) - |\fett{l}| |\fett{q}| (1-\cos \alpha)] \,,
\end{equation}
we see immediately that it has four $|\fett{q}'|$-roots at $\lbrace{0,0,R_1,R_2}\rbrace$, where
\begin{equation}
\begin{aligned}
R_1 &= \frac{|\fett{q}|+|\fett{l}|-\sqrt{|\fett{q}|^2+|\fett{l}|^2+2|\fett{q}||\fett{l}|\cos\alpha}}{2} =  \frac{|\fett{q}|+|\fett{l}|-|\fett{q}+\fett{l}|}{2} \,, \\
R_2 &= \frac{|\fett{q}|+|\fett{l}|+\sqrt{|\fett{q}|^2+|\fett{l}|^2+2|\fett{q}||\fett{l}|\cos\alpha}}{2} =  \frac{|\fett{q}|+|\fett{l}|+|\fett{q}+\fett{l}|}{2}\, ,
 \label{roots}
\end{aligned}
\end{equation}
and $R_2 \geq R_1 \geq 0$. Because the parabolic part of equation~(\ref{eq:explicit}) always opens downwards, the region enclosed by $R_1$ and $R_2$ is always positive. 
Thus, we can rewrite $\Theta(b_3^2-4a_3c_3)$ as
\begin{equation}
\Theta(b_3^2-4a_3c_3)= \Theta(|\fett{q}'|-R_1)\, \Theta(R_2-|\fett{q}'|) \, ,
\label{Theta0}
\end{equation}
so that the $|\fett{q}'|$-integral in~\eqref{C4 after dcostheta} evaluates trivially to
\begin{equation}
\begin{aligned}
\int \mathrm{d|\fett{q}'|} \, \Theta(|\fett{q}|+|\fett{l}|-|\fett{q}'|)  \, \Theta(b^2_3-4a_3c_3) 
&=  \int \mathrm{d|\fett{q}'|} \, \Theta(|\fett{q}'|-R_1)\, \Theta(R_2-|\fett{q}'|) = |\fett{q}+\fett{l}|,
\label{eq:r1r2}
\end{aligned}
\end{equation}
noting that the step function $\Theta(|\fett{q}|+|\fett{l}|-|\fett{q}'|)$ has become redundant because  $R_2 \leq |\fett{q}|+|\fett{l}|$ from equation~(\ref{roots}).

Substituting equation~(\ref{eq:r1r2})  into equation~(\ref{C4 after dcostheta}) then gives
\begin{equation}
\begin{aligned}
{\cal C}_{3}^{\nu \nu \leftrightarrow \nu \nu}[f] = & - \frac{ 3 \mathfrak{g}^4}{(2\pi)^{3}|\fett{q}|}F_{\nu}(\fett{k},\fett{q},\eta)\int \mathrm{d \cos\alpha} \, \mathrm{d|\fett{l}|} \bar{f}_{\nu}(|\fett{l}|,\eta) |\fett{l}| = -  \frac{6 \mathfrak{g}^4}{(2\pi)^{3}}X_{\nu} (|\fett{q}|,\eta) \,F_{\nu}(\fett{k},\fett{q},\eta) ,
\label{eq:c3final}
\end{aligned}
\end{equation}
where we have defined
\begin{equation}
X_{i=\nu, \phi} (|\fett{q}|,\eta)\equiv \frac{1}{|\fett{q}|}\int \mathrm{d|\fett{l}|} \bar{f}_i(|\fett{l}|,\eta) |\fett{l}| \, ,
\label{X_self}
\end{equation}
and reinstated the time label to remind the reader  that the background neutrino phase space distribution $\bar{f}_\nu$ appearing in the integrand is time-dependent.

\subsubsection{Reduction of \texorpdfstring{${{\cal C}_{2}^{\nu \nu \leftrightarrow \nu \nu}[f]}$}{C2}}
\label{C2_self}

The calculation of ${\cal C}_{2}^{\nu \nu \leftrightarrow \nu \nu}[f]$ is identical to that of ${\cal C}_{3}^{\nu \nu \leftrightarrow \nu \nu}[f]$  up to equation~(\ref{eq:c3final}), instead of which we find
\begin{equation}
 {\cal C}_{2}^{\nu \nu \leftrightarrow \nu \nu}[f]= - \frac{3 \mathfrak{g}^4}{(2\pi)^3|\fett{q}|}\bar{f}_{\nu}(|\fett{q}|,\eta)\int  \dd\cos\alpha \, \dd|\fett{l}| \, |\fett{l}| \, F_{\nu}(\fett{k},\fett{l},\eta) \, .
 \label{eq:c2fin}
\end{equation}
No further integration is possible, as $F_{\nu}(\fett{k},\fett{l},\eta)$ is a function of both $|\fett{l}|$ and $\cos \alpha$.

\subsubsection{Reduction of \texorpdfstring{${\cal C}_{1}^{{\nu \nu \leftrightarrow \nu \nu}}[f]$}{C1}}
\label{Computing C_2[f] and C_1[f]}

Since the integrand of ${\cal C}_{1}^{\nu \nu \leftrightarrow \nu \nu}[f]$ does not depend on $\fett{l}$, the Dirac delta can be simplified to
\begin{align}
 \!\!\int \frac{\mathrm{d^{3}}\fett{l}}{2 |\fett{l}|} \delta_\D^{(4)}(q+l-q'-l') = \delta_\D\!\left( \left[|\fett{q}'|+|\fett{l}'|-|\fett{q}| \right]^{2}-|\fett{q}'+\fett{l}'-\fett{q}|^{2}\right) \, \Theta(|\fett{q}'|+|\fett{l}'|-|\fett{q}|) \,.
\end{align}
Then, using the parameterisation
\begin{equation}
\begin{aligned} 
\label{param2}
\fett{q} &= |\fett{q}|(0,0,1) \,,  \\
\fett{q}' &= |\fett{q}'|(0, \sin\theta, \cos\theta) \,,  \\
 \fett{l}' &= |\fett{l}'|(\sin\beta \sin\alpha, \cos\beta \sin\alpha, \cos\alpha) \, ,
 \end{aligned}
 \end{equation}
 and following the arguments of section~\ref{Computing C_3[f]}, we find 
 \begin{equation}
 \begin{aligned}
 {\cal C}_{1}^{\nu \nu \leftrightarrow \nu \nu}[f]=  & \ \frac{12\mathfrak{g}^4}{(2\pi)^{4}|\fett{q}|} 
\int \, \dd\beta \, \dd\cos\theta \, \dd\cos\alpha \, \dd|\fett{l}'| \, \dd|\fett{q}'| \, \delta_\D(g(\cdots)) \\
&\hspace{25mm} \times \Theta(|\fett{q}'|+|\fett{l}'|-|\fett{q}|) |\fett{q}'||\fett{l}'| \bar{f}_{\nu}(| \fett{l}'|)\, F_{\nu}(\fett{k},\fett{q}')\,,
\label{C2 after d'l}
\end{aligned}
\end{equation}
where 
\begin{align}
g(\cdots)/2 = |\fett{q}'||\fett{l}'|(1-\sin\alpha \sin\theta \cos\beta - \cos\alpha \cos\theta)+|\fett{q}||\fett{q}'|(\cos\theta-1)+|\fett{l}'||\fett{q}|(\cos\alpha-1) \, 
\end{align}
is the argument of the Dirac delta distribution.

Again, because the Dirac delta alone depends on $\beta$, we can rewrite the $\beta$-integral as per equation~(\ref{beta}),
where in this case the step function has arguments
\begin{equation}
\left| \frac{\partial g}{\partial \beta} \right|^{2}_{\cos\beta_i} = a_1 \cos^{2}\alpha +  b_1 \cos\alpha +c_1 \,, 
\label{eq:dgdb1}
\end{equation}
with
\begin{equation}
\begin{aligned}
 a_1 &= -4|\fett{l}'|^{2}(|\fett{q}'|^{2}-2|\fett{q}||\fett{q}'|\cos\theta+|\fett{q}|^{2})=-4|\fett{l}'|^{2}|\fett{q}'-\fett{q}|^{2}<0 \,, \\
 b_1 &=  8 |\fett{l}'| ( |\fett{q}||\fett{q}'|(\cos\theta-1) + |\fett{q}'||\fett{l}'|-|\fett{l}'||\fett{q}| ) (|\fett{q}'| \cos\theta -|\fett{q}|) \,, \\
 c_1 &=  4|\fett{l}'|^2|\fett{q}'|^2(1-\cos^2\theta)-4 ( |\fett{q}||\fett{q}'|(\cos\theta-1)-|\fett{l}'||\fett{q}|+|\fett{l}'||\fett{q}'| )^2 \ ,
 \label{a-c}
\end{aligned}
\end{equation}
and $a_1$ is, again, always negative.  Following the reasoning of section~\ref{Computing C_3[f]}, equation~(\ref{C2 after d'l}) can be recast as
\begin{equation}
\begin{aligned}
{\cal C}_{1}^{\nu \nu \leftrightarrow \nu \nu}[f]&= \frac{12 \mathfrak{g}^4}{(2\pi)^{4}|\fett{q}|}\int \, \mathrm{d \cos\theta} \,   \mathrm{d|\fett{l}'|} \, \mathrm{d|\fett{q}'|} \, \Theta(|\fett{q}'|+|\fett{l}'|-|\fett{q}|)  \frac{ |\fett{q}'|}{|\fett{q}'-\fett{q}|^2} \bar{f}_{\nu}(| \fett{l}'|) \, F_{\nu}(\fett{k},\fett{q}') \\
&\hspace{22mm}\times \int_{y_-}^{y_+} \dd \cos \alpha \, \frac{1}{\sqrt{(y_+-\cos \alpha)(\cos \alpha -y_-)}} \, \Theta(b_1^2-4 a_1 c_1) \\
& =\frac{6 \mathfrak{g}^4}{(2\pi)^{3}|\fett{q}|}\int  \dd\cos\theta \, \dd|\fett{q}'|\frac{|\fett{q}'|}{\sqrt{|\fett{q}|^{2}-2|\fett{q}||\fett{q}'|\cos\theta+|\fett{q}'|^{2}}} \, F_{\nu}(\fett{k},\fett{q}')\\ 
&\hspace{22mm}  \times \int \dd|\fett{l}'| \, \Theta(|\fett{q}'|+|\fett{l}'|-|\fett{q}|) \, \bar{f}_{\nu} \, (|\fett{l}'|)  \, \Theta(b_1^2-4 a_1 c_1) \ ,
\label{C2 after dcosalpha}
\end{aligned}
\end{equation}
where $y_\pm$ are the real and nondegenerate roots of the quadratic~(\ref{eq:dgdb1}).

Looking at the step function $\Theta(b_1^2-4a_1c_1)$, we see that 
\begin{equation}
\label{eq:bbb}
b_1^2-4 a_1c_1 = 512 |\fett{l}'|^2 |\fett{q}'|^3 |\fett{q}| \cos^2 \left(\frac{\theta}{2} \right)\sin^4 \left(\frac{\theta}{2} \right)
[2|\fett{l}'|^2+ 2 |\fett{l}'|(|\fett{q}'|-|\fett{q}|)-|\fett{q}'| |\fett{q}|(1-\cos \theta)] 
\end{equation}
has four $|\fett{l}'|$-roots at $\lbrace 0,0,R_-,R_+\rbrace$, where
\begin{align}
 \label{r1+-}
R_\pm = \frac{|\fett{q}|-|\fett{q}'| \pm \sqrt{|\fett{q}'|^2-2|\fett{q}||\fett{q}'| \cos\theta+|\fett{q}|^2}}{2}
=\frac{|\fett{q}|-|\fett{q}'| \pm |\fett{q}-\fett{q}'|}{2}  \ ,
\end{align}
and $R_\pm \gtrless 0$ follows from the reverse triangular inequality. We discard the negative root~$R_-$ because it is unphysical.  For the remaining $R_+$, because 
the parabolic part of~(\ref{eq:bbb}) opens upwards, it serves as a lower limit on $|\fett{l}'|$, so that the step function is equivalently
\begin{equation}
\Theta(b_1^2-4a_1c_1)= \Theta(|\fett{l}'|-R_+) \,.
\end{equation}
Also, because $R_+ \geq |\fett{q}|-|\fett{q}'|$, the other step function $\Theta(|\fett{l}'|-(|\fett{q}|-|\fett{q}'|))$ in the collision integral~(\ref{C2 after dcosalpha})
is redundant. Therefore, we find
\begin{equation}
\begin{aligned}
{\cal C}_{1}^{\nu \nu \leftrightarrow \nu \nu}[f] 
=  \frac{6 \mathfrak{g}^4}{(2\pi)^{3}|\fett{q}|}\int  \dd\cos\theta \, \dd|\fett{q}'| \, |\fett{q}'|\,  K^0_\nu(|\fett{q}|,|\fett{q}'|,\cos \theta,\eta) \, F_{\nu}(\fett{k},\fett{q}',\eta) \, ,
 \label{C1_final}
\end{aligned}
\end{equation}
with 
\begin{equation}
K_{i=\nu,\phi}^0(|\fett{q}|,|\fett{q}'|,\cos \theta,\eta)
\equiv \frac{1}{\sqrt{|\fett{q}|^{2}-2|\fett{q}||\fett{q}'|\cos\theta+|\fett{q}'|^{2}}} 
\int_{R_+}^\infty \dd |\fett{l}'| \, \bar{f}_{i}(|\fett{l}'|,\eta),
\label{K1_self}
\end{equation} 
as the integration kernel.



\subsection{First-order \texorpdfstring{$\nu \nu \leftrightarrow \phi \phi$}{nu nu <-> phi phi} collision integrals}
\label{sec:nu,nu->phi,phi}

Using equation~\eqref{C[f]0}, the first-order collision integral for  the neutrinos is
\begin{equation}
\begin{aligned}
\left( \frac{\partial f_{\nu}}{\partial \eta}\right)^{(1)}_{\nu \nu \leftrightarrow \phi \phi}(\fett{k},\fett{q},\eta) 
 &= \frac{2\mathfrak{g}^4}{(2 \pi)^{5}|\fett{q}| } \int \frac{\mathrm{d^{3}} \fett{q}'}{2 | \fett{q}'|}  \int \frac{\mathrm{d^{3}}\fett{l}}{2 | \fett{l}|}  \int \frac{\mathrm{d^{3}}\fett{l}'}{2 | \fett{l}'|} \, \delta_\D^{(4)}(q+l-q'-l') \left( \frac{\tilde{s}^2}{\tilde{t}\tilde{u}} -4 \right)  \\
& \hspace{-3mm} \times  \Big[ 2 \bar{f}_{\phi}(| \fett{l}'|,\eta) \, F_{\phi}(\fett{k},\fett{q}',\eta)-\bar{f}_{\nu}(| \fett{q}|,\eta) \, F_{\nu}(\fett{k},\fett{l},\eta)-\bar{f}_{\nu}(| \fett{l}|,\eta) \, F_{\nu}(\fett{k},\fett{q},\eta) \Big] \\
&\equiv {\cal  C}_{1}^{\nu \nu \leftrightarrow \phi \phi}[f]+{\cal C}_{2}^{\nu \nu \leftrightarrow \phi \phi}[f]+{\cal C}_{3}^{\nu \nu \leftrightarrow \phi \phi}[f] \, ,
\label{Collision Integralannih1}
\end{aligned}
\end{equation}
where $\tilde{s},\tilde{t},\tilde{u}$ are the Mandelstam variables for the comoving 4-momenta~$q,l,q',l'$. 
Those parts that are proportional to the momentum-independent portion of the matrix element can be immediately generalised from the results of section \ref{Massless scalar}.  It remains for us to determine in this section those terms proportional to $\tilde{s}^2/(\tilde{t}\tilde{u})$.


\subsubsection{Reduction of \texorpdfstring{${{\cal C}_{3}^{\nu \nu \leftrightarrow \phi \phi}[f]}$}{C3}}  
\label{Computing C3_ann}

Here, $\fett{l}'$ is the redundant variable.
Then, using the parameterisation~\eqref{k-parametrisation}, we find
\begin{equation}
\label{eq:s2tu}
\frac{\tilde{s}^2}{\tilde{u} \tilde{t}}=\frac{|\fett{l}| |\fett{q}|(1-\cos \alpha)^2}{|\fett{q}'|^2(1- \cos\theta) \left( 1- \cos \beta \sin \alpha \sin \theta - \cos \alpha \cos \theta \right)}.
\end{equation}
Integrating this in $\mathrm{d \beta}$ following equations~\eqref{beta-relation} to \eqref{cosbi} gives
\begin{equation}
\int_{0}^{2\pi} \mathrm{d\beta} \, \delta_{\rm D}(g(\ldots,\beta,\ldots)) \: \frac{\tilde{s}^2}{\tilde{u} \tilde{t}} = 2 \left| \frac{\partial g}{\partial \beta}\right|_{\cos \beta_i}^{-1}
 \frac{ |\fett{l}|^2 \left( 1-\cos\alpha \right)^2}{|\fett{q}'|^2(1-\cos \theta)(\cos \theta-d_3)} \Theta\left( \left|\frac{\partial g}{\partial \beta}\right|_{\cos \beta_i}^{2}\right),
\end{equation}
where 
\begin{equation}
\label{eq:dminus}
d_3= 1-\frac{|\fett{l}|}{|\fett{q}'|}(1-\cos \alpha) .
\end{equation}
Then, combining with the momentum-independent part of $|\mathcal{M}_{\nu \nu \leftrightarrow \phi \phi}|^2$, we obtain, following the reasoning of section~\ref{Computing C_3[f]},
\begin{equation}
\begin{aligned}
\label{eq:c3nunuphiphi}
{\cal C}_{3}^{\nu \nu \leftrightarrow \phi \phi}[f] & = \frac{ 2  \mathfrak{g}^4}{(2\pi)^{3}}X_{\nu}(|\fett{q}|) \,F_{\nu}(\fett{k},\fett{q}) \\ &-\frac{  \mathfrak{g}^4}{2 (2 \pi)^4 |\fett{q}|} F_{\nu}(\fett{k},\fett{q}) \int \mathrm{d} \cos \alpha \, \mathrm{d} |\fett{l}| \, \mathrm{d} |\fett{q}'| \, \Theta \left( |\fett{q}|+|\fett{l}|-|\fett{q}'| \right) \, \frac{|\fett{l}|^3}{|\fett{q}'|^2 |\fett{l}+\fett{q}|}  \; \bar{f}_{\nu}(|\fett{l}|) \\
& \hspace{1.cm} \times \int_{x_-}^{x_+} \dd \cos \theta \,
\frac{ (1-\cos\alpha)^2}{(1-\cos \theta)(\cos \theta-d_3)} \frac{\Theta\left(b_3^2-4 a_3 c_3 \right) }{\sqrt{(x_+ - \cos \theta)(\cos \theta -x_-)} },
\end{aligned}
\end{equation}
where $a_3, b_3, c_3$ are given in equation~\eqref{eq:abc}, and $x_\pm$ in equation~\eqref{eq:roots}.

The condition $b_3^2-4a_3 c_3>0$ encoded in the step function represents a kinematic constraint.  Once it is satisfied, we can further establish that $x_+<1$ and $x_-> d_3$. Then, the $\cos \theta$-integral in equation~(\ref{eq:c3nunuphiphi}) has a simple analytic solution,
\begin{equation}
\begin{aligned}
\label{eq:divergent}
&\int^{x_+}_{x_-}  \frac{1}{(1-x)(x-d_3)} \frac{\dd x}{\sqrt{(x_+-x) (x-x_-)} }  \\
&\hspace{25mm} 
= \frac{\pi}{(1-d_3)} \left(
\frac{1}{\sqrt{(1-x_+)(1-x_-)}}+\frac{1}{\sqrt{(x_+-d_3)(x_--d_3)}}\right) \\
&\hspace{25mm}= \frac{\pi |\fett{q}'|^2 \, |\fett{l}+\fett{q}|}{ |\fett{l}|^2 (1-\cos \alpha)^2} \left(
\frac{1}{||\fett{q}|-|\fett{q}'||}+\frac{1}{||\fett{l}|-|\fett{q}'||}\right).
\end{aligned}
\end{equation}

Note that the expression is divergent at $|\fett{q}'|=|\fett{q}|$ and  $|\fett{q}'|=|\fett{l}|$, which arises because the $\nu\nu \leftrightarrow \phi \phi$ process occurs through the 
$t$- and $u$-channels mediated by a neutrino that is approximately massless.  We regularise these divergences  by introducing a comoving neutrino mass~$\tilde{m}_\nu \equiv a m_\nu$ in the denominator, i.e.,
$1/||\fett{q}|-|\fett{q}'|| \to 1/\sqrt{(|\fett{q}|-|\fett{q}'|)^2+ \tilde{m}_\nu^2}$, etc.   Similar regularisation tricks have also been employed in, e.g., leptogenesis calculations~\cite{HahnWoernle:2009qn}.

Feeding equation~(\ref{eq:divergent}) into~(\ref{eq:c3nunuphiphi}) and noting that the condition $b_3^2-4 a_3 c_3> 0$ amounts to limiting the $|\fett{q'}|$-integral to the region~$(R_1,R_2)$ given in equation~\eqref{roots},
we find
\begin{equation}
\begin{aligned}
{\cal C}_{3}^{\nu \nu \leftrightarrow \phi \phi}[f] =&  \frac{2  \mathfrak{g}^4}{(2\pi)^{3}} 
F_{\nu}(\fett{k},\fett{q},\eta)
\Bigg[X_{\nu}(|\fett{q}|,\eta) \\
&\hspace{5mm}-\frac{1}{|\fett{q}|}
  \int \, \mathrm{d} \cos \alpha \, \mathrm{d} |\fett{l}| \;   \bar{f}_{\nu}(|\fett{l}|,\eta) \; |\fett{l}|\; \Big( \kappa(|\fett{q}|,|\fett{l}|,\cos \alpha)
 + |\fett{q}| \leftrightarrow |\fett{l}|
 \Big)
   \Bigg] ,
\label{C3_dcostheta}
\end{aligned}
\end{equation}
where $|\fett{q}| \leftrightarrow |\fett{l}|$ indicates an exchange of variables in the argument of $\kappa$, defined as
\begin{equation}
\begin{aligned}
\kappa(|\fett{q}|,|\fett{l}|,\cos \alpha) & \equiv  \frac{1}{8}
\int_{(|\fett{l}|+ |\fett{q}|-|\fett{l}+\fett{q}|)/2} ^{(|\fett{l}|+ |\fett{q}|+|\fett{l}+\fett{q}|)/2} \mathrm{d |\fett{q}'|} \, 
 \frac{1}{\sqrt{(|\fett{l}|-|\fett{q}'|)^2+ \tilde{m}_\nu^2}} \\
&=  \frac{1}{8} \log 
\left[\frac{\sqrt{(|\fett{l}|-|\fett{q}|+|\fett{l}+\fett{q}|)^2+4 \tilde{m}_\nu^2} +|\fett{l}|-|\fett{q}|+|\fett{l}+\fett{q}|} 
{\sqrt{(|\fett{l}|-|\fett{q}|-|\fett{l}+\fett{q}|)^2+4 \tilde{m}_\nu^2} +|\fett{l}|-|\fett{q}|-|\fett{l}+\fett{q}|}
\right],
\label{K3_ann}
\end{aligned}
\end{equation}
and we note that the $\cos \alpha$-dependence is hidden in $S\equiv |\fett{l}+\fett{q}|=\sqrt{|\fett{l}|^2+2 |\fett{l}| |\fett{q}| \cos \alpha+|\fett{q}|^2}$.

Changing the $\cos \alpha$-integration variable according to
\begin{equation}
\label{eq:changeS}
\int^1_{-1} \dd \cos  \alpha\, (\cdots) \to \frac{1}{|\fett{l}| |\fett{q}|} \int_{||\fett{l}|-|\fett{q}||}^{|\fett{l}|+|\fett{q}|} \dd S \, S\, (\cdots)
\end{equation}
enables us to establish that  
\begin{equation}
\label{eq:kappaintegral}
\int \dd \cos \alpha \; \kappa(|\fett{q}|,|\fett{l}|,\cos \alpha) \to \frac{1}{4}
\left[ 
\log(4) -1 + \log\left(\frac{|\fett{l}| |\fett{q}|}{\tilde{m}_\nu^2}\right)
\right]
\end{equation}
in the limit $|\fett{l}|,|\fett{q}| \gg \tilde{m}_\nu$. The third term in \eqref{C3_dcostheta} with $|\fett{q}| \leftrightarrow |\fett{l}|$ gives the same contribution as \eqref{eq:kappaintegral}.
Then, defining 
\begin{equation}
Z_{i=\nu,\phi} (|\fett{q}|,\eta)\equiv \frac{1}{2 |\fett{q}|} \int  \mathrm{d |\fett{l}|} \, \bar{f}_i (|\fett{l}|,\eta) \, |\fett{l}| \;
\log\left(\frac{|\fett{l}| |\fett{q}|}{\tilde{m}_\nu^2}\right),
\label{X_ann}
\end{equation}
we can finally rewrite equation~(\ref{C3_dcostheta}) as
\begin{equation} \label{C3_nuann}
{\cal C}_{3}^{\nu \nu \leftrightarrow \phi \phi}[f] = \frac{2 \mathfrak{g}^4}{(2\pi)^{3}} \left[
\left( \frac{3}{2} - \frac{1}{2} \log(4) \right)
X_{\nu}(|\fett{q}|,\eta) - Z_\nu(|\fett{q}|,\eta) \right] F_{\nu}(\fett{k},\fett{q},\eta),
\end{equation}
where we note again that both $X_\nu$ and $Z_\nu$ depend on the time-dependent neutrino background phase space distribution. 


\subsubsection{Reduction of \texorpdfstring{${{\cal C}_{2}^{\nu \nu \leftrightarrow \phi \phi}[f]}$}{C3}}

Since ${\cal C}_{2}^{\nu \nu \leftrightarrow \phi \phi}[f]$ and ${\cal C}_{3}^{\nu \nu \leftrightarrow \phi \phi}[f]$ differ only in the exchange 
$\bar{f}_{\nu}(|\fett{l}|) \leftrightarrow \bar{f}_{\nu}{(|\fett{q}}|)$ and $F_{\nu}(\fett{k},\fett{q})\leftrightarrow F_{\nu}(\fett{k},\fett{l})$, we can immediately generalise equations~\eqref{eq:c2fin} and~\eqref{C3_dcostheta} to get
\begin{equation}
\begin{aligned}
{\cal C}_2^{\nu \nu \leftrightarrow \phi \phi}[f] =  \frac{2 \mathfrak{g}^4}{(2\pi)^3|\fett{q}|}\bar{f}_{\nu}(|\fett{q}|,\eta) \! \int  \dd\cos\alpha \, \dd|\fett{l}|  \, |\fett{l}| 
\left[ \frac{1}{2} - \Big( \kappa(|\fett{q}|,|\fett{l}|,\cos \alpha) \!+\! |\fett{l}| \leftrightarrow|\fett{q}|\Big)  \right] F_{\nu}(\fett{k},\fett{l},\eta) ,
\end{aligned}
\end{equation}
where the time-independent kernel $\kappa(|\fett{q}|,|\fett{l}|,\cos \alpha)$ is given in equation~(\ref{K3_ann}).


\subsubsection{Reduction of \texorpdfstring{${{\cal C}_{1}^{\nu \nu \leftrightarrow \phi \phi}[f]}$}{C3}}

Here, $\fett{l}$ is the redundant variable.  Then, using  the parameterisation~(\ref{param2}) and integrating in $\dd \beta$ as in section~\ref{Computing C_2[f] and C_1[f]}, we find 
\begin{equation}
\int_{0}^{2\pi} \mathrm{d\beta} \, \delta_{\rm D}(g(\ldots,\beta,\ldots)) \: \frac{\tilde{s}^2}{\tilde{u} \tilde{t}} = 2 \left| \frac{\partial g}{\partial \beta}\right|_{\cos \beta_i}^{-1}
 \frac{|\fett{l}'|}{|\fett{q}'| (1-\cos \theta)}
  \frac{(\cos \alpha -d_1)^2}{(1-\cos \alpha)} \, \Theta\left( \left|\frac{\partial g}{\partial \beta}\right|_{\cos \beta_i}^{2}\right),
\end{equation}
where
\begin{equation}
d_1\equiv 1 + \frac{|\fett{q}'|}{|\fett{l}'|} (1- \cos \theta).
\label{eq:dplus}
\end{equation}
Thus, similar to section~\ref{Computing C_2[f] and C_1[f]}, the
collision integral ${\cal C}_{1}^{\nu \nu \leftrightarrow \phi \phi}[f]$ is given by
\begin{equation}
\begin{aligned}
{\cal C}_{1}^{\nu \nu \leftrightarrow \phi \phi}[f] &=  -\frac{2\mathfrak{g}^4}{(2\pi)^{3} |\fett{q}|}\int  \dd\cos\theta \, \dd|\fett{q}'| \,  |\fett{q'}|\, K_{\phi}^0(|\fett{q}|,|\fett{q}'|,\cos \theta,\eta) \, F_{\phi}(\fett{k},\fett{q}') \\ 
&\hspace{-5mm} +\frac{\mathfrak{g}^4}{ (2 \pi)^4 |\fett{q}|} \int \mathrm{d \cos \theta} \; \mathrm{d |\fett{l}'|} \, \mathrm{d| \fett{q}'|} \, \Theta \left( |\fett{q}'|+|\fett{l}'|-|\fett{q}| \right) 
\frac{|\fett{l}'|}{|\fett{q}'-\fett{q}|} F_{\phi}(\fett{k},\fett{q}') \bar{f}_{\phi}(|\fett{l}'|) \\
& \hspace{5mm} \times \int_{y_-}^{y_+} \dd \cos \alpha \,
 \frac{1}{(1-\cos \theta)}
  \frac{(\cos \alpha -d_1)^2}{(1-\cos \alpha)}
\frac{\Theta\left( b_1^2-4 a_1 c_1 \right) }{\sqrt{(y_+-\cos \alpha)(\cos \alpha-y_-)} },
\label{C1_ann}
\end{aligned}
\end{equation}
where we have used equation~(\ref{C1_final}) in the first line, with $K_{\phi}^0$ defined in~\eqref{K1_self}, $a_1,b_1,c_1$ are given in~(\ref{a-c}), and $y_\pm$ are the real and nondegenerate roots of the quadratic~(\ref{eq:dgdb1}).

The kinematic constraint $b_1^2 -4 a_1 c_1>0$ guarantees that $y_+<1$ and $d_1>y_+$.  Then, the $\cos \alpha$-integral has the solution 
\begin{equation}
\begin{aligned}
&\int^{y_+}_{y_-} \frac{(y-d_1)^2}{(1-y)} \frac{\dd y}{\sqrt{(y_+-y) (y-y_-)} }  
= \pi \left( \frac{(1-d_1)^2}{\sqrt{(1-y_+) (1-y_-)}}  + 2 d_1 -1 - \frac{y_++y_-}{2}  \right) \\
 &\hspace{5mm} =  \frac{\pi |\fett{q}'| |\fett{q}'-\fett{q}| (1-\cos \theta) }{|\fett{l}'|} 
\left( \frac{1 }{ ||\fett{l}'|\!-\! |\fett{q}||}
 + \frac{3}{2} \frac{1}{|\fett{q}- \fett{q}'|}
 + \frac{(2 |\fett{l}'| -|\fett{q}|+|\fett{q}'|)(|\fett{q}|+|\fett{q}'|)}
  {2 |\fett{q}-\fett{q}'|^3} \right).
 \end{aligned}
\end{equation}
Then, noting that the condition $b_1^2-4 a_1 c_1>0$ also translates into a lower integration limit $R_+$ on $|\fett{l}'|$, given in equation~(\ref{r1+-}), 
the collision integral can be written as
\begin{equation}
\begin{aligned}
{\cal C}_{1}^{\nu \nu \leftrightarrow \phi \phi}[f]=   -\frac{2\mathfrak{g}^4}{(2\pi)^{3}|\fett{q}|}\int  \dd\cos\theta \, \dd|\fett{q}'| \, |\fett{q}'|
\left(\frac{5}{8}K_\phi^0-K^u_\phi-K^t_\phi \right)(|\fett{q}|,|\fett{q}'|,\cos \theta,\eta) \;
F_{\phi}(\fett{k},\fett{q}',\eta),
\label{eq:C1_ann}
\end{aligned}
\end{equation}
where 
\begin{equation}
\begin{aligned}
K_{i=\nu,\phi}^u (|\fett{q}|,|\fett{q}'|,\cos \theta,\eta) & \equiv \frac{1}{4}   \int_{R_+}^{\infty} \mathrm{d |\fett{l}'|} \, \bar{f}_{i}(|\fett{l}'|,\eta)  \: \frac{1}{\sqrt{(|\fett{l}'|-|\fett{q}|)^2+\tilde{m}_\nu^2}}, \\
K_{i=\nu,\phi}^t (|\fett{q}|,|\fett{q}'|,\cos \theta,\eta) & \equiv 
\frac{1}{8}  \frac{|\fett{q}|+|\fett{q}'|}{ |\fett{q}-\fett{q}'|^3} , \int_{R_+}^{\infty} \mathrm{d |\fett{l}'|} \, \bar{f}_{i}(|\fett{l}'|,\eta) \:
 (2 |\fett{l}'| -|\fett{q}|+|\fett{q}'|),
 \label{eq:kut}
\end{aligned}
\end{equation}
and we have inserted a comoving neutrino mass $\tilde{m}_\nu$ to regularise the divergence at $|\fett{l}'| = |\fett{q}|$.


\subsection{First-order \texorpdfstring{$\nu \phi \leftrightarrow \nu \phi$}{nu phi <-> nu phi} collision integrals}
\label{Neutrino scalar}

The neutrino population has the first-order collision integral 
\begin{equation}
\begin{aligned}
\left( \frac{\partial f_{\nu}}{\partial \eta}\right)^{(1)}_{\nu \phi \leftrightarrow \nu \phi}(\fett{k},\fett{q},\eta) 
 & = \frac{8\mathfrak{g}^4}{(2 \pi)^{5}|\fett{q}|} \int \frac{\mathrm{d^{3}} \fett{q}'}{2 | \fett{q}'|}  \int \frac{\mathrm{d^{3}}\fett{l}}{2 | \fett{l}|}  \int \frac{\mathrm{d^{3}}\fett{l}'}{2 | \fett{l}'|} \, \delta_\D^{(4)}(q+l-q'-l') \;  \left(4- \frac{\tilde{t}^2}{\tilde{s}\tilde{u}} \right)\\
 &\hspace{15mm} \times   \Big[  \bar{f}_{\nu}(| \fett{q}'|,\eta) \, F_{\phi}(\fett{k},\fett{l}',\eta)+\bar{f}_{\phi}(| \fett{l}'|,\eta) \, F_{\nu}(\fett{k},\fett{q}',\eta)\\
 &\hspace{25mm}-\bar{f}_{\nu}(| \fett{q}|,\eta) \, F_{\phi}(\fett{k},\fett{l},\eta)-\bar{f}_{\phi}(| \fett{l}|,\eta) \, F_{\nu}(\fett{k},\fett{q},\eta) \Big] \\
& \equiv {\cal  C}_{1a}^{\nu \phi \leftrightarrow \nu \phi}[f]+{\cal C}_{1b}^{\nu \phi \leftrightarrow \nu \phi}[f]+{\cal C}_{2}^{\nu \phi \leftrightarrow \nu \phi}[f] + {\cal C}_{3}^{\nu \phi \leftrightarrow \nu \phi}[f] \, . 
\label{Collision Integralscat1}
\end{aligned}
\end{equation}
Again, we need only to deal with the momentum-dependent component of the  squared matrix element; the momentum-independent part can be generalised from the results of section~\ref{Massless scalar}.


\subsubsection{Reduction of \texorpdfstring{${{\cal C}_{3}^{\nu \phi \rightarrow \nu \phi}[f]}$}{C3}}  

Integrating in $\dd^3 \fett{l}'$ and using the parameterisation~\eqref{k-parametrisation} for the remaining 3-momenta,
we find,  following equations~\eqref{beta-relation} to \eqref{cosbi}, 
\begin{equation}
\int_{0}^{2\pi} \mathrm{d\beta} \, \delta_{\rm D}(g(\ldots,\beta,\ldots)) \: \frac{\tilde{t}^2}{\tilde{s} \tilde{u}} = 2 \left| \frac{\partial g}{\partial \beta}\right|_{\cos \beta_i}^{-1}
\frac{|\fett{q}'|}{|\fett{l}|(1-\cos \alpha)}
\frac{\left( 1-\cos\theta \right)^2}{(\cos \theta - d_3)}
 \Theta\left( \left|\frac{\partial g}{\partial \beta}\right|_{\cos \beta_i}^{2}\right),
\end{equation}
where $d_3$ is given in equation~(\ref{eq:dminus}).
The collision integral is therefore
\begin{equation}
\begin{aligned}
{\cal C}_{3}^{\nu \phi \leftrightarrow \nu \phi}[f] = & - \frac{8\mathfrak{g}^4}{(2\pi)^{3}}X_\phi(|\fett{q}|) \,F_{\nu}(\fett{k},\fett{q}) \\ & +\frac{2\mathfrak{g}^4}{(2 \pi)^4 |\fett{q}|} F_{\nu}(\fett{k},\fett{q}) \int  \mathrm{d \cos \alpha} \, \mathrm{d |\fett{l}|} \, \mathrm{|d \fett{q}'|} \, \Theta \left( |\fett{q}|+|\fett{l}|-|\fett{q}'| \right) \frac{|\fett{q}'|}{|\fett{l}+\fett{q}|} \,  \bar{f}_{\phi}(|\fett{l}|) \\
& \hspace{5mm} \times \int_{x_-}^{x_+} \dd \cos \theta \; \frac{1}{(1-\cos \alpha)}
\frac{(1-\cos\theta)^2}{(\cos \theta - d_3)} \frac{\Theta\left(b_3^2-4 a_3 c_3 \right) }{ \sqrt{(x_+-\cos \theta)(\cos \theta - x_-)} },
\end{aligned}
\end{equation}
where $X_\phi(|\fett{q}|,\eta)$ is defined in equation~(\ref{X_self}), 
$a_3,b_3, c_3$ given in~\eqref{eq:abc}, and  $x_\pm$  in~\eqref{eq:roots}.

The kinematic condition $b_3^2-4 a_3 c_3>0$ encoded in the step function  ensures  that $x_-> d_3$.  Then, integrating in  $\dd \cos \theta$ over the interval~$[x_-,x_+]$ we find
\begin{equation}
\begin{aligned}
 & \int^{x_+}_{x_-} \frac{(1-x)^2}{(x - d_3)} \frac{\dd x}{ \sqrt{(x_+-x)(x-x_-)} } 
 =\pi \left( \frac{(1-d_3)^2}{\sqrt{(x_+-d_3) (x_--d_3)}}  + d_3- 2 + \frac{x_++x_-}{2}  \right) 
 \\ &\hspace{5mm} 
 = \frac{\pi |\fett{l}|  |\fett{l}+\fett{q}| (1-\cos \alpha)}{|\fett{q}'|}  \left( \frac{1 }{ ||\fett{l}|-|\fett{q}'||}
 - \frac{3}{2} \frac{1}{|\fett{l}+\fett{q}|} + 
  \frac{(|\fett{l}| + |\fett{q}| - 2 |\fett{q}'|)  (|\fett{l}|-|\fett{q}|)}
  {2 |\fett{l}+\fett{q}|^3}
   \right) ,
   \label{eq:ellipticc4}
 \end{aligned}
\end{equation}
which has  a divergence at $|\fett{q}'|=|\fett{l}|$, to be regularised as discussed in section~\ref{Computing C3_ann}.
Using the $|\fett{q}'|$-integration limits $(R_1,R_2)$  of equation~(\ref{roots}), 
the collision integral becomes
\begin{equation}
\begin{aligned}
\label{eq:nuphinuphi}
{\cal C}_{3}^{\nu \phi \leftrightarrow \nu \phi}[f] =  - \frac{8\mathfrak{g}^4}{(2\pi)^{3}}
\left[X_\phi(|\fett{q}|) - \frac{1}{|\fett{q}|}\int  \dd \cos \alpha \, \mathrm{d |\fett{l}|} \,  \bar{f}_{\phi}(|\fett{l}|) \, |\fett{l}| \, \Lambda (|\fett{q}|,|\fett{l}|,\cos \alpha)\right] F_{\nu}(\fett{k},\fett{q}),
\end{aligned}
\end{equation}
with
\begin{equation}
\begin{aligned}
 \Lambda (|\fett{q}|,|\fett{l}|,\cos \alpha)&=  \kappa(|\fett{q}|,|\fett{l}|,\cos \alpha)- \frac{1}{8 } \int_{R_1}^{R_2} \dd |\fett{q}'| 
  \left(\frac{3}{2} \frac{1}{|\fett{l}+\fett{q}|} -
  \frac{(|\fett{l}| + |\fett{q}| - 2 |\fett{q}'|)  (|\fett{l}|-|\fett{q}|)}
  {2 |\fett{l}+\fett{q}|^3} \right) \\
 &= \kappa(|\fett{q}|,|\fett{l}|,\cos \alpha)
 - \frac{3}{16},
 \label{eq:lambda2}
\end{aligned}
\end{equation}
where $\kappa(|\fett{q}|,|\fett{l}|,\cos \alpha)$ is given in equation~(\ref{K3_ann}). 

To integrate equation~(\ref{eq:lambda2}) in $\cos \alpha$ we use the result of equation~(\ref{eq:kappaintegral}), valid  in the limit $|\fett{l}|,|\fett{q}| \gg \tilde{m}_\nu$.  This gives
\begin{equation}
\int \dd \cos \alpha \; \Lambda(|\fett{q}|,|\fett{l}|,\cos \alpha) =  \frac{1}{4}
\left[ 
\log(4) -\frac{5}{2} + \log\left(\frac{|\fett{l}| |\fett{q}|}{\tilde{m}_\nu^2}\right)
\right],
\end{equation}
which we can then substitute into equation~(\ref{eq:nuphinuphi}) to get
\begin{equation}
\begin{aligned}
{\cal C}_{3}^{\nu \phi \leftrightarrow \nu \phi}[f] =  - \frac{8\mathfrak{g}^4}{(2\pi)^{3}}
\left[ \left( \, \frac{13}{8} - \frac{1}{4} \log(4) \right)
X_\phi(|\fett{q}|,\eta) - \frac{1}{2} Z_\phi(|\fett{q}|,\eta) \right] F_{\nu}(\fett{k},\fett{q},\eta),
\label{eq:C3elasticfinal}
\end{aligned}
\end{equation}
where $Z_\phi(|\fett{q}|,\eta)$ is given in equation~(\ref{X_ann}).


\subsubsection{Reduction of \texorpdfstring{${{\cal C}_{2}^{\nu \phi \leftrightarrow \nu \phi}[f]}$}{C3}}

As ${{\cal C}_{2}^{\nu \phi \leftrightarrow \nu \phi}[f]}$ and ${{\cal C}_{3}^{\nu \phi \leftrightarrow \nu \phi}[f]}$
differ only in the exchange $\bar{f}_{\phi}(|\fett{l}|) \rightarrow  \bar{f}_{\nu}(|\fett{q}|)$ and  $F_{\nu}(\fett{k},\fett{q})\rightarrow F_{\phi}(\fett{k},\fett{l})$, we may simply
generalise equations~\eqref{eq:c2fin} and~(\ref{eq:nuphinuphi}) to obtain
\begin{equation}
\begin{aligned}
{\cal C}_2^{\nu \phi \leftrightarrow \nu \phi}[f] = -\frac{8 \mathfrak{g}^4}{(2\pi)^3|\fett{q}|}\bar{f}_{\nu}(|\fett{q}|,\eta)\int  \dd\cos\alpha \, \dd|\fett{l}| \, |\fett{l}|
\left[\frac{11}{16} - \kappa(|\fett{q}|,|\fett{l}|,\cos \alpha)  \right] F_{\phi}(\fett{k},\fett{l},\eta),
\end{aligned}
\end{equation}
where the time-independent kernel $\kappa$ is given in equation~(\ref{K3_ann}).


\subsubsection{Reduction of \texorpdfstring{${{\cal C}_{1b}^{\nu \phi \leftrightarrow \nu \phi}[f]}$}{C3}}

We begin by integrating in $\dd^3 \fett{l}$.  Then, using the parameterisation~(\ref{param2}) and integrating in $\dd \beta$ as in section~\ref{Computing C_2[f] and C_1[f]} gives
\begin{equation}
\int_{0}^{2\pi} \mathrm{d\beta} \, \delta_{\rm D}(g(\ldots,\beta,\ldots)) \: \frac{\tilde{t}^2}{\tilde{s} \tilde{u}} = 2 \left| \frac{\partial g}{\partial \beta}\right|_{\cos \beta_i}^{-1}
 \frac{|\fett{q}'|^2(1-\cos \theta)^2}{|\fett{l}'|^2(\cos \alpha-1)(\cos \alpha-d_1)}
 \Theta\left( \left|\frac{\partial g}{\partial \beta}\right|_{\cos \beta_i}^{2}\right),
\end{equation}
where $d_1$ is given in equation~(\ref{eq:dplus}).
The  collision integral ${\cal C}_{1b}^{\nu \phi \leftrightarrow \nu \phi}[f]$ is then
\begin{equation}
\begin{aligned}
{\cal C}_{1b}^{\nu \phi \leftrightarrow \nu \phi}[f] &= \frac{4\mathfrak{g}^4}{(2\pi)^{3}  |\fett{q}|} \int  \dd\cos\theta \, \dd|\fett{q}'| \,  |\fett{q'}| \, K_{\phi}^0(|\fett{q}|,|\fett{q}'|,\cos \theta) \, F_{\nu}(\fett{k},\fett{q}') \\ 
&\hspace{-10mm}-\frac{2\mathfrak{g}^4}{ (2 \pi)^4 |\fett{q}|} \int \mathrm{d \cos \theta} \,\mathrm{d |\fett{l}'|} \, \mathrm{d| \fett{q}'|} \, \Theta \left( |\fett{q}'|+|\fett{l}'|-|\fett{q}| \right) \, \frac{|\fett{q}'|^3}{|\fett{l}'|^2|\fett{q}'-\fett{q}|} F_{\nu}(\fett{k},\fett{q}') \bar{f}_{\phi}(|\fett{l}'|) \\
& \hspace{5mm} \times \int_{y_-}^{y_+} \dd \cos \alpha \,   \frac{(1-\cos \theta)^2}{(\cos \alpha-1)(\cos \alpha-d_1)}
\frac{\Theta\left(b_1^2-4 a_1 c_1 \right) }{\sqrt{(y_+- \cos \alpha)(\cos \alpha - y_-)} },
\label{C1_elastic}
\end{aligned}
\end{equation}
where we have used \eqref{C1_final} in the first line, $a_1,b_1,c_1$ are given in equation~(\ref{a-c}), and  $b_1^2 -4 a_1 c_1>0$ ensures that $y_+<1$ and $d_1>y_+$, with $y_\pm$  the roots of the quadratic~(\ref{eq:dgdb1}).

Then, as in equation~(\ref{eq:divergent}), the $\cos \alpha$-integral has the solution
\begin{equation}
\begin{aligned}
\int^{y_+}_{y_-}  \frac{\dd y}{(y-1)(y-d_1)} \frac{1}{\sqrt{(y_+-y)(y-y_-)} } 
= \frac{\pi |\fett{l}'|^2 |\fett{q}'-\fett{q}|}{ |\fett{q}'|^2 (1-\cos \theta)^2} \left(
\frac{1}{||\fett{l}'|-|\fett{q}||} - \frac{1}{|\fett{l}'|+|\fett{q}'|}\right). \nonumber
\end{aligned}
\end{equation}
Substituting this into equation~(\ref{C1_elastic}) and  noting that  the $|\fett{l}'|$-integral has now a lower limit $R_+ = (|\fett{q}|-|\fett{q}'|+\sqrt{|\fett{q}|^2-2 |\fett{q}| |\fett{q}'| \cos \theta + |\fett{q}'|^2})/2$, we find
\begin{equation}
\begin{aligned}
{\cal C}_{1b}^{\nu \phi \leftrightarrow \nu \phi}[f] =
\frac{4\mathfrak{g}^4}{(2\pi)^{3}|\fett{q}|} \int  \dd\cos\theta \, \dd|\fett{q}'| \, |\fett{q}'|  \left(K_{\phi}^0 -K_\phi^u +K_\phi^s\right)(|\fett{q}|,|\fett{q}'|,\cos \theta,\eta) \;
F_{\nu}(\fett{k},\fett{q}',\eta),  
\label{eq:C1b_elastic}
\end{aligned}
\end{equation}
where
\begin{equation}
\begin{aligned}
K_{i=\nu,\phi}^s \equiv & \frac{1}{4}
\int_{R_+} ^{\infty} \mathrm{d |\fett{l}'|} \, f_i(|\fett{l}'|,\eta) \;
\frac{1}{|\fett{l}'|+|\fett{q}'|}\, ,
\label{eq:ks}
 \end{aligned}
\end{equation}
whereas $K_{\phi}^0$ and $K_\phi^u$ are defined in equations~(\ref{K1_self}) and~(\ref{eq:kut}), respectively.


\subsubsection{Reduction of \texorpdfstring{${{\cal C}_{1a}^{\nu \phi \leftrightarrow \nu \phi}[f]}$}{C3}}

The ${\cal C}_{1a}^{\nu \phi \leftrightarrow \nu \phi}[f]$ term is formally identical to ${\cal C}_{1b}^{\nu \phi \leftrightarrow \nu \phi}[f]$ 
under the exchange of the 4-momenta $q'\leftrightarrow l'$ and the labels  $\nu \leftrightarrow \phi$ of $\bar{f}_i$ and $F_i$.  The 4-momentum exchange also demands that we modify the momentum-dependent part of the squared matrix element according to
\begin{equation}
\frac{\tilde{t}^2}{\tilde{s} \tilde{u}} \to \frac{\tilde{u}^2}{\tilde{s} \tilde{t}} = \frac{|\fett{l}'| |\fett{q}| (1-\cos \alpha)^2}
{|\fett{q}'|^2(1-\cos \theta)(\cos \alpha \cos \theta + \cos \beta \sin \alpha \sin \theta -1)},
\end{equation}
using the parameterisation~(\ref{param2}).  Integration in $\dd \fett{l}$ and $\dd \beta$ following section~\ref{Computing C_2[f] and C_1[f]} yields
\begin{equation}
\int_{0}^{2\pi} \mathrm{d\beta} \, \delta_{\rm D}(g(\ldots,\beta,\ldots)) \: \frac{\tilde{t}^2}{\tilde{s} \tilde{u}} = 2 \left| \frac{\partial g}{\partial \beta}\right|_{\cos \beta_i}^{-1}
 \frac{|\fett{l}'|}{|\fett{q}'| (1-\cos \theta)} \frac{ (1-\cos \alpha)^2}{(d_1- \cos \alpha)}
 \Theta\left( \left|\frac{\partial g}{\partial \beta}\right|_{\cos \beta_i}^{2}\right),
\end{equation}
with $d_1$ is defined in equation~(\ref{eq:dplus}), which can then be  substituted in to equation~(\ref{C1_elastic}) followed by an exchange of the $\nu$ and $\phi$ labels to give 
\begin{equation}
\begin{aligned}
{\cal C}_{1a}^{\nu \phi \leftrightarrow \nu \phi}[f] &= \frac{4\mathfrak{g}^4}{(2\pi)^{3}{|\fett{q}|}} \int  \dd\cos\theta \, \dd|\fett{q}'| \, {|\fett{q}'|}\, K_{\nu}^0(|\fett{q}|,|\fett{q}'|,\cos \theta) \, F_{\phi}(\fett{k},\fett{q}') \\ 
&\hspace{-2mm}-\frac{2\mathfrak{g}^4}{ (2 \pi)^4 |\fett{q}|} \int \mathrm{d \cos \theta} \, \mathrm{d |\fett{l}'|} \, \mathrm{d| \fett{q}'|} \, \Theta \left( |\fett{q}'|+|\fett{l}'|-|\fett{q}| \right) \, \frac{|\fett{l}'|}{|\fett{q}'-\fett{q}|} \; F_{\phi}(\fett{k},\fett{q}') \bar{f}_{\nu}(|\fett{l}'|) \\
& \hspace{5mm} \times  \int_{y_-}^{y_+} \dd \cos \alpha \, \frac{1}{(1-\cos \theta)} \frac{ (1-\cos \alpha)^2}{(d_1- \cos \alpha)}
\frac{\Theta\left(b_1^2-4 a_1 c_1 \right) }{\sqrt{(y_+-\cos \alpha)(\cos \alpha-y_-)} },
\label{eq:c1elastic}
\end{aligned}
\end{equation}
where $d_1>y_+$ if the kinematic constraint $b_1^2-4 a_1 c_1>0$ is satisfied.

Integrating in $\dd \cos \alpha$ in the interval $[y_-,y_+]$ gives, similarly to equation~(\ref{eq:ellipticc4}), 
\begin{align}
 &\int^{y_+}_{y_-}  \frac{(1-y)^2}{(d_1-y)} \frac{\dd y}{ \sqrt{(y_+-y)(y-y_-)} }  \nonumber \\
 & = \frac{\pi |\fett{q}'| |\fett{q}'-\fett{q}|  (1-\cos \theta)}{ |\fett{l}'|}  \left( \frac{1 }{ |\fett{l}'|+|\fett{q}'|}
 -  \frac{3}{2} \frac{1}{|\fett{q}'-\fett{q}|} +  \frac{(2 |\fett{l}'|-|\fett{q}|+|\fett{q}'|)(|\fett{q}|+|\fett{q}'|)}{2 |\fett{q}'-\fett{q}|^3}
    \right).
\end{align}
Folding this back into equation~(\ref{eq:c1elastic}), we obtain
\begin{equation}
\begin{aligned}
{\cal C}_{1a}^{\nu \phi \leftrightarrow \nu \phi}[f] =
\frac{4\mathfrak{g}^4}{(2\pi)^{3}|\fett{q}|} \int  \dd\cos\theta \, \dd|\fett{q}'| \, |\fett{q}'| \left( \frac{11}{8}  K_{\nu}^0-K_\nu^s - K_\phi^t \right) (|\fett{q}|,|\fett{q}'|,\cos \theta,\eta) \;
F_{\phi}(\fett{k},\fett{q}',\eta),  
\label{eq:C1a_elastic}
\end{aligned}
\end{equation}
where $K_\nu^s$ and $K_\nu^t$ are given in equations~(\ref{eq:ks}) and~(\ref{eq:kut}) respectively.


\subsection{Zeroth-order collision integrals}
\label{app:zeroth}

The zeroth-order collision integral for the neutrinos undergoing the process $\nu j \leftrightarrow k l$ is 
\begin{equation}
\begin{aligned} 
\left(\frac{\partial f_\nu}{\partial \eta} \right)^{(0)}_{\nu j \leftrightarrow kl}(\fett{q},\eta)  = & \ \frac{g_j g_k g_l }{2 (2 \pi)^{5} |\fett{q}|  } \int \frac{\mathrm{d^{3}}\fett{q}'}{2 | \fett{q}'|}  \int \frac{\mathrm{d^{3}}\fett{l}}{2 | \fett{l}|}  \int \frac{\mathrm{d^{3}} \fett{l}'}{2 | \fett{l}'|} \, \delta_\D^{(4)}(q+l-q'-l')  \\
 &\hspace{10mm}\times   |{\cal M}_{\nu j\leftrightarrow kl}|^2 \Big( \bar{f}_k(| \fett{l}'|,\eta) \, \bar{f}_l(|\fett{q}'|,\eta)  - 
 \bar{f}_\nu(| \fett{q}|,\eta) \, \bar{f}_j(|\fett{l}|,\eta) \Big) \\
 = & \ {\cal D}^{\nu j \leftrightarrow k l}_1[f] + {\cal D}^{\nu j \leftrightarrow k l}_2[f],
\end{aligned}
\end{equation}
which splits into two terms, to be dealt with separately below.


\subsubsection{Reduction of ${\cal D}^{\nu j \leftrightarrow k l}_2[f]$}

The computation of the zeroth-order ${\cal D}^{\nu j \leftrightarrow k l}_2[f]$ is  procedurally identical to that of the first-order ${\cal C}^{\nu j \leftrightarrow k l}_3[f]$ discussed in the previous sections, save for the replacement of the perturbed phase space $F_j(\fett{k},\fett{q}, \eta)$ with the background distribution~$\bar{f}_j(|\fett{l}|,\eta)$.
Thus, we can immediately generalise equations~(\ref{eq:c3final}), (\ref{C3_nuann})  and~(\ref{eq:C3elasticfinal}) to obtain 
\begin{equation}
\begin{aligned}
{\cal D}_{2}^{\nu \nu \leftrightarrow \nu \nu}[f] = &  -  \frac{6 \mathfrak{g}^4}{(2\pi)^{3}|\fett{q}|} \, \bar{f}_{\nu}(|\fett{q}|,\eta)
\int \mathrm{d|\fett{l}|} \,  |\fett{l}|\,  \bar{f}_\nu(|\fett{l}|,\eta),  \\
{\cal D}_{2}^{\nu \nu \leftrightarrow \phi \phi}[f] = &\frac{2 \mathfrak{g}^4}{(2\pi)^{3} |\fett{q}|}\,  \bar{f}_{\nu}(|\fett{q}|,\eta) 
\int \dd |\fett{l}| \, |\fett{l} |\left[
\left( \frac{3}{2} - \frac{1}{2} \log(4) \right)- \frac{1}{2} \log\left(\frac{|\fett{l}| |\fett{q}|}{\tilde{m}_\nu^2}\right)
\right] \bar{f}_\nu (|\fett{l}|,\eta), \\
{\cal D}_{2}^{\nu \phi \leftrightarrow \nu \phi}[f] =&  - \frac{8\mathfrak{g}^4}{(2\pi)^{3}|\fett{q}|} \, \bar{f}_\nu (|\fett{q}|,\eta)
\int \dd |\fett{l}|\, |\fett{l}|
\left[ \left( \frac{13}{8}  - \frac{1}{4} \log(4) \right)- \frac{1}{4}
 \log\left(\frac{|\fett{l}| |\fett{q}|}{\tilde{m}_\nu^2}\right)
\right] \bar{f}_\phi(|\fett{l}|,\eta),
\end{aligned}
\end{equation}
where we remind the reader that $\tilde{m}_\nu$ is the comoving neutrino mass.


\subsubsection{Reduction of ${\cal D}^{\nu j \leftrightarrow k l}_1[f]$}

Except for the replacement of the perturbed phase space $F_l(\fett{k},\fett{q}', \eta)$ with the background distribution~$\bar{f}_l(|\fett{q}'|,\eta)$, 
the derivation of  ${\cal D}^{\nu j \leftrightarrow k l}_1[f]$ parallels that of  the first-order ${\cal C}^{\nu j \leftrightarrow k l}_1[f]$ described previously. 
Then, following from equations~(\ref{C2 after dcosalpha}), (\ref{C1_ann}), (\ref{eq:C1b_elastic}) and (\ref{eq:C1a_elastic}), 
we can immediately write down as a general form,

\begin{equation}
{\cal D}^{\nu j \leftrightarrow k l}_1[f] =\int_0^{\infty} \dd |\fett{q}'| \, \bar{f}_l(|\fett{q}'|,\eta) \int_{-1}^{1} \dd \cos \theta \int_{R_+}^\infty \dd |\fett{l}'| \; 
(\cdots) \,
\bar{f}_k(|\fett{l}'|,\eta)
\label{eq:dgeneral}
\end{equation}
where $R_+ \equiv (|\fett{q}|-|\fett{q}'|+ |\fett{q}-\fett{q}'|)/2$.

Since the dynamical variables, $\bar{f}_l(|\fett{q}'|,\eta)$ and $\bar{f}_k(|\fett{l}'|,\eta)$, are independent of $\cos \theta$, we proceed to integrate the rest of the integrand~$(\cdots)$ in $\dd \cos \theta$.   This can be accomplished first by a change of  variables from $\cos \theta$ to $P \equiv |\fett{q}-\fett{q}'| = \sqrt{|\fett{q}|^2- 2 |\fett{q}| |\fett{q}'| \cos \theta + |\fett{q}'|^2}$, so that
\begin{equation}
\int_{-1}^{1} \dd \cos \theta \, (\cdots) \to \frac{1}{|\fett{q}| \, |\fett{q}'|} \int_{||\fett{q}|-|\fett{q}'||}^{|\fett{q}|+|\fett{q}'|} \dd P\, P \, (\cdots).
\end{equation}
Secondly, we rewrite the integration limits of~(\ref{eq:dgeneral}) in terms of step functions; in particular,
\begin{equation}
\int_{||\fett{q}|-|\fett{q}'||}^{|\fett{q}|+|\fett{q}'|} \dd P \int_{R_+}^\infty \dd |\fett{l}'| \to \Theta \left(P-||\fett{q}|-|\fett{q}'|| \right) \Theta\left(|\fett{q}|+|\fett{q}'|-P\right)
\Theta\left(2 |\fett{l}'|-|\fett{q}|+|\fett{q}'|-P \right).
\end{equation}
Then, using the relations
\begin{equation}
\begin{aligned}
&\Theta \left(P-||\fett{q}|-|\fett{q}'|| \right) \Theta\left(2 |\fett{l}'|-|\fett{q}|+|\fett{q}'|-P \right) = \\
&\hspace{10mm} \Theta\left(|\fett{q}|- |\fett{q}'| \right) \Theta \left(P-|\fett{q}|+|\fett{q}'| \right)  \Theta \left(|\fett{l}'|-|\fett{q}|+|\fett{q}'| \right)
+ \Theta\left(|\fett{q}'|- |\fett{q}| \right) \Theta \left(P-|\fett{q}'|+|\fett{q}| \right), \\
&\Theta \left(|\fett{q}|+|\fett{q}'|-P \right)\Theta\left(2 |\fett{l}'|-|\fett{q}|+|\fett{q}'|-P \right) =\\
&\hspace{20mm} \Theta \left( |\fett{l}'|-|\fett{q}|\right)\Theta \left(|\fett{q}|+|\fett{q}'|-P \right) +
\Theta\left(|\fett{q}|-|\fett{l}'| \right) \Theta\left(2 |\fett{l}'|-|\fett{q}|+|\fett{q}'|-P \right) ,
\end{aligned}
\end{equation}
we find that the integral~(\ref{eq:dgeneral}) splits into three parts, with new integration limits given by
\begin{equation}
\begin{aligned}
&\int_0^\infty \dd |\fett{q}'|  \int_{||\fett{q}|-|\fett{q}'||}^{|\fett{q}|+|\fett{q}'|} \dd P \int_{R_+}^\infty \dd |\fett{l}'| \to 
\\ & \hspace{15mm} 
\int_0^{|\fett{q}|} \dd |\fett{q}'|  \left[\int_{|\fett{q}|-|\fett{q}'|}^{|\fett{q}|} \dd |\fett{l}'| 
 \int_{|\fett{q}|-|\fett{q}'|}^{2|\fett{l}'|-|\fett{q}|+|\fett{q}'|}  \dd P
+\int_{|\fett{q}|}^\infty \dd |\fett{l}'| 
 \int_{|\fett{q}|-|\fett{q}'|}^{|\fett{q}|+|\fett{q}'|}  \dd P \right] \\
&  \hspace{25mm} + \int_{|\fett{q}|}^\infty  \dd |\fett{q}'| \left[ \int_0^{|\fett{q}|} \dd |\fett{l}'| 
 \int_{|\fett{q}'|-|\fett{q}|}^{2|\fett{l}'|-|\fett{q}|+|\fett{q}'|}  \dd P +\int_{|\fett{q}|}^\infty \dd |\fett{l}'| 
 \int_{|\fett{q}'|-|\fett{q}|}^{|\fett{q}|+|\fett{q}'|}  \dd P \right],
 \label{eq:newlimits}
 \end{aligned}
 \end{equation}
so that the $P$-integration can now be performed before the $|\fett{l}'|$-integration.

Applying this new integration procedure~(\ref{eq:newlimits}) to the kernels of equations~(\ref{C1_final}), (\ref{C1_ann}), (\ref{eq:C1b_elastic}) and (\ref{eq:C1a_elastic}), we obtain 
\begin{equation}
\begin{aligned}
{\cal D}^{\nu \nu \leftrightarrow \nu \nu}_1[f] =& \frac{3\mathfrak{g}^4}{(2\pi)^{3}|\fett{q}|^2} \sum_{n=1}^4 \int_{ {\cal I}_n} \!  \dd |\fett{q}'| \, \dd |\fett{l}'| \, \bar{f}_{\nu}(|\fett{q}'|,\eta)\, \bar{f}_{\nu}(|\fett{l}'|,\eta) \, k^0_n(|\fett{q}|,|\fett{q}'|,|\fett{l}'|), \\
{\cal D}^{\nu \nu \leftrightarrow \phi \phi}_1[f] =& \! - \! \frac{\mathfrak{g}^4}{(2\pi)^{3}|\fett{q}|^2} \sum_{n=1}^4 \int_{ {\cal I}_n} \!  \dd |\fett{q}'| \, \dd |\fett{l}'| \, \bar{f}_{\phi}(|\fett{q}'|,\eta)\, \bar{f}_{\phi}(|\fett{l}'|,\eta)  \left( \frac{5}{8} k^0_n -k^u_n -k^t_n \right)(|\fett{q}|,|\fett{q}'|,|\fett{l}'|), \\
{\cal D}^{\nu \phi \leftrightarrow \nu \phi}_1[f] =& \frac{4\mathfrak{g}^4}{(2\pi)^{3}|\fett{q}|^2} \sum_{n=1}^4 \int_{ {\cal I}_n} \! \dd |\fett{q}'| \, \dd |\fett{l}'| \, \bar{f}_{\nu}(|\fett{q}'|,\eta)\, \bar{f}_{\phi}(|\fett{l}'|,\eta) \, \left( k^0_n -k^u_n +k^s_n \right)(|\fett{q}|,|\fett{q}'|,|\fett{l}'|),
\label{D1_zeroth}
\end{aligned}
\end{equation}
where the integral kernels are
\begin{equation}
\begin{aligned}
k^0_{n=1,3}(|\fett{q}|,|\fett{q}'|,|\fett{l}'|)= & 2|\fett{l}'|-|\fett{q}|+|\fett{q}'|-||\fett{q}|-|\fett{q}'||, \\
k^0_{n=2,4}(|\fett{q}|,|\fett{q}'|,|\fett{l}'|)=  & |\fett{q}|+|\fett{q}'|-||\fett{q}|-|\fett{q'}||, \\
k^s_{n=1,3}(|\fett{q}|,|\fett{q}'|,|\fett{l}'|)=& \frac{|\fett{l}'|(|\fett{l'}|-|\fett{q}|+|\fett{q}'|)}{2(|\fett{l}'|+|\fett{q'}|)}, \quad
k^s_{n=2,4}(|\fett{q}|,|\fett{q}'|,|\fett{l}'|)= \frac{|\fett{q}||\fett{q}'|}{2(|\fett{l}'|+|\fett{q'}|)}, \\
k^u_{n=1,3}(|\fett{q}|,|\fett{q}'|,|\fett{l}'|)= & \frac{|\fett{l}'|(|\fett{l}'|-|\fett{q}|+|\fett{q}'|)}{2\sqrt{(|\fett{l}'|-|\fett{q}|)^2+\tilde{m}^2_{\nu}}}, \quad
k^u_{n=2,4}(|\fett{q}|,|\fett{q}'|,|\fett{l}'|)= \frac{|\fett{q}||\fett{q}'|}{2\sqrt{(|\fett{l}'|-|\fett{q}|)^2+\tilde{m}^2_{\nu}}}, \\
k^t_{n=1,3}(|\fett{q}|,|\fett{q}'|,|\fett{l}'|)= & \frac{1}{8} \left( 2|\fett{l}'|-|\fett{q}|+|\fett{q}'| \right) \left( |\fett{q}| +|\fett{q}'| \right) \left( \frac{1}{|\fett{q}|-|\fett{q}'|-2|\fett{l}'|} +\frac{1}{||\fett{q}|-|\fett{q}'||} \right), \\
k^t_{n=2,4}(|\fett{q}|,|\fett{q}'|,|\fett{l}'|)= & \frac{1}{8} \left( 2|\fett{l}'|-|\fett{q}|+|\fett{q}'| \right) \left( |\fett{q}| +|\fett{q}'| \right) \left(\frac{1}{||\fett{q}|-|\fett{q}'||}  -\frac{1}{|\fett{q}|+|\fett{q}'|}  \right), 
\label{eq:IntKernelsZero}
\end{aligned}
\end{equation}
and the integral domains ${\cal I}_n$ can be found in equation~(\ref{eq:domains}).
Note that to obtain ${\cal D}^{\nu \phi \leftrightarrow \nu \phi}_1[f]$ we can start with either the expression for ${\cal C}^{\nu \phi \leftrightarrow \nu \phi}_{1a}[f]$ (equation~\eqref{eq:C1a_elastic})
 or ${\cal C}^{\nu \phi \leftrightarrow \nu \phi}_{1b}[f]$ (equation~\eqref{eq:C1b_elastic}).  Both lead to the same result.

\section{Collision integral reduction: massive scalar limit}
\label{app:massive}

\subsection{First-order \texorpdfstring{$\nu \nu \leftrightarrow \nu \nu$}{nu nu -> nu nu} collision integrals}

Neutrino self-interaction is the only cosmologically relevant process in the case of a very massive scalar particle.
The neutrino collision integral can again be split into three parts,
\begin{equation}
\begin{aligned}
\left( \frac{\partial f_\nu}{\partial \eta}\right)^{(1)}_{\rm coll,m}(\fett{k},\fett{q}, \eta)
 &= \frac{2 \mathfrak{g}^4}{a^4 m_{\phi}^4 (2 \pi)^{5}|\fett{q}|} \! \int \frac{\mathrm{d^{3}} \fett{q}'}{2 | \fett{q}'|}  \int \frac{\mathrm{d^{3}}\fett{l}}{2 | \fett{l}|}  \int \frac{\mathrm{d^{3}}\fett{l}'}{2 | \fett{l}'|}  \delta_\D^{(4)}(q+l-q'-l')  \left(\tilde{s}^2+\tilde{t}^2+\tilde{u}^2 \right) \\
& \hspace{2mm} \times 
\Big[  2\bar{f}^{\rm eq}_{\nu}(| \fett{l}'|) \, F_{\nu}(\fett{k},\fett{q}',\eta)-\bar{f}^{\rm eq}_{\nu}(| \fett{q}|) \, F_{\nu}(\fett{k},\fett{l},\eta)-\bar{f}^{\rm eq}_{\nu}(| \fett{l}|) \, F_{\nu}(\fett{k},\fett{q},\eta) \Big] \\
&\equiv {\cal  C}_{1}^{\rm m}[f]+{\cal C}_{2}^{\rm m}[f]+{\cal C}_{3}^{\rm m}[f] \, .
\label{Collision Integralmassive}
\end{aligned}
\end{equation}
 In this limit, the background neutrino phase space distribution has no time dependence.


\subsubsection{Reduction of \texorpdfstring{${\cal C}_{3}^{\rm m}[f]$}{C3m}}

Here, it is convenient to rewrite the matrix element as
$\tilde{s}^2+\tilde{t}^2+\tilde{u}^2 = 2\left(\tilde{s}^2+ \tilde{t}^2 + \tilde{s} \tilde{t}\right)$,
exploiting the relation $\tilde{s}+\tilde{t}+\tilde{u}=0$ for massless incoming and outgoing particles.  Then, using the parameterisation~(\ref{k-parametrisation}) for the 3-vectors, we find
\begin{equation}
\tilde{s}^2+\tilde{t}^2+\tilde{u}^2=2\left(\tilde{s}^2+ \tilde{t}^2 + \tilde{s} \tilde{t}\right) =8 |\fett{q}|^2 |\fett{q}'|^2 \left(\cos^2 \theta + B_3 \cos \theta +C_3\right)  ,
\end{equation}
where 
\begin{equation}
\begin{aligned}
B_3 & =  -2 + \frac{|\fett{l}|}{|\fett{q}'|} (1-\cos \alpha)  ,\\
C_3 & = 1 + \frac{|\fett{l}|^2}{|\fett{q}'|^2} (1-\cos \alpha)^2 -\frac{|\fett{l}|}{|\fett{q}'|} (1-\cos \alpha)\, .
\end{aligned}
\end{equation}  
The remaining steps of the reduction of ${\cal C}_3^{\rm m}[f]$ are  then functionally identical to those applied to its ``massless'' counterpart ${\cal C}_3^0[f]$ outlined in section~\ref{Computing C_3[f]}
up to equation~(\ref{C4 after dbeta}):
\begin{equation}
\begin{aligned}
{\cal C}_3^{\rm m}[f]&=-\frac{4 \mathfrak{g}^4|\fett{q}|}{a^4 m_\phi^4(2\pi)^{4}}F_{\nu}(\fett{k},\fett{q})\int \,   \mathrm{d\cos\alpha} \, \mathrm{d|\fett{l}|} \, \mathrm{d|\fett{q}'|} \, \Theta(|\fett{q}|+|\fett{l}|-|\fett{q}'|) \, \frac{|\fett{q}'|^2 |\fett{l}|}{|\fett{l}+\fett{q}|} \, \bar{f}^{\rm eq}_{\nu}(| \fett{l}|) \\
&\hspace{15mm}\times \int_{x_-}^{x_+} \dd \cos \theta \frac{\cos^2 \theta + B_3 \cos \theta +C_3}{\sqrt{(x_+-\cos \theta)(\cos \theta - x_-)}} \Theta(b_3^2-4 a_3 c_3) \, ,
\label{eq:blah}
\end{aligned}
\end{equation}
where $a_3,b_3,c_3$ are given in equation~(\ref{eq:abc}), and the roots~$x_\pm$ in equation~(\ref{eq:roots}).

The $\cos \theta$-integral has solution
\begin{equation}
\begin{aligned}
\label{eq:elliptic}
\int_{x_-}^{x_+} \mathrm{d}x \frac{ x^2+ B_3 x + C_3 }{\sqrt{(x_+-x)(x-x_-)}} & = \pi \left[
\frac{3b_3^2-4a_3c_3}{8 a_3^{2}}
- \frac{B_3 b_3}{2 a_3}
+C_3
\right] \, .
\end{aligned}
\end{equation}
Then, equation~(\ref{eq:blah}) can be written as
\begin{equation}
\label{eq:c3massive}
{\cal C}_3^{\rm m}[f]=-\frac{2 \mathfrak{g}^4}{a^4 m_\phi^4(2\pi)^{3}}F_{\nu}(\fett{k},\fett{q})\int    \mathrm{d\cos\alpha} \, \mathrm{d|\fett{l}|}  \, \bar{f}^{\rm eq}_{\nu}(| \fett{l}|) \, \Omega (|\fett{q}|, |\fett{l}|, \cos \alpha) \, ,
\end{equation}
where the integration kernel is
\begin{equation}
\begin{aligned}
\Omega (|\fett{q}|, |\fett{l}|, \cos \alpha) = \frac{|\fett{q}|  |\fett{l}|}{|\fett{l}+\fett{q}|} \int^{R_2}_{R_1} \!  \mathrm{d|\fett{q}'|}  |\fett{q}'|^2  \left[
\frac{3b_3^2-4a_3c_3}{8 a_3^{2}}
- \frac{B_3 b_3}{2 a_3}
+C_3
\right] = \frac{5 |\fett{q}| |\fett{l}|^3}{6} (1-\cos \alpha)^2,
\label{eq:k3}
\end{aligned}
\end{equation}
using the integration limits $R_{1,2} \equiv (|\fett{q}|+|\fett{l}|\pm |\fett{q}+\fett{l}|)/2$.
Integrating  equation~(\ref{eq:c3massive}) now in $\dd \cos \alpha$ over the interval~$[-1,1]$  gives
\begin{equation}
{\cal C}_3^{\rm m}[f]=-\frac{40 \mathfrak{g}^4|\fett{q}|}{9 a^4 m_\phi^4(2\pi)^{3}}F_{\nu}(\fett{k},\fett{q})\int \,   \mathrm{d|\fett{l}|}  \, \bar{f}^{\rm eq}_{\nu}(| \fett{l}|) \, 
|\fett{l}|^3 = -\frac{80  \mathrm{N} \mathfrak{g}^4  |\fett{q}|\, T_{\nu,0}^4}{ 3 a^4 m_\phi^4(2\pi)^{3}} \, F_{\nu}(\fett{k},\fett{q}),
\end{equation}
where in the last equality we have further performed the $|\fett{l}|$-integration using a Maxwell--Boltzmann distribution for 
$\bar{f}^{\rm eq}_\nu(\fett{l})$.


\subsubsection{Reduction of \texorpdfstring{${\cal C}_2^{\rm m}[f]$}{C2m}}

Equation~(\ref{eq:c3massive}) holds also for ${\cal C}_2^{\rm m}[f]$, except for the replacements $\bar{f}_{\nu}(|\fett{l}|) \to \bar{f}_{\nu}(|\fett{q}|)$ and $F_{\nu}(\fett{k},\fett{q}) \to F_{\nu}(\fett{k},\fett{l})$, i.e.,
\begin{equation}
\label{eq:c2massive}
{\cal C}_2^{\rm m}[f]=-\frac{5 \mathrm{N} \mathfrak{g}^4 }{3 a^4 m_\phi^4(2\pi)^{3}} \, \e^{-|\fett{q}|/T_{\nu,0}} 
\int \,   \mathrm{d\cos\alpha} \, \mathrm{d|\fett{l}|} \, 
|\fett{q}|\,  |\fett{l}|^3\, (1-\cos \alpha)^2 \; F_{\nu}(\fett{k},\fett{l}) \, ,
\end{equation}
where we have again  used a Maxwell--Boltzmann background distribution.


\subsubsection{Reduction of \texorpdfstring{${\cal C}_1^{\rm m}[f]$}{C1m}}

Using the parameterisation~(\ref{param2}) for the 3-vectors, we find it convenient to rewrite the matrix element here as
\begin{equation}
\tilde{s}^2+\tilde{t}^2+\tilde{u}^2 = 2\left(\tilde{t}^2+ \tilde{u}^2 + \tilde{u} \tilde{t}\right)=8|\fett{q}|^2|\fett{l}'|^2\left(\cos^2 \alpha + B_1 \cos \alpha +C_1\right) \, ,
\end{equation}
where 
\begin{equation}
\begin{aligned}
B_1 & = - 2 -\frac{|\fett{q}'|}{|\fett{l}'|}  (1-\cos \theta) \, ,\\
C_1 & = 1+ \frac{|\fett{q}'|^2}{|\fett{l}'|^2} (1-\cos \theta)^2 + \frac{|\fett{q}'|}{ |\fett{l}'|} (1-\cos \theta)\, .
\end{aligned}
\end{equation}  
Then, the reduction of ${\cal C}_1^{\rm m}[f]$ proceeds in the same way as for its ``massless'' counterpart ${\cal C}_1^0[f]$ up to 
equation~(\ref{C2 after dcosalpha}):
\begin{equation}
\begin{aligned}
{\cal C}_1^{\rm m}[f]&= \frac{8 \mathfrak{g}^4|\fett{q}|}{a^4 m_\phi^4 (2\pi)^{4}}\int \, \mathrm{d \cos\theta} \,   \mathrm{d|\fett{l}'|} \, \mathrm{d|\fett{q}'|} \, \Theta(|\fett{q}'|+|\fett{l}'|-|\fett{q}|) \, \frac{ |\fett{q}'|\, |\fett{l}'|^2}{|\fett{q}'-\fett{q}|}  \, \bar{f}_{\nu}(| \fett{l}'|) \, F_{\nu}(\fett{k},\fett{q}') \\
&\hspace{27mm}\times \int_{y_-}^{y_+} \dd \cos \alpha \; 
 \frac{ \cos^2 \alpha + B_1 \cos \alpha +C_1}{\sqrt{(y_+-\cos \alpha)(\cos \alpha -y_-)}} \Theta(b_1^2-4 a_1 c_1) \, ,
 \label{eq:c1massive0}
\end{aligned}
\end{equation}
where $a_1,b_1,c_1$ are the coefficients of the quadratic~(\ref{eq:dgdb1}) with  real and degenerate roots~$y_\pm$.

Integrating equation~(\ref{eq:c1massive0})  in $\dd \cos \alpha$ using the relation~(\ref{eq:elliptic}) yields
\begin{equation}
{\cal C}_1^{\rm m}[f]=\frac{4\mathrm{N} \mathfrak{g}^4}{a^4 m_\phi^4 (2\pi)^{3}|\fett{q}|}\int \, \mathrm{d \cos\theta} \, \mathrm{d|\fett{q}'|} \, |\fett{q}'|  \, K^{\rm m}(|\fett{q}|, |\fett{q}'|, \cos \theta) \,
F_{\nu}(\fett{k},\fett{q}')\, ,
 \label{eq:c1massive2}
\end{equation}
where, given $R_+=(|\fett{q}|-|\fett{q}'|+ |\fett{q}-\fett{q}'|)/2$, the integration kernel is 
\begin{equation}
\begin{aligned}
K^{\rm m}(|\fett{q}|, |\fett{q}'|, \cos \theta) \equiv& \frac{|\fett{q}|^2  }{|\fett{q}-\fett{q}'|} \int^\infty_{R_+}   \mathrm{d|\fett{l}'|} \,  |\fett{l}'|^2\;
\e^{ -|\fett{l}'|/T_{\nu,0}}
\left[\frac{(3b_1^2-4a_1c_1)}{8 a_1^{2}}
-  \frac{B_1b_1}{2 a_1}
+C_1
\right], \\
= & \frac{1}{16 P^5}\; \e^{-(Q_-+P)/(2 T_{\nu,0})} \; T_{\nu,0}  \left(Q_-^2-P^2 \right)^2 \\
& \times \Big[ P^2 \left(3 P^2 - 2 PT_{\nu,0}-4 T^2_{\nu,0}\right)+ Q_+^2 \left( P^2 + 6 P T_{\nu,0} + 12 T_{\nu,0}^2 \right)
 \Big],
\label{eq:k1}
\end{aligned}
\end{equation}
and we have defined $P \equiv |\fett{q}-\fett{q}'|$ and $Q_\pm \equiv |\fett{q}|\pm |\fett{q}'|$.


\section{Legendre decomposition}
\label{app:legendre}

A  subset of first-order collision integrals comes in the general form
\begin{equation}
\begin{aligned}
\left( \frac{\partial f}{\partial \eta}\right)^{(1)}(\fett{k},\fett{q},\eta) 
=  \int  \dd\cos\theta \, \dd|\fett{q}'|\, |\fett{q}'|\, K (|\fett{q}|,|\fett{q}'|,\cos \theta,\eta) \, F(\fett{k},\fett{q}',\eta).
\label{eq:LPexample}
\end{aligned}
\end{equation}
To perform a Legendre decomposition, we first write $F_{j}(\fett{k},\fett{q}')$ as a Legendre series in $\LP (\hat{k} \cdot \hat{q}')$  following equation~(\ref{eq:legendre}).
Then, decomposing equation~(\ref{eq:LPexample}) in terms of $\LP (\hat{k} \cdot \hat{q})$, we find
\begin{equation}
\begin{aligned}
\label{eq:c2ugly}
& \frac{\ii^\ell}{2}  \int_0^{2 \pi} \frac{\dd \psi}{2 \pi}\int^1_{-1} \dd \cos \epsilon \, \LP_\ell (\cos \epsilon)\, 
\left( \frac{\partial f}{\partial \eta}\right)^{(1)}
 =  \int \dd |\fett{q}'| \,  |\fett{q}'|
  \sum_{\ell'=0}^\infty 
  \ii^\ell (-\ii)^{\ell'} (2 \ell'+1)\,  F_{\ell'}(|\fett{k}|,|\fett{q}'|) 
  \\&\hspace{40mm} \times 
\frac{1}{4\pi}  \int \dd \Omega \int_{-1}^{1} \dd \cos \theta \;   K(|\fett{q}|,|\fett{q}'|,\cos \theta)\,  \LP_\ell(\hat{ k}\cdot \hat{ q})\,  \LP_{\ell'}(\hat{ k} \cdot \hat{ q}') \, ,
 \end{aligned}
 \end{equation}
where $\int \dd \Omega \equiv \int_0^{2 \pi} \dd \psi \int_{-1}^1 \dd \cos \epsilon$, and 
the integration $\int_0^{2 \pi} \dd \psi/(2 \pi)$ has been added to eliminate any residual dependence on the  azimuthal angle between $\fett{k}$ and $\fett{q}$.  See discussion in section~\ref{sec:hierarchy}.

To simplify equation~(\ref{eq:c2ugly}) we apply on $\LP_\ell(\hat{ k}\cdot \hat{ q})$ and $\LP_{\ell'}(\hat{ k} \cdot \hat{ q}')$
the relation 
\begin{equation}
\label{eq:sphericalharmonics}
\LP_\ell(\hat{ x}\cdot \hat{ y}) = \frac{4 \pi}{2 \ell + 1}  \sum_{m=-\ell}^\ell Y_{\ell m} (\hat{{x}}) Y^*_{\ell m} (\hat{{y}}) \,,
\end{equation}
where $Y_{\ell m}(\hat{ x})$ is a spherical harmonic, followed by the orthogonality condition
\begin{equation}
\label{eq:orthogonality}
\int \dd \Omega \; Y_{\ell m}(\hat{ x}) \, Y^*_{\ell' m'}(\hat{ x}) = \delta_{\ell \ell'} \delta_{m m'} \,.
\end{equation}
Then, we can write equation~(\ref{eq:c2ugly}) as
\begin{equation}
\begin{aligned}
 \frac{\ii^\ell}{2}  \int_0^{2 \pi} \frac{\dd \psi}{2 \pi}\int^1_{-1} \dd \cos \epsilon \, \LP_\ell (\cos \epsilon)\, 
 \left( \frac{\partial f}{\partial \eta}\right)^{(1)} = \int \dd |\fett{q}'|  \, |\fett{q}'|\,  K_\ell (|\fett{q}|,|\fett{q}'|, \eta)\, F_{\ell} (|\fett{k}|,|\fett{q}'|,\eta) ,
  \end{aligned}
 \end{equation}
where
\begin{equation}
K_{\ell}(|\fett{q}|,|\fett{q}'|,\eta) \equiv  \int_{-1}^1 \dd \cos \theta \,
   K (|\fett{q}|,|\fett{q}'|,\cos \theta,\eta) \,  \LP_\ell(\cos \theta)\ 
   \label{eq:clapp}
\end{equation}
is $2 \ii^{-\ell}$ times  the Legendre transform of the kernel function~$K$.



\end{document}